\documentclass[preprint]{aastex}
\usepackage{natbib,graphicx,latexsym}

\newcommand{\Spitzer}{{\sl Spitzer}}
\newcommand{\Hipparcos}{{\sl Hipparcos}}
\newcommand{\HST}{{\sl HST}}

\newcommand{\ROSAT}{{\sl ROSAT}}
\newcommand{\Msun}{\mbox{$M_{\sun}$}}

\newcommand{\Lsun}{\mbox{$L_{\sun}$}}
\newcommand{\Rsun}{\mbox{$R_{\sun}$}}
\newcommand{\Mjup}{\mbox{$M_{\rm Jup}$}}

\newcommand{\degree}{\mbox{$^{\circ}$}}
\newcommand{\perpix}{\mbox{pixel$^{-1}$}}

\newcommand{\kms}{\hbox{km~s$^{-1}$}}

\newcommand{\Hc}{\mbox{$H_{\rm cont}$}}
\newcommand{\Kc}{\mbox{$K_{\rm cont}$}}
\newcommand{\Kp}{\mbox{$K^{\prime}$}}
\newcommand{\Ks}{\mbox{$K_S$}}
\newcommand{\Mtot}{\mbox{$M_{\rm tot}$}}
\newcommand{\Lbol}{\mbox{$L_{\rm bol}$}}

\newcommand{\Teff}{\mbox{$T_{\rm eff}$}}
\newcommand{\logg}{\mbox{$\log(g)$}}

\newcommand{\Lp}{\mbox{${L^\prime}$}}
\newcommand{\Mp}{\mbox{${M^\prime}$}}

\newcommand{\lhsint}{\hbox{LHS~1901}}
\newcommand{\lhsAB}{\hbox{LHS~1901AB}}
\newcommand{\lhsA}{\hbox{LHS~1901A}}
\newcommand{\lhsB}{\hbox{LHS~1901B}}
\newcommand{\lpint}{\hbox{LP~349-25}}
\newcommand{\lpAB}{\hbox{LP~349-25AB}}
\newcommand{\lpA}{\hbox{LP~349-25A}}
\newcommand{\lpB}{\hbox{LP~349-25B}}
\newcommand{\glint}{\hbox{Gl~569B}}
\newcommand{\glAB}{\hbox{Gl~569Bab}}
\newcommand{\glA}{\hbox{Gl~569Ba}}
\newcommand{\glB}{\hbox{Gl~569Bb}}
\newcommand{\orbit}{\hbox{\tt ORBIT}}

\shorttitle{Dynamical Masses of Late-M Dwarfs}
\shortauthors{Dupuy~et~al.}

\begin{document}

\title{Studying the Physical Diversity of Late-M Dwarfs with Dynamical
  Masses\altaffilmark{*,\dag,\ddag}}

\author{Trent J.\ Dupuy,\altaffilmark{1}
        Michael C.\ Liu,\altaffilmark{1} 
        Brendan P.\ Bowler,\altaffilmark{1}
        Michael C.\ Cushing,\altaffilmark{2} \\
        Christiane Helling,\altaffilmark{3}
        Soeren Witte,\altaffilmark{4}
        and Peter Hauschildt\altaffilmark{4}}

      \altaffiltext{*}{Some of the data presented herein were obtained
        at the W.\ M.\ Keck Observatory, which is operated as a
        scientific partnership among the California Institute of
        Technology, the University of California, and the National
        Aeronautics and Space Administration. The Observatory was made
        possible by the generous financial support of the W.\ M.\ Keck
        Foundation.}

      \altaffiltext{\dag}{Based partly on observations obtained at the
        Canada-France-Hawaii Telescope (CFHT) which is operated by the
        National Research Council of Canada, the Institut National des
        Sciences de l'Univers of the Centre National de la Recherche
        Scientifique of France, and the University of Hawaii.}
      
      \altaffiltext{\ddag}{Based partly on observations made with ESO
        Telescopes at the Paranal Observatory under program IDs
        073.C-0155, 077.C-0783, and 077.C-0441.}

      \altaffiltext{1}{Institute for Astronomy, University of Hawai`i,
        2680 Woodlawn Drive, Honolulu, HI 96822}

      \altaffiltext{2}{NASA Jet Propulsion Laboratory, 4800 Oak Grove
        Drive, Mail Stop 264-723, Pasadena, CA 91109}

      \altaffiltext{3}{SUPA, School of Physics and Astronomy,
        University of St.\ Andrews, North Haugh, St.\ Andrews KY16
        9SS, UK}

      \altaffiltext{4}{Hamburger Sternwarte, Gojenbergsweg 112, 21029
        Hamburg, Germany}

\begin{abstract}

  We present a systematic study of the physical properties of late-M
  dwarfs based on high-quality dynamical mass measurements and
  near-infrared (NIR) spectroscopy. We use astrometry from Keck
  natural and laser guide star adaptive optics imaging to determine
  orbits for the late-M binaries \lpAB\ (M7.5+M8), \lhsAB\
  (M6.5+M6.5), and \glAB\ (M8.5+M9).  We find that \lpAB\ ($\Mtot =
  0.120^{+0.008}_{-0.007}~\Msun$) is a pair of young brown dwarfs for
  which Lyon and Tucson evolutionary models jointly predict an age of
  $140\pm30$~Myr, consistent with the age of the Pleiades.  However,
  at least the primary component seems to defy the empirical Pleiades
  lithium depletion boundary, implying that the system is in fact
  older (if the parallax is correct) and that evolutionary models
  underpredict the component luminosities for this magnetically active
  binary. We find that \lhsAB\ is a pair of very low-mass stars
  ($\Mtot = 0.194^{+0.025}_{-0.021}~\Msun$) with evolutionary
  model-derived ages consistent with the old age ($>6$~Gyr) implied by
  its lack of activity.  Our improved orbit for \glAB\ results in a
  higher mass for this binary ($\Mtot =
  0.140^{+0.009}_{-0.008}~\Msun$) compared to previous work
  \citep[$0.125\pm0.007~\Msun$][]{2006ApJ...644.1183S}.
  We use these mass measurements along with our published results for
  2MASS~J2206$-$2047AB (M8+M8) to test four sets of ultracool model
  atmospheres currently in use.  Fitting these models to our NIR
  integrated-light spectra provides temperature estimates warmer by
  $\approx$250~K than those derived independently from Dusty
  evolutionary models given the measured masses and luminosities.  We
  propose that model atmospheres are more likely to be the source of
  this discrepancy, as it would be difficult to explain a uniform
  temperature offset over such a wide range of masses, ages, and
  activity levels in the context of evolutionary models.  This
  contrasts with the conclusion of \citet{qk10} that model-predicted
  masses (given input \Teff\ and \Lbol) are at fault for differences
  between theory and observations.  In addition, we find an opposite
  (and smaller) mass discrepancy from what they report when we adopt
  their model-testing approach: masses are too high rather than too
  low because our \Teff\ estimates derived from fitting NIR spectra
  are $\approx$650~K higher than their values from fitting broadband
  photometry alone.

\end{abstract}



\section{Introduction}

While the spectra of late-M dwarfs are by definition very similar,
their underlying nature can vary widely from youthful brown dwarfs to
old low-mass stars. The first field dwarf with a spectral type later
than M6 was discovered by \citet{1944AJ.....51...61V}, yet even 50
years later the nature of such objects was open to debate, with
\citet{1994ApJS...94..749K} suggesting that the latest type M dwarfs
may all be young enough to be substellar. \citet{1992ApJ...389L..83R}
proposed a method for discriminating between brown dwarfs and low-mass
stars by using the \ion{Li}{1} doublet at 6708~\AA, as models predict
that stars and the highest mass brown dwarfs ($M \gtrsim 0.06~\Msun$)
should deplete their initial lithium rapidly. Such model predictions
of lithium depletion have also been used to age-date open clusters
\citep[e.g.,][]{1998ApJ...499L.199S, 1999ApJ...522L..53B}. However,
the underlying theory still remains unconstrained by direct mass
measurements for ultracool dwarfs with lithium measurements near the
predicted depletion boundary.  In recent years, spectroscopic
signatures of low surface gravity have been used to identify very
young ($\sim$10--100~Myr) ultracool dwarfs among the field population
of late-M and L dwarfs \citep{2004ApJ...600.1020M,
  2007ApJ...657..511A, 2008ApJ...689.1295K, 2010ApJS..186...63R}, thus
enabling a means of discriminating between stars and brown dwarfs if
the objects are young enough.

Dynamical mass measurements provide a direct means for studying the
diversity in the physical properties of late-M dwarfs. We present here
masses for the late-M binaries \lpAB, \lhsAB, and \glAB\ based on Keck
natural guide star (NGS) and laser guide star (LGS) adaptive optics
(AO) imaging from our ongoing orbital monitoring program targeting
ultracool binaries. We combine our new measurements with published
results to perform a systematic study of late-M dwarfs.  To that end,
we also present medium-resolution ($R \approx 2000$) near-infrared
(NIR) spectra for this sample of late-M binaries.  The resulting
combination of high-quality dynamical masses and spectra enables
strong tests of model atmospheres over a wide range of masses, ages,
and activity levels.

LP~349-25 was first identified as a nearby M8 dwarf by
\citet{2000AJ....120.1085G}, and \citet{2005A&A...435L...5F}
discovered its binary nature after being initially unresolved by
\citet{2003ApJ...587..407C}.  \citet{2009AJ....137..402G} measured a
trigonometric parallax of $75.8\pm1.6$~mas for \lpint, corresponding
to a distance of $13.2\pm0.3$~pc, which is somewhat more distant than
the photometric estimate of $10.1\pm1.2$~pc
\citep{2005A&A...435L...5F}.  \citet{2009ApJ...705.1416R} obtained
high-resolution integrated-light spectroscopy using Keck/HIRES and
found no evidence for lithium (equivalent width $\lesssim0.5$~\AA).
\citet{qk10} recently published a dynamical mass of
$0.122\pm0.009$~\Msun\ based on Keck LGS AO data. We derive here a
dynamical mass of $0.120_{-0.007}^{+0.008}$~\Msun, based on an
independent set of astrometry.  Thus, this binary is a pair of brown
dwarfs.

LHS~1901 was first identified as a nearby M6.5 dwarf by
\citet{2003AJ....126.3007R}, and \citet{2009AJ....137.4109L} have
independently classified it as M7.0. \citet{2006A&A...460L..19M}
discovered \lhsint\ to be a $0\farcs2$ binary, which was also later
identified by \citet{2008MNRAS.384..150L}.
\citet{2009AJ....137.4109L} measured the parallax of \lhsint\ to be
$77.8\pm3.0$~mas, corresponding to a distance of $12.9\pm0.5$~pc. Our
dynamical mass of $0.194^{+0.025}_{-0.021}$~\Msun\ indicates that this
is a pair of very low-mass stars in a highly eccentric orbit ($e =
0.830\pm0.005$).

Gl~569B was discovered as a companion to the chromospherically active
M2 star Gl~569 by \citet{1988ApJ...330L.119F}, and
\citet{2000ApJ...529L..37M} resolved it as an M8.5+M9 binary using
Keck AO.  There are three published orbit determinations for this
binary \citep{2004astro.ph..7334O, 2006ApJ...644.1183S, qk10} that all
give consistent values of the total mass ($0.125\pm0.007$~\Msun).
However, these orbits are all based largely on astrometry from 1999 to
2001 obtained with the Keck NIR cameras KCAM and SCAM (the
slit-viewing camera for NIRSPEC). Since these instruments are not
astrometrically well-calibrated, we have derived a new orbit for
\glAB\ based solely on data from Keck/NIRC2 and the \textsl{Hubble
  Space Telescope} (\HST) Space Telescope Imaging Spectrograph
(STIS). In addition, we use the revised \Hipparcos\ parallax for
Gl~569A of $103.59\pm1.72$~mas \citep{2007hnrr.book.....V} rather than
the original value of $101.91\pm1.67$~mas \citep{1997ESASP1200.....P}
used in all previous orbit determinations. Our improved measurement of
the total dynamical mass for \glAB\ is
$0.140^{+0.009}_{-0.008}$~\Msun, indicating that the binary is more
massive than previously thought but still with at least one
unambiguously substellar component.

In addition to these three binaries, we include the M8+M8 binary
2MASS~J22062280$-$2047058AB (hereinafter 2MASS~J2206$-$2047AB) in our
study as it also has a well-determined mass
\citep[$0.15^{+0.05}_{-0.03}$~\Msun;][]{me-2206} and integrated-light
spectroscopy.\footnote{The only other late-M dwarf binary with a
  precise dynamical mass is LHS~1070BC \citep[$\Mtot =
  0.157\pm0.009~\Msun$][]{2001A&A...367..183L, 2008A&A...484..429S},
  but it is only $1\farcs5$ from the bright primary LHS~1070A (M5.5)
  making it difficult to obtain spectroscopy of the late-M dwarfs.  We
  also do not consider four other late-M dwarf binaries from
  \citet{qk10}, as their dynamical masses are poorly constrained with
  60--190\% errors.}


\section{Observations \label{sec:obs}}

\subsection{Keck/NIRC2 AO \label{sec:keck}}

We have monitored \lpAB, \lhsAB, and \glAB\ with the AO system at the
Keck~II Telescope on Mauna Kea, Hawaii, using the facility NIR camera
NIRC2 in its narrow field-of-view mode. At each epoch, we obtained
data in one or more filters covering the standard atmospheric windows
from the Mauna Kea Observatories (MKO) filter consortium
\citep{mkofilters1, mkofilters2}. For \lpAB, we used LGS AO
\citep{2006PASP..118..297W, 2006PASP..118..310V} when observing
conditions allowed, and NGS AO otherwise (see Table~\ref{tbl:obs}).
The LGS brightness, as measured by the flux incident on the AO
wavefront sensor, was equivalent to a $V$~$\approx$~9.5--10.4~mag
star. The tip-tilt correction and quasi-static changes in the image of
the LGS as seen by the wavefront sensor were measured
contemporaneously by a second, lower bandwidth wavefront sensor
monitoring \lpint, which saw the equivalent of an
$R$~$\approx$~14.3--14.6~mag star. For \lhsAB, we used NGS AO
\citep{2000PASP..112..315W, 2004ApOpt..43.5458V}, and the incident
flux on the wavefront sensor from \lhsint\ was equivalent to a
$V$~$\approx$~13.0--13.4~mag star.  For \glAB, the primary star
Gl~569A ($R = 9.4$~mag) located 5$\farcs$0 from the science target
provided the NGS AO correction.

Our procedure for obtaining, reducing, and analyzing our images is
described in detail in our previous work
\citep[e.g.,][]{2009ApJ...692..729D, me-2397a}. Table~\ref{tbl:obs}
summarizes our observations of \lpAB, \lhsAB, and \glAB.  The binary
separation, position angle (PA), and flux ratio were determined using
the same three-component Gaussian representation of the point-spread
function (PSF) as described in \citet{me-2397a}.  At epochs where the
binary was sufficiently well-separated, we were also able to use the
StarFinder software package \citep{2000A&AS..147..335D} to
simultaneously solve for the PSF and binary parameters.  As described
in \citet{me-2397a}, we assessed systematic errors in both PSF-fitting
procedures by applying them to artificial binary images constructed
from images of PSF reference stars with similar FWHM and Strehl.  We
found good agreement between both methods, with StarFinder typically
giving equivalent or smaller errors.  We adopted the astrometric
calibration of \citet{2008ApJ...689.1044G}, with a pixel scale of
9.963$\pm$0.005~mas~\perpix\ and an orientation for the detector's
$+y$-axis of $+$0$\fdg$13$\pm$0$\fdg$02 east of north.  We applied the
distortion correction developed by B. Cameron (2007, private
communication), which changed our astrometry below the 1$\sigma$
level.  The resulting relative astrometry and flux ratios for \lpAB,
\lhsAB, and \glAB\ are given in Table~\ref{tbl:astrom}.

For \glAB, we also used our Keck images to measure relative photometry
between the primary star Gl~569A and \glAB\ at $JHK$, as there is no
photometry published in the MKO system.  We summed the flux in
circular apertures with radii of 0$\farcs$3 to 0$\farcs$5.  We used
NIRC2 in subarray mode to reduce the minimum allowed exposure time and
thus avoid saturating the primary, which limited the size of the
largest aperture we could use.  However, in experimenting with even
smaller apertures we found the same relative photometry to within
1$\sigma$.  We used a portion of the array minimally affected by light
from either the bright primary star or the binary to measure the
median sky level, which was subtracted from our summed aperture
fluxes.  The relative flux of the binary in integrated light compared
to Gl~569A was $\Delta{J} = 4.15\pm0.05$~mag, $\Delta{H} =
4.14\pm0.03$~mag, and $\Delta{K} = 3.86\pm0.03$~mag.  To compute
absolute photometry for \glint\ we applied these flux ratios to MKO
photometry for Gl~569A, which was derived by converting its Two Micron
All Sky Survey (2MASS) photometry to the MKO system using offsets
computed from synthetic photometry of a SpeX prism spectrum of Gl~569A
(A.\ Burgasser 2010, private communication.  The conversions from
2MASS to MKO for $J$, $H$, and $K$ were $-$0.03~mag, $+$0.02~mag, and
$-$0.01~mag, respectively (all $\leq 1.3\sigma$ compared to the 2MASS
errors).  Thus, the final MKO integrated-light photometry for \glint\
was $J = 10.75\pm0.06$~mag, $H = 10.15\pm0.04$~mag, $K =
9.62\pm0.03$~mag.

\subsection{\HST/STIS Imaging \label{sec:hst}}

Gl~569Bab was observed by \HST/STIS on 2002 June 26 UT as part of a
program to obtain resolved optical spectroscopy of ultracool binaries
(GO-9499; PI Mart\'{i}n).  During acquisition, the STIS CCD imager
took two images with the long-pass filter F28X50LP in which the binary
is well-detected and the primary star Gl~569A is
saturated.\footnote{F28X50LP has a cut-on wavelength around
  0.55~\micron\ and extends as red as 1.0~\micron.} We analyzed these
images using our PSF-fitting routine based on the TinyTim model of the
\HST\ PSF \citep{1995ASPC...77..349K}, as described in our previous
work \citep[e.g.,][]{2008ApJ...689..436L, me-2206}.  We adopted the
pixel scale and distortion solution from the STIS Instrument Science
Report 2001-02
\citep[][$50.725\pm0.056$~mas~\perpix]{2001stis.rept....3A}.

We tested for systematic errors in our best-fit binary parameters by
applying our PSF-fitting routine to artificial binary images created
from STIS CCD images of single stars from \HST\ calibration programs
in Cycles 8 and 10 (CAL/STIS-8422 and 8924).  To preserve the
undersampled nature of the PSF, we created only artificial binaries
separated by an integer number of pixels.  The separation and
instrumental PA of \glAB\ in the STIS images were 1.92~pixels and
3$\fdg$1, so artificial binaries with $\Delta(x,y) = (0,2)$ were an
excellent representation of the science data.  The single stars were
$\approx$10$\times$ brighter in the STIS images than \glA, so we
scaled them to match the science data and then injected them into the
actual science image to accurately simulate the noise.  The binary
images were injected at a distance of 5$\arcsec$ from Gl~569A, chosen
to be comparable to Gl~569B's separation.  For the final adopted
errors we used the larger of the rms from the artificial binaries or
the rms of the best-fit parameters to the science data.  From our
artificial binary tests, we found systematic offsets in separation
(3.0~mas, 2.3$\sigma$), PA (0$\fdg$2, 0.2$\sigma$) and flux ratio
(0.15~mag, 5$\sigma$), which we applied to our best-fit binary
parameters for Gl~569Bab.  The final separation, PA, and flux ratio we
derived were $97.3\pm1.3$~mas, $94\fdg0\pm0\fdg9$, and
$0.99\pm0.03$~mag (Table~\ref{tbl:astrom}).

\subsection{VLT/NACO \label{sec:vlt}}

We retrieved archival images of \lpAB\ obtained with the Very Large
Telescope (VLT) at Paranal Observatory on five epochs spanning six
months in 2006. These data were taken with the NACO adaptive optics
system \citep{2003SPIE.4841..944L, 2003SPIE.4839..140R} using the
N90C10 dichroic and S13 camera.  We adopt the pixel scale of
$13.221\pm0.017$~mas~\perpix\ derived by B.\ Sicardy (2008, private
communication)\footnote{\texttt{\url{http://www.eso.org/sci/facilities/paranal/instruments/naco/doc/VLT-MAN-ESO-14200-2761\_v83.3.pdf}}}
from observations of Pluto's motion compared to its known ephemeris,
and an orientation of the detector of $0\fdg0\pm0\fdg1$
\citep{2005A&A...435L...5F}.  We followed the same procedure for
analyzing these data as in our previous work with VLT images
\citep{me-2206, me-2397a}.  We registered, sky-subtracted, and
performed cosmic ray rejection on the raw archival images. We used the
same analytic PSF-fitting routine as was used for the Keck data to fit
the VLT images, with uncertainties determined from the standard
deviation of measurements from individual dithers.  The resulting
best-fit parameters are summarized in Table~\ref{tbl:astrom}.

\subsection{Previously Published Astrometry \label{sec:prev}}

\citet{2005A&A...435L...5F} published two epochs of astrometry for
\lpAB, and we investigated the possibility of improving the precision
of their measurements by applying our PSF-fitting procedure to their
data, which is in public archives. The CFHT observations of \lpAB\ on
2004 July 3 UT were not available in the CFHT archive because they
were obtained during engineering time (T.\ Forveille 2010, private
communication), but the VLT/NACO observations from 2004 September 24
UT are in the VLT archive.  We measured a separation of
$111.2\pm0.3$~mas, a PA of $5.66\pm0.12\degree$, and a flux ratio at
$H$ band of $0.40\pm0.03$~mag.  Our flux ratio and separation are in
good agreement with those measured by \citet{2005A&A...435L...5F},
$0.38\pm0.05$~mag and $107\pm10$~mas, while our PA of
$5.66\pm0.12\degree$ is somewhat inconsistent with their
$7.1\pm0.5\degree$ (at 2.8$\sigma$). We tried using each of these
measurements to fit the orbit and found that they gave consistent
dynamical masses with a very similar $\chi^2$ for the best-fit.  Our
higher precision measurement would enable the uncertainty in some of
the orbital parameters to be improved by up to a factor of $\approx$2.
However, we conservatively chose to use the published astrometry of
\citet{2005A&A...435L...5F} as their larger errors and PA offset may
result from analysis of astrometric calibration data unavailable to us
that was taken along with their observations.

\citet{2006A&A...460L..19M} published three epochs of astrometry for
\lhsAB, and the first two data sets are available in the CFHT archive.
Unfortunately, not all of the astrometric calibration data was
available in the archive, so independent measurements from these data
cannot be used in the orbit fit because, e.g., the orientation of the
detector at the time of those observations is not accurately known.
\citet{2006A&A...460L..19M} found uncertainties of 5~mas and
0.5\degree\ for all three data sets. The rms scatter of our
PSF-fitting measurements is comparable or slightly lower (3--4~mas and
0.4--0.6\degree), and thus even with the needed calibration data a
drastic improvement in precision seems unlikely.  Thus, we use the
published astrometry of \citet{2006A&A...460L..19M} when fitting the
orbit of \lhsAB\ (Table~\ref{tbl:astrom}).

For \glAB, we do not use any of the previously published astrometry.
We performed our own analysis on the original Keck/NIRC2 images from
\citet{2006ApJ...644.1183S} following the method described in
Section~\ref{sec:keck}.  \citet{qk10} published one epoch of
astrometry for \glAB\ which is contemporaneous with our 2009 May data;
we conservatively choose to exclude their data from our orbit fit,
although it is consistent with our best-fit orbit.  We do not include
any of the Keck/KCAM or SCAM astrometry published by
\citet{2001ApJ...560..390L} or \citet{2004astro.ph..7334O}, as those
cameras have not been rigorously astrometrically calibrated, unlike
NIRC2 and the STIS CCD.

\subsection{IRTF/SpeX Spectroscopy \label{sec:spex}}

In addition to astrometry, which we use to determine orbits and thus
dynamical masses, we have also obtained NIR spectra to perform tests
of model atmospheres, as discussed in Sections \ref{sec:results} and
\ref{sec:tests}.  Using the NASA Infrared Telescope Facility (IRTF)
spectrograph SpeX \citep{2003PASP..115..362R}, we obtained spectra of
\lpint, \lhsint, and \glint\ on 2008 July 6, 2010 January 30, and 2010
March 3 UT, respectively.  We used SpeX in cross-dispersed (SXD) mode,
with five orders spanning 0.81--2.42~\micron. For \lpint\ we used a
$0\farcs5$ slit ($R = 1200$), and for \lhsint\ and \glint\ we used a
$0\farcs3$ slit ($R = 2000$). We calibrated, extracted, and
telluric-corrected the data using the SpeXtool software package
\citep{2003PASP..115..389V, 2004PASP..116..362C}. We used
integrated-light photometry from 2MASS to flux-calibrate our spectra
of \lpint\ and \lhsint.  For \glint, we used our own integrated-light
photometry derived from our Keck images (Section~\ref{sec:keck}) to
flux calibrate its spectrum.  We used the mean of the $JHK$ scaling
factors computed for each bandpass to place our spectra on an absolute
flux scale.


\section{Results \label{sec:results}}

\subsection{Orbit Determination and Dynamical Masses \label{sec:orbit}}

\begin{figure}
\centerline{\includegraphics[width=2.7in,angle=0]{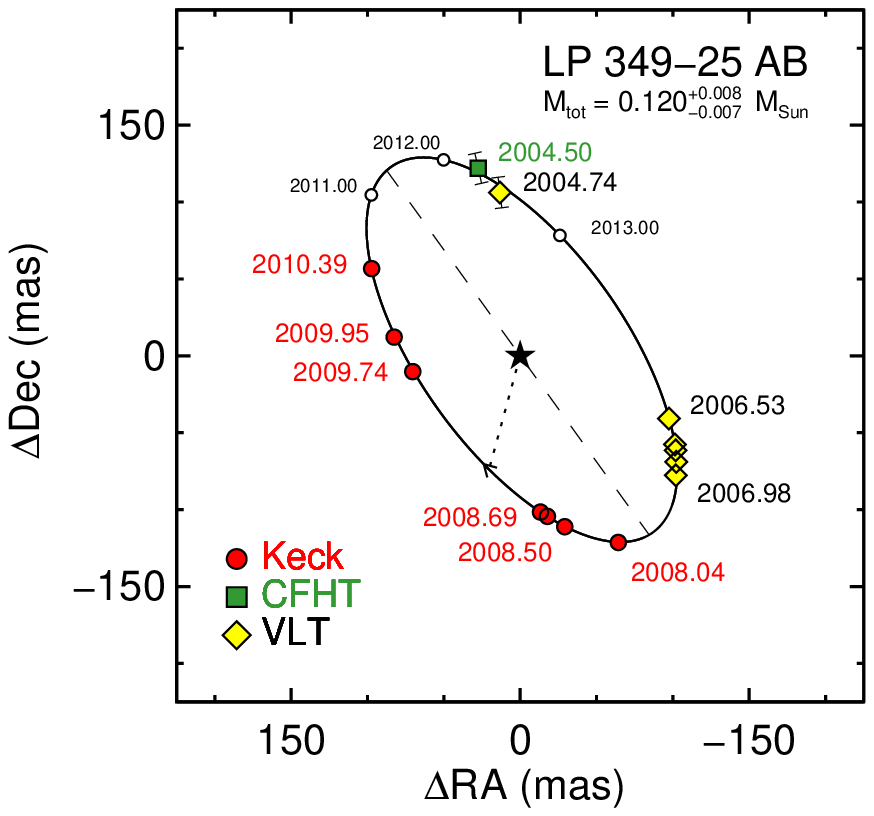}}
\vskip -0.15in
\centerline{\includegraphics[width=2.7in,angle=0]{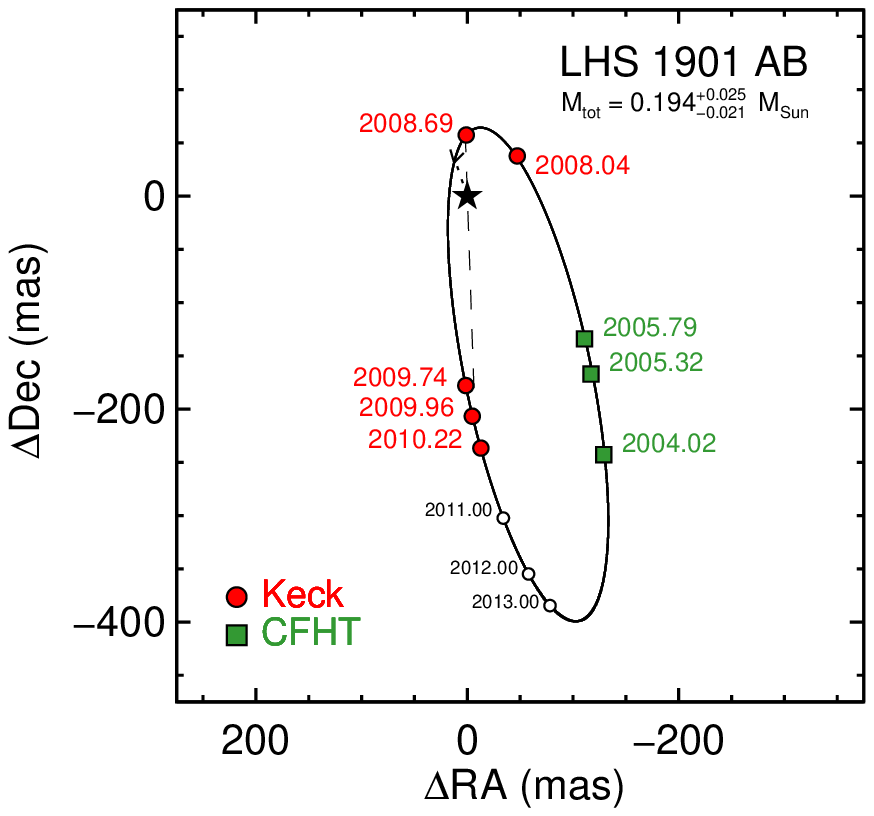}}
\vskip -0.15in
\centerline{\includegraphics[width=2.7in,angle=0]{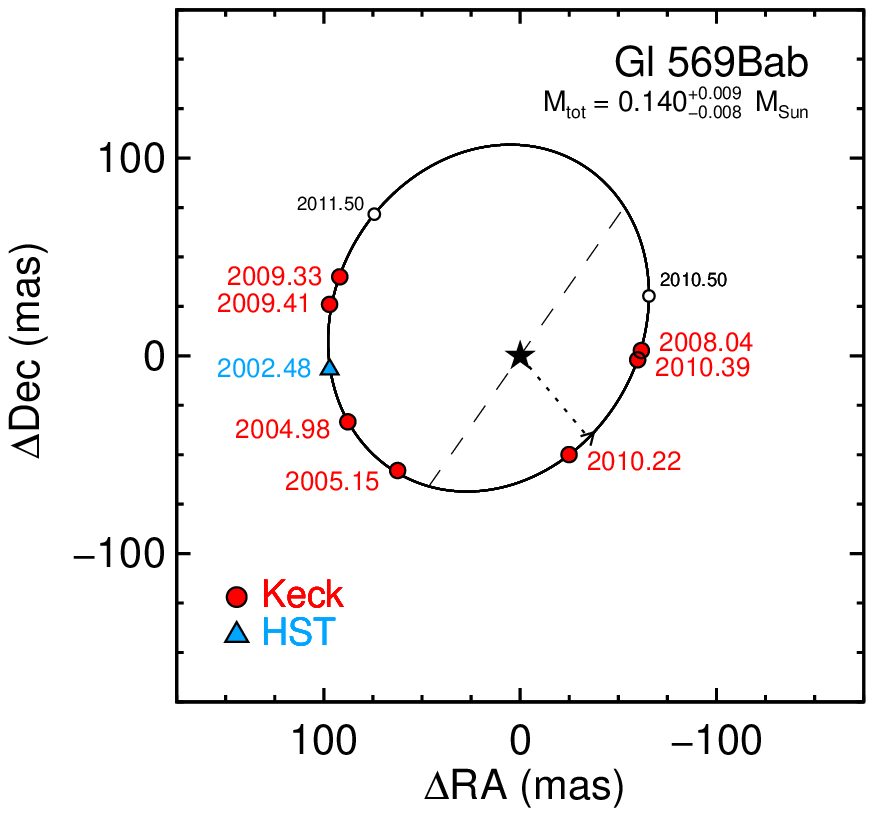}}

\caption{\normalsize Relative astrometry for \lpAB, \lhsAB, and \glAB\
  along with the best-fitting orbits. The dotted line and arrow
  indicate the time of periastron passage, the dashed line shows the
  line of nodes, and the empty circles show predicted future
  locations.  Error bars are typically smaller than the plotting
  symbols. All orbits are sufficiently well constrained that the
  uncertainties in the total masses are dominated by the parallax
  errors. \label{fig:orbit-skyplot}}

\end{figure}

\begin{figure}
\centerline{
\includegraphics[width=1.7in,angle=0]{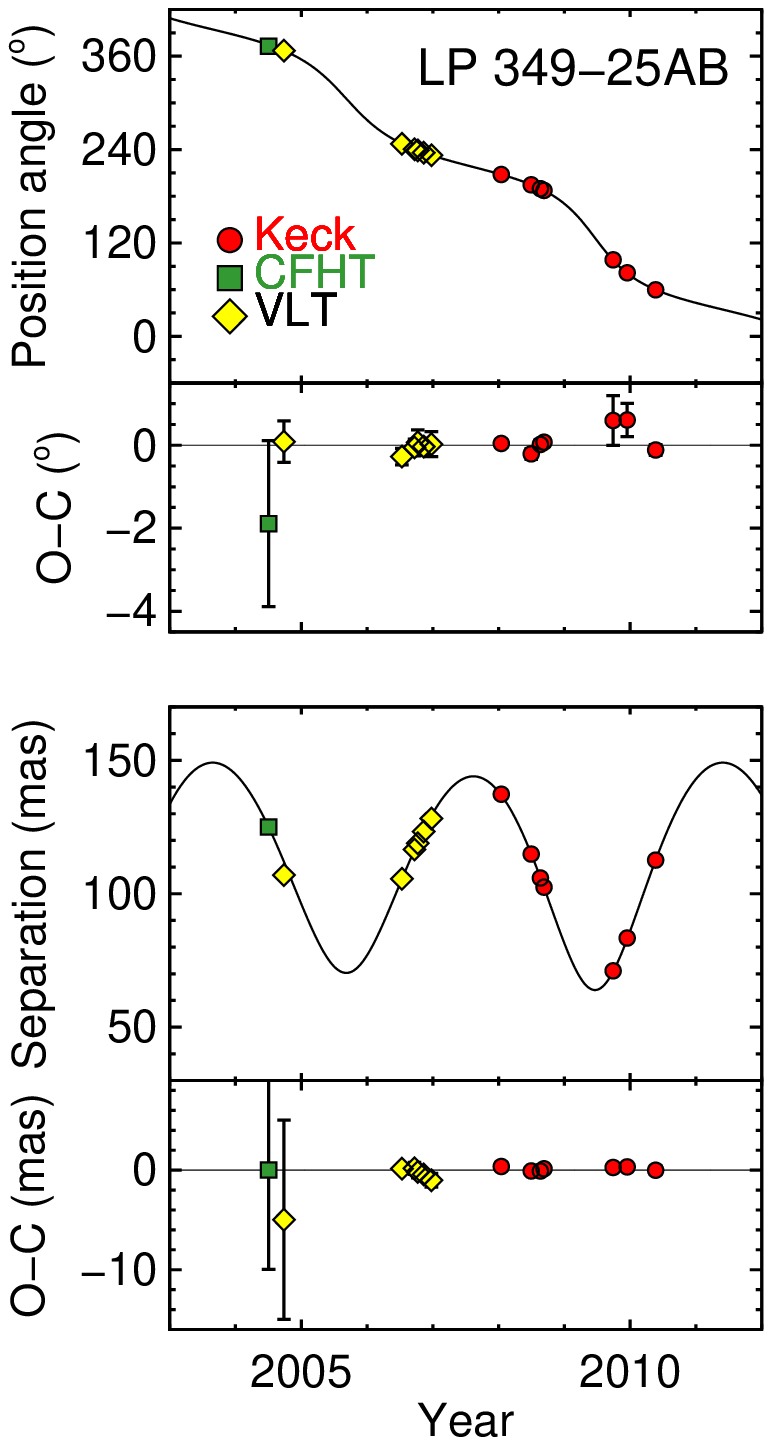}
\hskip 0.4in
\includegraphics[width=1.7in,angle=0]{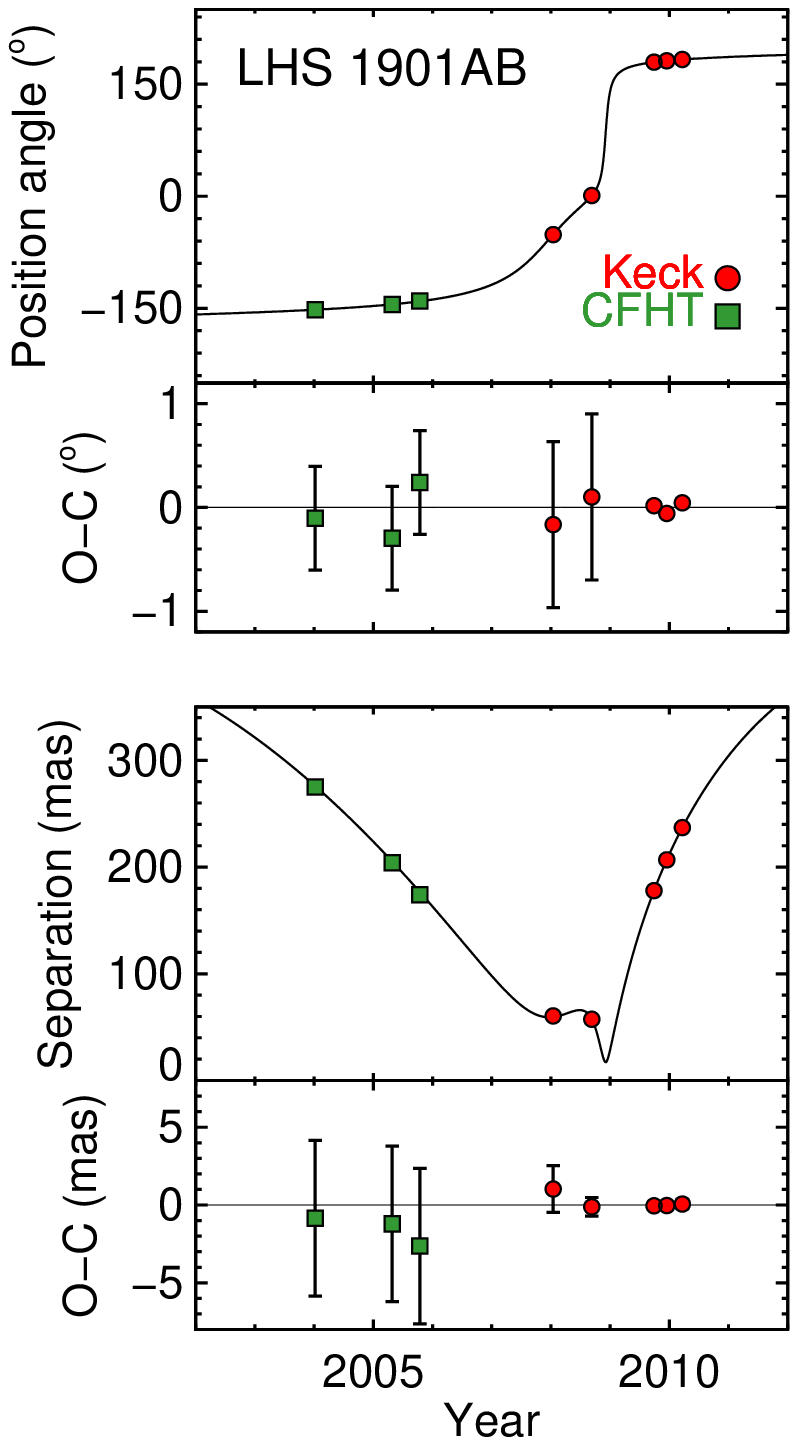}
\hskip 0.4in
\includegraphics[width=1.7in,angle=0]{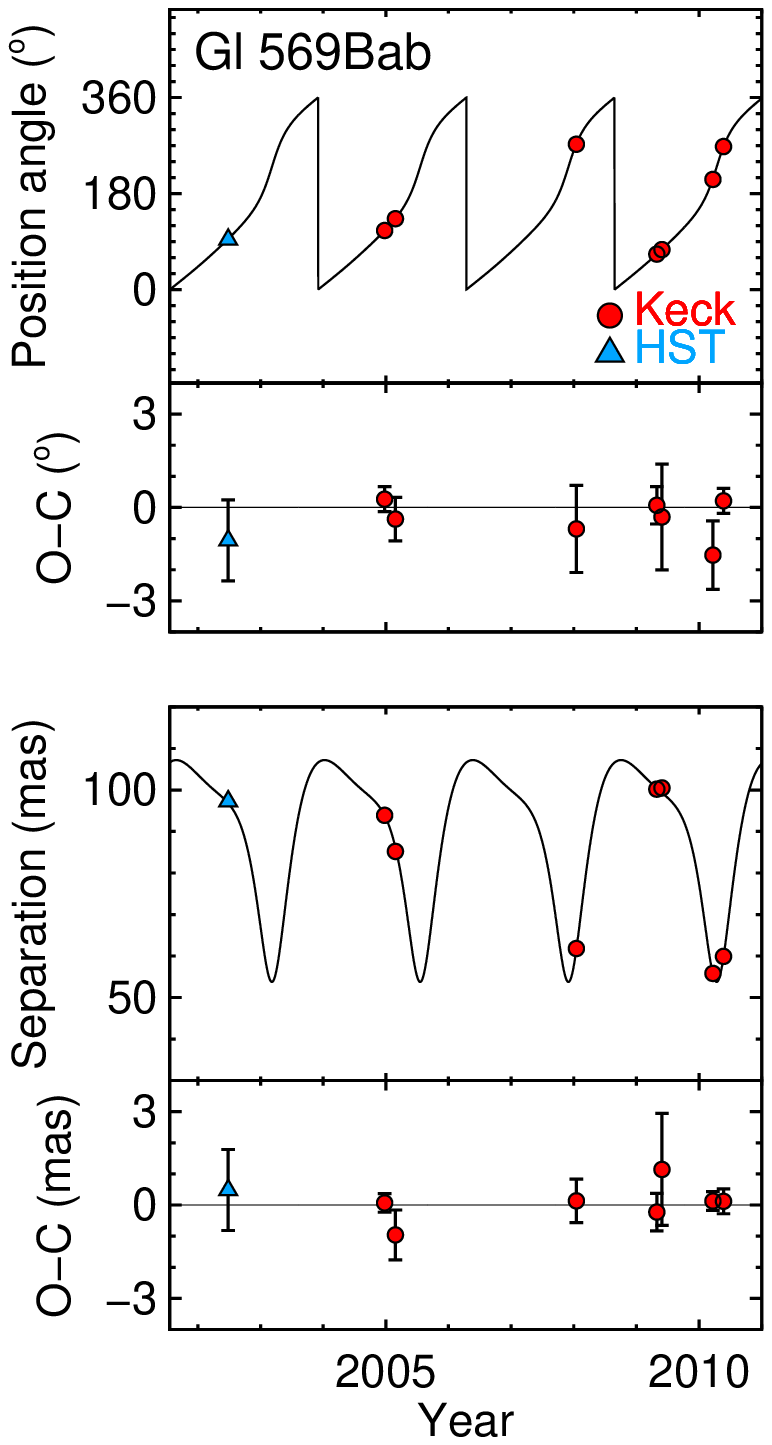}}

\caption{\normalsize Measurements of the projected separation and PA
  of \lpAB\ (\emph{left}), \lhsAB\ (\emph{middle}), and \glAB\
  (\emph{right}).  The best-fit orbits from our MCMC analysis are
  shown as solid lines.  The bottom panel of each plot shows the
  observed minus computed ($O-C$) measurements with observational
  error bars. \label{fig:orbit-sep-pa}}

\end{figure}

Combining our Keck AO monitoring with published discovery data, we
have observations spanning 5.9 years for \lpAB, 5.9 years for \lhsAB,
and 7.9 years for \glAB. Following our previous studies, we have
determined the orbital parameters and their uncertainties using both a
Markov Chain Monte Carlo (MCMC) approach, described in detail by
\citet{2008ApJ...689..436L}, as well as the least-squares minimization
routine \orbit\ \citep{1999A&A...351..619F}.  In addition, we have
developed our own visual binary orbit fitting routine that utilizes
the Levenberg-Marquardt algorithm as implemented in the MPFIT routine
for IDL \citep{2009ASPC..411..251M}.  To determine uncertainties on
the best-fit parameters from this routine, we have performed Monte
Carlo simulations in which Gaussian noise is added to the data
corresponding to their measurement uncertainties. We used the
resulting distributions to determine the medians and standard
deviations of the best-fit parameters.  When multiple astrometric
measurements were available from different bandpasses at a single
epoch, we used the highest precision measurement in the orbit fit (see
Table~\ref{tbl:astrom}).  The best-fit orbits are shown in
Figures~\ref{fig:orbit-skyplot} and \ref{fig:orbit-sep-pa}.

We found essentially symmetric uncertainties in the MCMC-derived
orbital parameters for \lpAB. These parameters agreed to within
$<0.5\%$ of those derived by our custom fitting routine, and the
errors agreed to within $<15\%$. As an additional test of our
Levenberg-Marquardt fitting routine, we found that the best-fit
parameters from \orbit\ also agreed to within $<1\%$, and the errors
derived from our Monte Carlo approach agreed to within $<10\%$ of the
errors computed from the full covariance matrix in \orbit. The orbital
parameters and errors derived by our MCMC analysis and custom fitting
routine are given in Table~\ref{tbl:orbit}.

We experimented with including the five epochs of relative astrometry
from \citet{qk10} when fitting the orbit of \lpAB. If all of their
data are included, the $\chi^2$ becomes unrealistically large:
$\chi^2$ of 58.2 for 31 degrees of freedom (DOF).  This large $\chi^2$
value is due to the fact that two of their measurements from 2007
December and 2008 May are extreme outliers from the best-fit orbit.
(Indeed, their own orbit fit is inconsistent with these data as the
\citealt{qk10} orbit has a reduced $\chi^2$ of 2.15, which is
unrealistically large.)  By excluding their most discrepant epoch
(2008 May 30 UT), we were able to achieve a reasonable $\chi^2$ (33.1
for 29 DOF).  Neither the orbital parameters nor their uncertainties
change significantly after including the \citet{qk10} data.  This is
because their astrometry does not offer additional unique constraints
on the orbit, as it was obtained contemporaneously with ours and is of
lower precision by a factor of $\approx$4 to 10.  Thus, we
conservatively choose to exclude the \citet{qk10} astrometry from our
analysis.  Our best-fit total mass (\Mtot) for \lpAB\ is
$0.1205\pm0.0007$~\Msun\ (0.6\%), and after including the 2.1\% error
in the parallax this becomes $0.120_{-0.007}^{+0.008}$~\Msun\ (6\%).

\begin{figure}
\centerline{
\includegraphics[width=2.7in,angle=0]{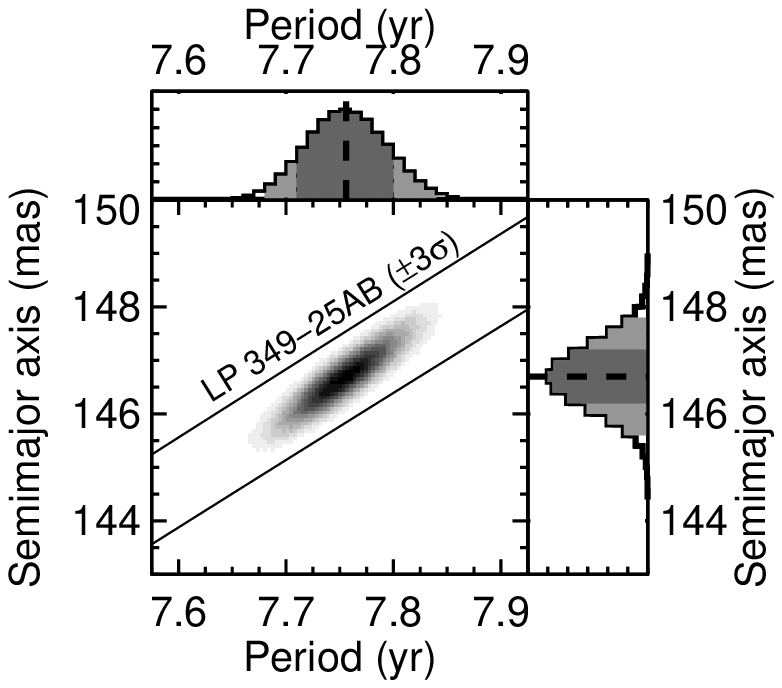}
\hskip 0.1in
\includegraphics[width=2.7in,angle=0]{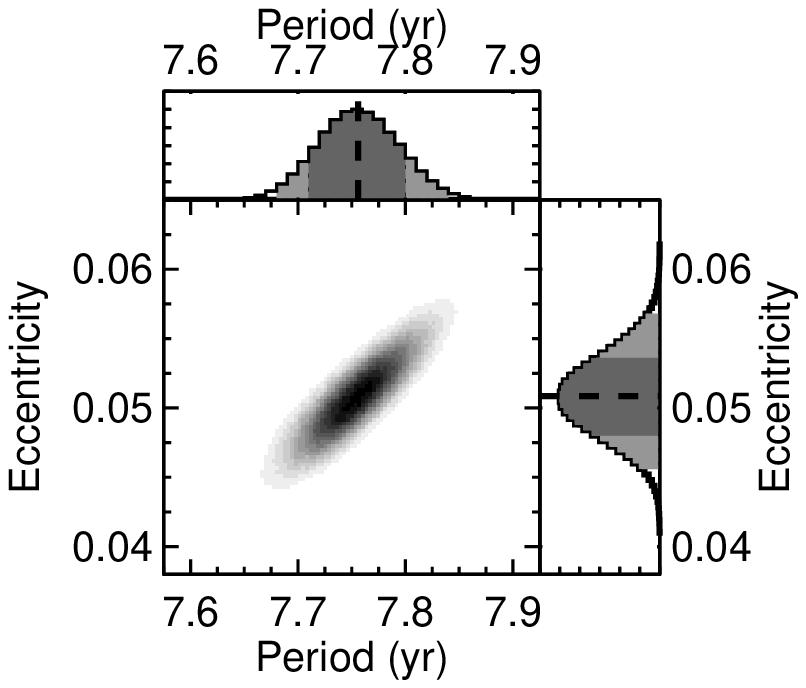}}
\vskip 0.2in
\centerline{
\includegraphics[width=2.7in,angle=0]{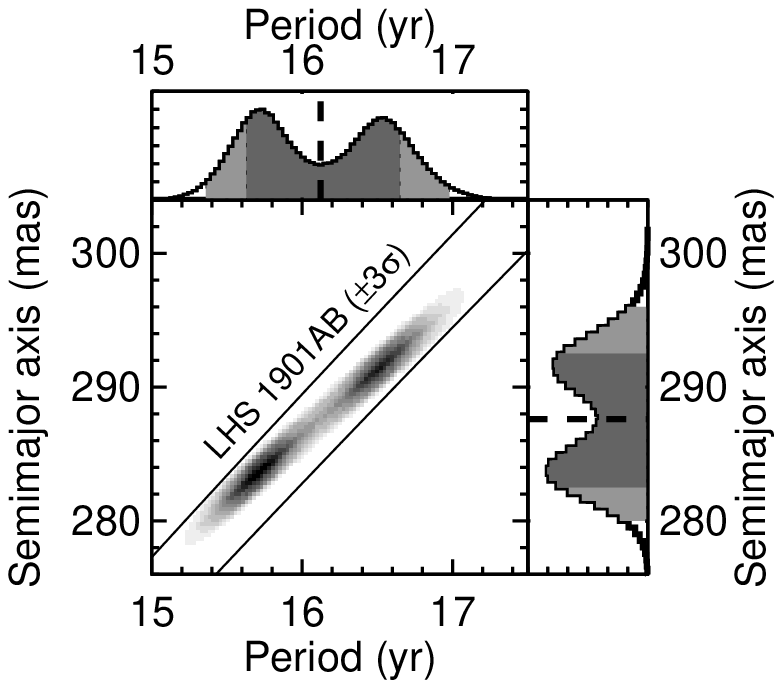}
\hskip 0.1in
\includegraphics[width=2.7in,angle=0]{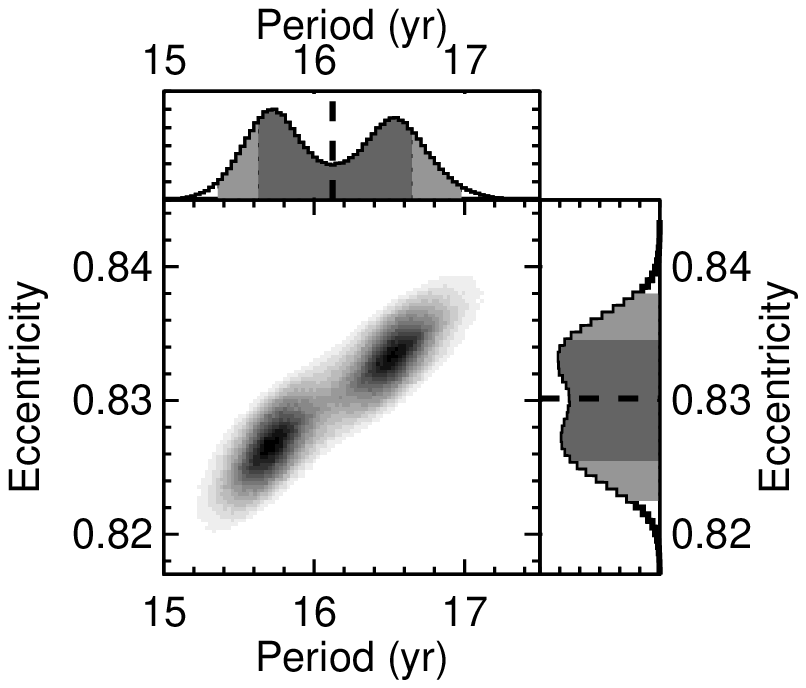}}
\vskip 0.2in
\centerline{
\includegraphics[width=2.7in,angle=0]{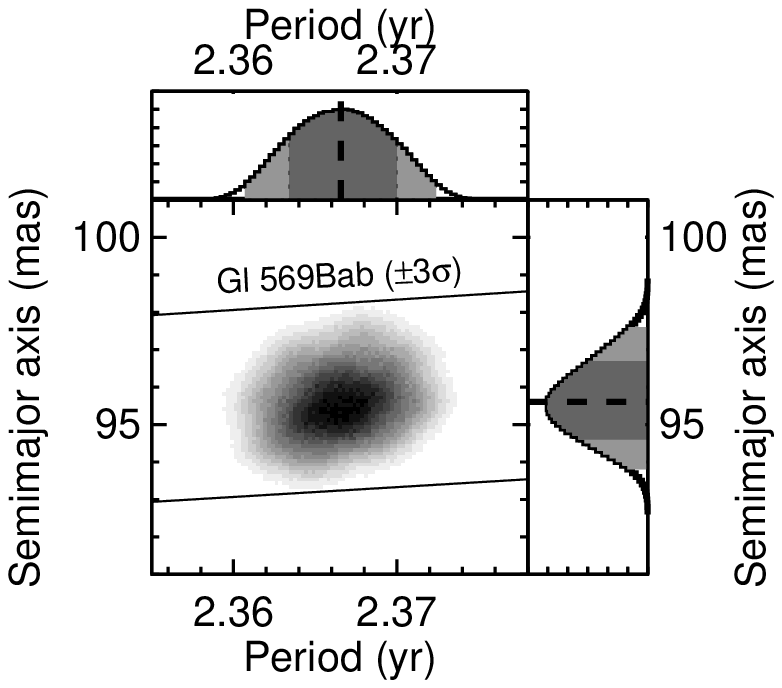}
\hskip 0.1in
\includegraphics[width=2.7in,angle=0]{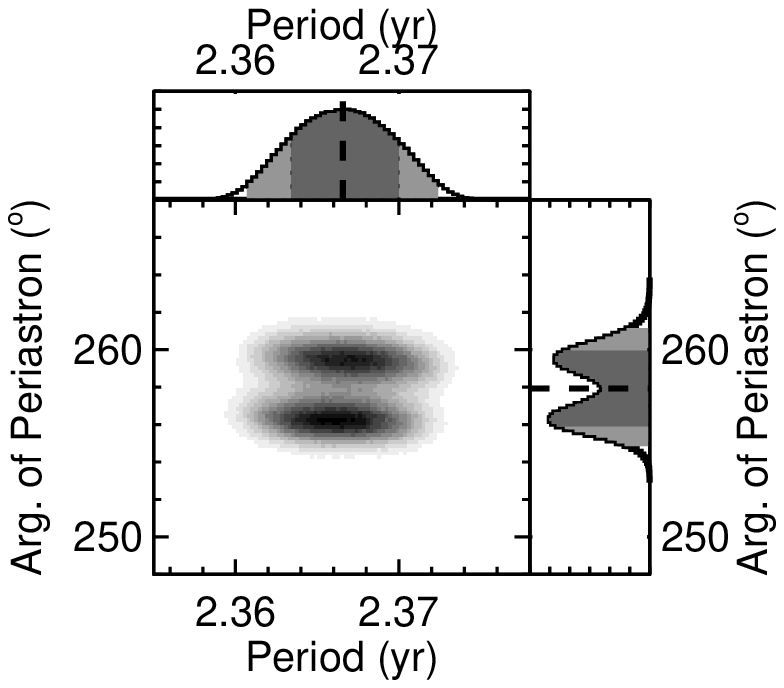}}

\caption{\normalsize The covariance between the orbital parameters for
  \lpAB\ (\emph{top}), \lhsAB\ (\emph{middle}), and \glAB\
  (\emph{bottom}) shown in grayscale using all $2\times10^6$ saved
  steps from our MCMC chains of length $2\times10^8$.  For \lpAB\ and
  \lhsAB\ the correlation between $P$ and $a$ enables the total mass
  to be determined more precisely than from simple propagation of
  errors, as illustrated by lines drawn demarcating the 3$\sigma$
  range for the total mass (prior accounting for the distance
  uncertainty).  The double-peaked nature of the MCMC parameter
  distributions for \lhsAB\ and \glAB\ also reveal slightly degenerate
  orbit solutions. \label{fig:pdist}}

\end{figure}

From our MCMC analysis of \lhsAB, we found significant asymmetries in
the posterior probability density distributions of the orbital period,
semimajor axis, and eccentricity. The confidence limits on all orbital
parameters are given in Table~\ref{tbl:orbit}, and the distributions
of these three parameters are shown in Figure~\ref{fig:pdist}. The
orbital period and semimajor axis display a high degree of covariance
as expected \citep[e.g., see Section~3.2 of][]{2009ApJ...692..729D}.
There appear to be two slightly degenerate solutions, with one having
a shorter period, lower eccentricity, and higher mass than the other.
Our MCMC analysis enables us to treat this degeneracy properly by
accounting for it in the confidence limits of these orbital parameters
and the resulting dynamical mass. The difference in mass between these
two solutions is very small (about 1.6\%), insignificant compared to
the 12\% error contributed by the distance uncertainty. Our best-fit
total mass for \lhsAB\ is $0.1944\pm0.0028$~\Msun\ (1.4\%), and after
including the $^{+4.0}_{-3.7}$\% error in the parallax this becomes
$0.194^{+0.025}_{-0.021}$~\Msun\ (12\%).

The orbital parameters we find from \glAB\ are all consistent with
previously published orbits, with a singular exception.  The semimajor
axis we find ($95.6^{+1.1}_{-1.0}$~mas) is 4.3$\sigma$ higher than the
value of $90.4\pm0.7$~mas derived by \citet{2006ApJ...644.1183S}, and
the \citet{2004astro.ph..7334O} and \citet{qk10} values are similarly
discrepant (by 1.6$\sigma$ and 3.7$\sigma$, respectively).  Since all
orbit determinations to date have been dominated by the nine epochs of
astrometry from Keck KCAM and SCAM, the semimajor axis discrepancy
implies that the pixel scale of one or both of these instruments was
underestimated by $\approx$5\%, resulting in systematically low
semimajor axes.  This is supported by the fact that the previously
published NIRC2 astrometry from \citet{2006ApJ...644.1183S} and
\citet{qk10} is in good agreement with our best-fit orbit.  Also, the
\HST/STIS measurement would be highly inconsistent with a value for
the semimajor axis as low as previously found, showing that our
semimajor axis is not simply due to a problem with the NIRC2
calibration.\footnote{We also note that a calibration problem with
  NIRC2 would have been apparent in our previously published binary
  orbits, which typically have data from NIRC2 as well as other
  well-calibrated instruments such as \HST/WFPC2-PC, \HST/ACS-HRC, and
  VLT/NACO.}  Since we find the same orbital period as in previous
work but a semimajor axis that is 5.4\% larger, our dynamical total
mass is significantly higher than previously found.  This would be a
16\% effect, but it is moderated somewhat by the revised \Hipparcos\
parallax that is 1.6\% (about 1$\sigma$) higher than the original
value, which brings the mass down by 4.8\% for a net change of 11\%.
Our best-fit total mass for \glAB\ is $0.140_{-0.004}^{+0.005}$~\Msun,
and after accounting for the 1.6\% parallax error this becomes
$0.140_{-0.008}^{+0.009}$~\Msun.  Our MCMC analysis also reveals a
slightly degenerate orbit solution with double-peaked distributions
for the argument of periastron ($\omega$) and PA of the ascending node
($\Omega$) as shown in Figure~\ref{fig:pdist}.  The degeneracy in
these values are apparently not correlated with any other orbital
parameters.

\subsection{Spectral Types \label{sec:spt}}

\citet{2000AJ....120.1085G} used the integrated-light optical spectrum
of \lpint\ to derive a spectral type (SpT) of M$8.0\pm0.5$. Comparing
our near-infrared spectrum of \lpint\ to spectra of M dwarf standards
from the IRTF Spectral Library \citep{2005ApJ...623.1115C,
  2009ApJS..185..289R}, we find an infrared spectral type of M8.
Without resolved spectroscopy of the binary, we cannot directly
determine the spectral types of the individual components. However, we
have used our photometry and the empirical SpT--absolute magnitude
relations of \citet[][$J$ band only]{2003AJ....126.2421C} and
\citet[][$L^{\prime}$ band only]{gol04} to estimate spectral types. We
used the method described by \citet{me-2397a}, which accounts for both
the measurement uncertainties and the intrinsic scatter in the
empirical absolute magnitude relations. We found that the range of
$J$-band absolute magnitudes allowed by our measurement errors
exceeded the bounds of the \citet{2003AJ....126.2421C} relation, and
thus we base our estimates only on the \Lp-band relation. We estimate
spectral types of M$7.5\pm1.0$ for the primary and
M$8.0^{+1.5}_{-1.0}$ for the secondary, which are consistent with the
published integrated-light spectral type.

\citet{2003AJ....126.3007R} measured the integrated-light spectral
type of \lhsint\ to be M6.5, and \citet{2009AJ....137.4109L} measured
M7.0. Our near-infrared spectrum is better matched by the M dwarf
standard vB~8 (M7) than GJ~1111 (M6.5), thus we find an infrared type
of M7 for \lhsint. Using the method described above, the $J$-band
absolute magnitudes of the individual components provide an estimated
spectral type of M$6.5^{+1.0}_{-0.5}$ for both the primary and
secondary, consistent with measured integrated-light spectral
types. The difference of zero subtype between the components is also
consistent with our measured flux ratios and late-M dwarfs with known
parallaxes. For example, the rms scatter among $K$-band absolute
magnitudes for M6.5 dwarfs is 0.12~mag (cf.\ $\Delta{K} =
0.107\pm0.007$~mag for \lhsAB).

For \glAB, we adopt the spectral types of M$8.5\pm0.5$ and
M$9.0\pm0.5$ derived by \citet{2001ApJ...560..390L}.  They used their
resolved near-infrared spectra of the individual components to compute
spectral indices and compared these values to field dwarfs of known
spectral type.  They found a difference of 0.5 subtypes between the
two components and assigned the integrated-light spectral type of M8.5
from \citet{1990ApJ...354L..29H} to the primary, resulting in a type
of M9.0 for the secondary.  These spectral types are consistent with
those found by \citet[][M9+M9]{2006A&A...456..253M} using resolved
optical spectroscopy of \glAB\ from \HST/STIS.

\subsection{Bolometric Luminosities \label{sec:lbol}}

We computed the total bolometric luminosities (\Lbol) of our sample
binaries from their integrated-light spectra and photometry.  In
addition to our SpeX spectra, we used the \Lp\ and \Mp\ band
photometry from \citet{gol04} and the 24~\micron\ \Spitzer\ photometry
from \citet{2007ApJ...667..527G} for \lpint, and we used the \Lp\ band
photometry from \citet{1988ApJ...330L.119F} for \glint.  For \lhsint,
we estimated a $K-\Lp$ color of $0.56\pm0.07$~mag from the empirical
relation of \citet{me-2206} to extend its SED to longer wavelengths.
To accurately account for the SED at shorter wavelengths, we appended
optical spectra, using the overlapping region of our SpeX spectra to
determine the flux scaling. For \lpint\ and \glint\ we used optical
spectra from \citet{2003AJ....126.3007R} and
\citet{1990ApJ...354L..29H}, respectively.  For \lhsint\ we used the
optical spectrum of the M7 standard vB~8 \citep{1990ApJ...354L..29H}.
We numerically integrated the spectra and photometry points,
interpolating between the gaps in the data, extrapolating to zero flux
at zero wavelength, and assuming a Rayleigh-Jeans tail beyond the last
photometric point. We accounted for errors as described in
\citet{me-2206}, finding a total $\log(\Lbol/\Lsun)$ of
$-2.804^{+0.026}_{-0.019}$~dex for \lpint, $-2.67^{+0.04}_{-0.03}$~dex
for \lhsint, and $-3.212\pm0.018$~dex for \glint.

We apportioned these total bolometric fluxes to the individual
components by using the measured $K$-band flux ratio and the
bolometric correction--SpT relation derived by \citet{gol04}, properly
accounting for uncertainties in the component spectral type estimates.
The resulting individual luminosities are given in
Table~\ref{tbl:meas}.  For the components of \glAB, our \Lbol\ values
are in good agreement with the estimates from
\citet{2001ApJ...560..390L}.  As in our previous work, we correctly
account for the covariance between the individual luminosities (via
the flux ratio) in the following analysis, allowing more precise
determinations of relative quantities (e.g., $\Delta\Teff$ and the
mass ratio).

\subsection{\Teff\ and \logg\ from Model Atmosphere
  Fitting \label{sec:atm-fit}}

We used a variety of solar-metallicity model atmosphere grids to fit
our integrated-light IRTF/SpeX spectra, summarized in
Table~\ref{tbl:atm-ref}.  The two publicly available grids we used are
Ames-Dusty \citep{2001ApJ...556..357A} and PHOENIX-Gaia
\citep{2005ESASP.576..565B}.  We also used an improved version of
Ames-Dusty that employs the same treatment of dust but with updated
line lists for some important opacity sources such as FeH, CrH, and
H$_2$O (see Section~3.2 of \citealp{2010ApJS..186...63R}); this
``Gaia-Dusty'' model grid is the same as that employed by
\citet{qk10}.  Our fourth model grid Drift-PHOENIX
\citep{2009A&A...506.1367W} relies on yet another gas opacity data set
and, more importantly, features improved dust opacities.  The model
treats the dynamics of the dust in a presently unrivaled level of
detail by calculating the rate of seed formation and solving an actual
growth rate equation system for gravitational settling of composite
grains \citep{2003A&A...399..297W, 2006A&A...455..325H,
  2008A&A...485..547H}.

We fit model atmospheres to our spectra using the procedure outlined
by \citet{2009ApJ...706.1114B}.  For every model spectrum, we found
the optimal scaling factor to match the observed spectrum by
minimizing the $\chi^2$ statistic.  The model spectrum with the lowest
resulting $\chi^2$ value gives the best-fit parameters for \Teff\ and
\logg. For each spectrum, we fit individual wavelength ranges
corresponding to the standard near-infrared bandpasses ($YJHK$).  We
also performed fits to the entire SED, both with and without the
segment of the spectrum between 0.81 and 0.95~\micron.  This region of
the spectrum is often excluded when fitting models to L and T dwarfs
\citep[e.g.,][]{2008ApJ...678.1372C} because it is strongly affected
by the broad wings of the resonant \ion{K}{1} doublet at 0.77~\micron,
which have long been difficult to model
\citep[e.g.,][]{2003ApJ...583..985B} and can be impacted by dust
modeling \citep{2008MNRAS.385L.120J}. Such alkali lines are present
but weaker in late-M dwarfs, and they may contribute to observed
discrepancies with models between $\approx$0.7 and 0.9~\micron\ for
objects as early as L0 \citep{2007A&A...473..245R}.

The results of our model atmosphere fitting procedure are given in
Table~\ref{tbl:atm-fit}.  We report fits both for our full wavelength
range (0.81--2.42~\micron) and for the NIR only
(0.95--2.42~\micron). We prefer the NIR SED fits for the following
reasons: (1)~the NIR SED fits are more consistent with the individual
$YJHK$ band fits (typically within one model grid step of 100~K,
0.5~dex in \logg); (2)~the full SED fits are systematically cooler by
$\approx$100~K compared to the individual band fits, implying that the
lowest $S/N$ portion of the spectrum from 0.81 to 0.95~\micron\ has a
disproportionate influence on the results; and (3)~by adopting a
0.95~\micron\ cutoff here, our results will be directly comparable to
future work on L and T dwarfs that require such a cutoff because of
the problems with \ion{K}{1} lines.

\begin{figure}
\hskip -0.1in
\centerline{
\includegraphics[height=6.0in,angle=90]{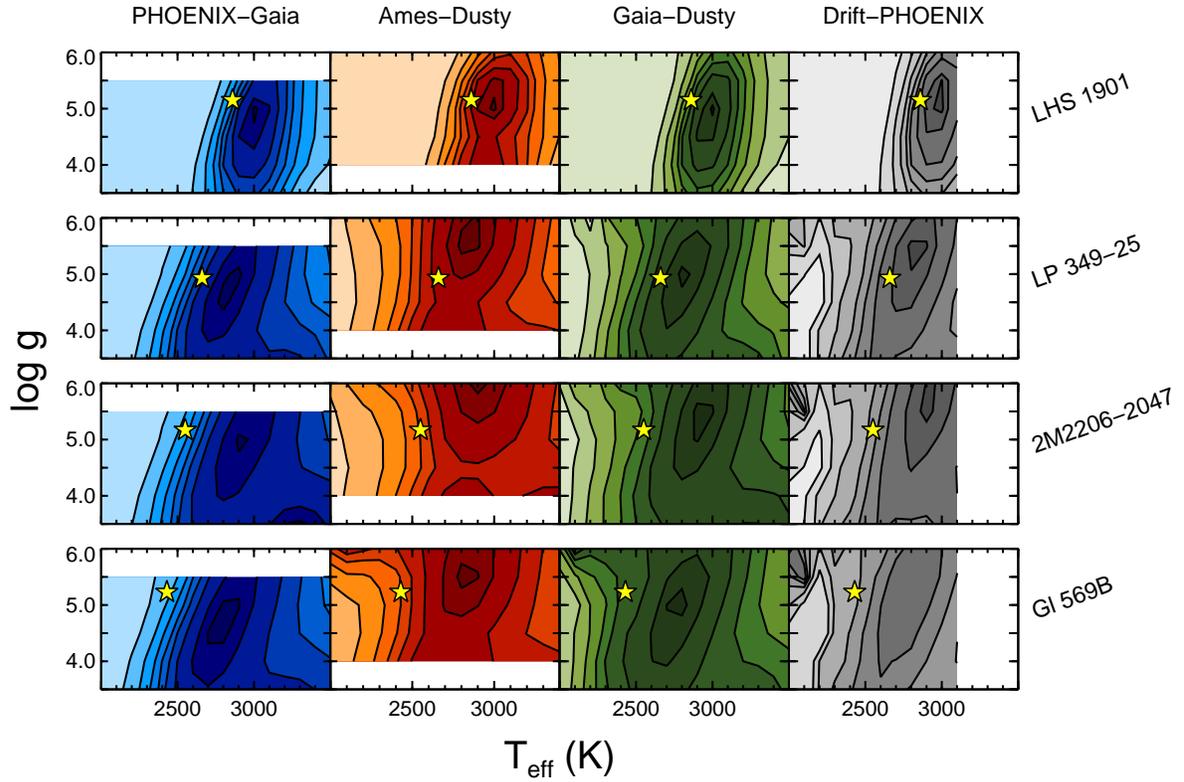}}

\caption{ \normalsize Contours of the $\chi^2$ values
    from model atmosphere fits to our sample binaries' observed
    integrated-light 0.95--2.42~\micron\ spectra (drawn at 1.02, 1.1,
    1.3, 1.5, 1.7, 2.0, 2.5, and 3.5 times the minimum $\chi^2$).
    \Teff\ is generally better constrained than \logg, as indicated by
    contours that are elongated in the \logg\ direction. White space
    shows where models are not present in our grids. For comparison,
    the gold stars show Lyon Dusty evolutionary model-derived
    properties, with \logg\ in general agreement but \Teff\
    systematically lower than the best-fit model atmospheres (see
    Section~\ref{sec:teff-logg}). \label{fig:chi2}}

\end{figure}

Figure~\ref{fig:chi2} shows the $\chi^2$ contours of the fits to our
observed NIR spectral energy distributions (SEDs) using all model
spectra. The elongation of the contours in the \logg\ direction
indicates that gravity is less well-constrained by our fits than
effective temperature. We also note that there are occasionally local
$\chi^2$ minima in these plots that correspond to physically
implausible parameters. For Drift-PHOENIX, the deviant $\chi^2$
minimum occurs at very low temperatures (2000~K) and high gravities
($\logg = 6.0$, which for masses of $<$0.08~\Msun\ correspond to radii
of $<$0.045~\Rsun, more than a factor of two lower than evolutionary
model predicted radii). For the other three model grids, $\chi^2$
minima extend to higher temperatures ($\approx$3300~K) and extremely
low gravities ($\logg = 3.5$, which for masses of $>$0.06~\Msun\
correspond to radii of $>$0.7~\Rsun).  In the rare cases when the
global minimum is found in one of these physically implausible regions
($\approx$3\% of fits, see Table~\ref{tbl:atm-fit}), we select the
second best fitting model instead, and this always results in best-fit
parameters from the prevailing solutions of \Teff~=~2700--3000~K.

The different grids of model atmospheres generally give very similar
best-fit parameters despite their significantly different input
assumptions.  For example, the treatment of dust in the Drift-PHOENIX
models is much more realistic than in other models, yet for our sample
it yields the same \Teff\ values as other grids.  Thus, it appears
that the effects of dust at the warm temperatures of late-M dwarfs
only subtly impact the resultant model spectra for our purposes.  In
fact, for the NIR SED fits (0.95--2.42~\micron), all four sets of
models give exactly the same \Teff.

Finally, we investigated how binarity impacts the atmospheric
parameters derived from integrated-light spectra. We constructed
artificial binary integrated-light spectra by summing individual
IRTF/SpeX SXD spectra, both from our own spectra (e.g., summing the
spectra of \lpint\ and \glint) and from publicly available spectra of
the M7, M8, and M9 spectral standards vB~8, vB~10, and LHS~2924
\citep{2005ApJ...623.1115C, 2009ApJS..185..289R}. We fit these spectra
both individually and summed, after scaling down the later type
spectrum to best match the measured $JHK$ flux ratios of \lpAB\ and
\glAB.\footnote{\lhsint\ and 2MASS~J2206$-$2047 both have flux ratios
  within $\approx$0.1~mag of unity across $JHK\Lp$ bandpasses and thus
  have components that are likely to be spectrally identical. For
  example, \lhsint\ has a bolometric flux ratio of
  $0.044\pm0.015$~dex, and thus the difference in \Teff\ between the
  individual components is expected to be 75~K (i.e., less than one
  model grid step) for an assumed radius ratio of unity.} Comparing
the summed and input ``primary'' spectrum fits, we found that the
best-fit values from individual bands were nearly identical (at most
different by one grid step of 100~K or 0.5~dex). In some cases of we
found that the summed spectra gave systematically lower temperatures
of 100~K as compared to the input ``primary'' spectrum. However, this
was not the case when the input spectra had $JHK$ colors comparable to
the components of \lpAB\ and \glAB.  We note that within each binary
the components have very similar $JHK$ colors, so we might expect that
simulated component and summed spectra would give the same results.
For example, the integrated-light $J-K$ color of \lpint\ is 1.02~mag,
and its components are only different by $\approx$0.03~mag ($J-K =
1.00$~mag and 1.05~mag) despite the binary flux ratio of
$\approx$0.3~mag (the $H-K$ and $J-H$ colors are even closer,
$\Delta{\rm color} \approx 0.01$~mag).

In summary, our simulations show that the individual band fits should
be unaffected by contamination from the spectrum of the secondaries in
our sample binaries, and even if the SED fits are affected they will
give \Teff\ values systematically lower by only 100~K.  Given that our
best-fit values of the NIR SEDs show no such systematic offsets from
the band fits, we conclude that our fits to the integrated-light
spectra accurately find the best-fit parameters of the primary
component's spectrum.

\subsection{Age Constraints from Kinematics and Activity
  \label{sec:age}}

We have considered whether the space motion and activity of our sample
binaries can provide useful constraints on their ages. For \lpint,
there are two discrepant measurements of its radial velocity (RV) in
the literature: (1)~\citet{2009ApJ...705.1416R} derived
$-16.8\pm2.0$~\kms\ from cross-correlation of their Keck/HIRES
spectrum with the RV standard Gl~406 (M6); and (2)~\citet{qk10} found
a center-of-mass velocity of $-8.0\pm0.5$~\kms\ by fitting an orbit to
their multi-epoch near-infrared RVs.\footnote{The discrepancy between
  these two RV measurements may be due to the fact that
  \citet{2009ApJ...705.1416R} measured the RV of the primary
  component, which would have been near the nadir in its orbital RV
  curve. Only 65 days prior to the HIRES measurement ($\Delta$phase =
  0.02), \citet{qk10} measured a primary RV of $-11\pm3$~\kms, which
  is more consistent (at 1.6$\sigma$) with the
  \citet{2009ApJ...705.1416R} value.} Combining these RVs with the
parallax and proper motion measured by \citet[][assuming 4~mas and
4\degree\ errors in their proper motion amplitude and PA,
respectively]{2009AJ....137..402G}, we derive heliocentric velocities
for \lpint\ of $(U, V, W)$~=~($+4.8\pm0.8$, $-17.2\pm1.4$,
$+5.2\pm1.3$)~\kms\ and ($+1.9\pm0.4$, $-11.2\pm0.4$,
$-0.5\pm0.4$)~\kms, respectively. We adopted the sign convention for
$U$ that is positive toward the Galactic center and accounted for the
errors in the parallax, proper motion, and RV in a Monte Carlo
fashion. These space motions place \lpint\ only 0.7$\sigma$ or
0.9$\sigma$ away from the mean of the ellipsoid defined by all other
ultracool dwarfs with $UVW$ measurements.  (This sample is described
in detail in Section~3.4 of \citealt{me-2397a}.) Using the same method
as in our previous work, we have also assessed \lpint's membership in
the Galactic thin and thick disk populations using the Besan\c{c}on
model of the Galaxy \citep{2003A&A...409..523R}, finding at most a
0.3\% probability of thick disk membership using the
\citet{2009ApJ...705.1416R} RV and $<0.1\%$ for the \citet{qk10} RV.
Thus, although the derived space motions are different, the conclusion
that \lpint\ is a normal thin disk member is unaffected.  As a likely
thin disk member, the age of \lpint\ is essentially unconstrained by
its kinematics. Finally, the fact that \lpint\ is chromospherically
active \citep[$\log(L_{\rm H\alpha}/L_{\rm
  bol})$~=~$-$4.52;][]{2000AJ....120.1085G} could potentially provide
an age constraint, as the activity of M~dwarfs changes with
age. However, the long lifetime of activity in such late-M dwarfs
found by \citet{2008AJ....135..785W} places only a weak constraint of
$\lesssim9$~Gyr for the age of the \lpAB\ system.  Given the measured
mass and luminosities of \lpAB, it is in fact likely to be much
younger, as discussed in detail in the following sections.

For \lhsint, there is no published radial velocity, precluding the
kinematic analysis described above. The proper motion and parallax of
\citet{2009AJ....137.4109L} provide a measurement of its tangential
velocity ($V_{\rm tan} = 41.1\pm1.6$~\kms), indicating that \lhsint\
is typical of late-M dwarfs within 20~pc \citep[$V_{\rm tan} =
29\pm21$~\kms; ][]{2009AJ....137....1F} and not likely to be a member
of the thick disk or halo populations. Both
\citet{2003AJ....126.3007R} and \citet{2009AJ....137.4109L} have shown
that \lhsint\ is not active, which is uncommon for nearby, late-M
dwarfs. \citet{2008AJ....135..785W} found that activity lifetime
increases monotonically with M~dwarf spectral type, and for M6 and M7
dwarfs these lifetimes are $7.0\pm0.5$~Gyr and
$8.0^{+0.5}_{-1.0}$~Gyr, respectively. Thus, the lack of activity seen
in \lhsint\ (M6.5) indicates that its age is likely to be at least as
old as 6~Gyr, the 2$\sigma$ lower limit for the activity lifetime of
M6 and M7 dwarfs.

The age of the M2.5 primary star in the Gl~569 triple system has been
discussed extensively in the literature. \citet{2004astro.ph..7334O}
find that the space motion of Gl~569A places it close to, but not
likely a member of, the Ursa Major moving group
\citep[300~Myr;][]{1993AJ....105.2299S}.  However, they suggest that
it may be a member of a stream of young A and F stars identified by
\citet{1999A&AS..135....5C}, which have ages of 300--800~Myr as
derived from Str\"{o}mgren photometry. \citet{2006ApJ...644.1183S}
prefer a younger age of 100--125~Myr because Gl~569A seems to lie on
the Pleiades sequence in the color--magnitude diagram of
\citet{2005ApJ...628L..69L}.  We have considered whether the X-ray
emission from Gl~569A may also place an independent constraint on the
age of the system. We computed the X-ray flux of Gl~569A from its
\ROSAT\ count rate, using the \citet{1995ApJ...450..392S} conversion
factor, finding $F_X =
(3.03\pm0.30)\times10^{-12}$~erg~cm$^{-2}$~s$^{-1}$ ($\log{L_X} =
28.54$). This lies between the median values of $L_X$ for M dwarfs in
the Pleiades (125~Myr) and Hyades (625~Myr) as determined by
\citet{2005ApJS..160..390P}. The X-ray luminosity of Gl~569A is higher
than $\approx$75\% of Hyades objects and is fainter than $\approx$70\%
of Pleiades objects. This suggests an intermediate age of
$\sim$300--450~Myr for the Gl~569 system, although its X-ray
luminosity is consistent with the full range of $\sim$125--625~Myr at
1$\sigma$.


\section{Tests of Models \label{sec:tests}}

Directly measured dynamical masses enable unique tests of theoretical
models of very low-mass stars and brown dwarfs. We consider two
independent sets of evolutionary models from the Tucson
\citep{1997ApJ...491..856B} and Lyon groups
\citep[Dusty;][]{2000ApJ...542..464C}.  Following the approach
developed in our previous work \citep[e.g.,][]{2008ApJ...689..436L,
  2009ApJ...692..729D}, we use the most directly measured properties
of our sample binaries (i.e., total mass and individual component
luminosities) to derive additional properties (e.g., \Teff) from
evolutionary models. These model-derived properties are then compared
to other available constraints (e.g., from atmospheric model fitting).
We note that currently available evolutionary models do not
incorporate many of the theoretical advances used in the latest model
atmospheres, such as detailed dust cloud models and more accurate
molecular and dust opacities.  However, our line of analysis is
necessarily centered on evolutionary models because we directly
measure mass and \Lbol, not the model atmosphere parameters \Teff\ and
\logg.  In addition to this analysis, we also explore the approach
recently utilized by \citet{qk10} in order to directly compare our
results to their findings for the same models.

\subsection{Model-derived Ages \label{sec:modelage}}

\begin{figure}
\centerline{\includegraphics[width=3.5in,angle=0]{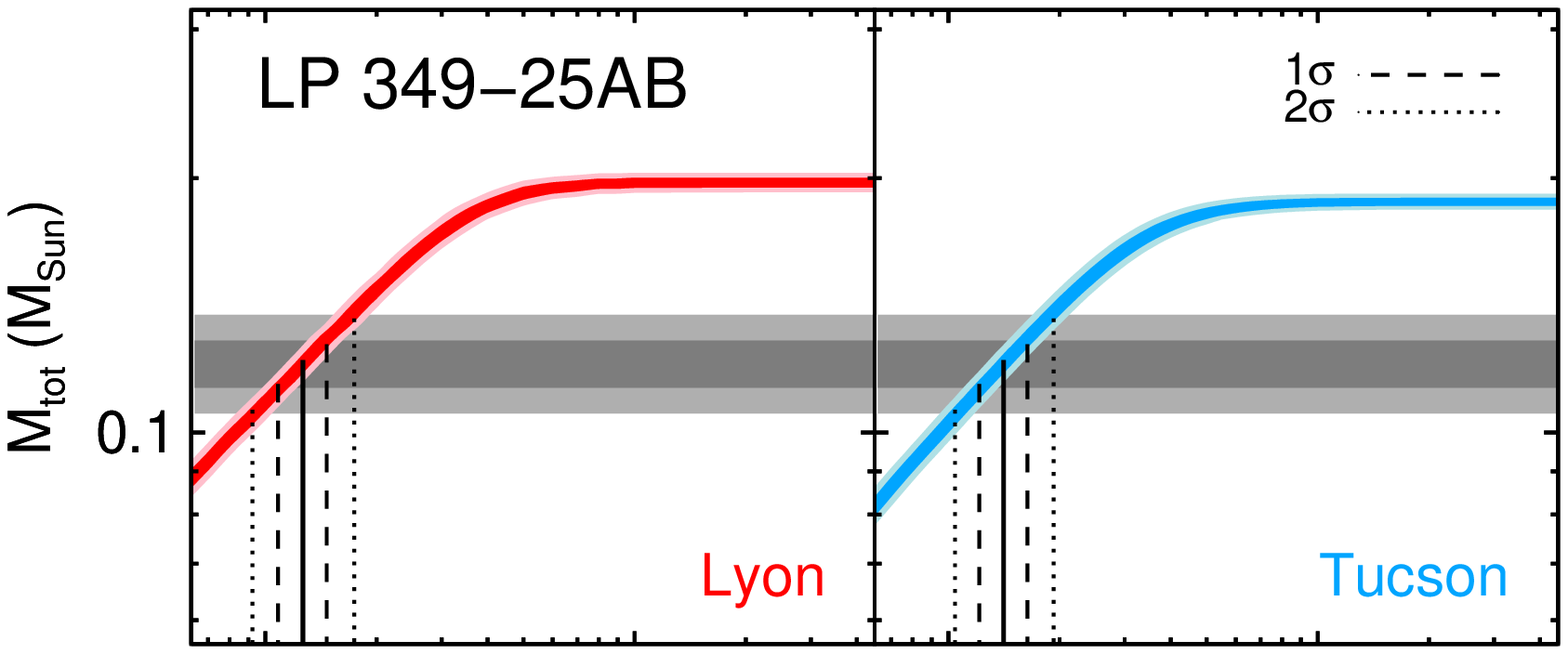}}
\vskip -0.3in
\centerline{\includegraphics[width=3.5in,angle=0]{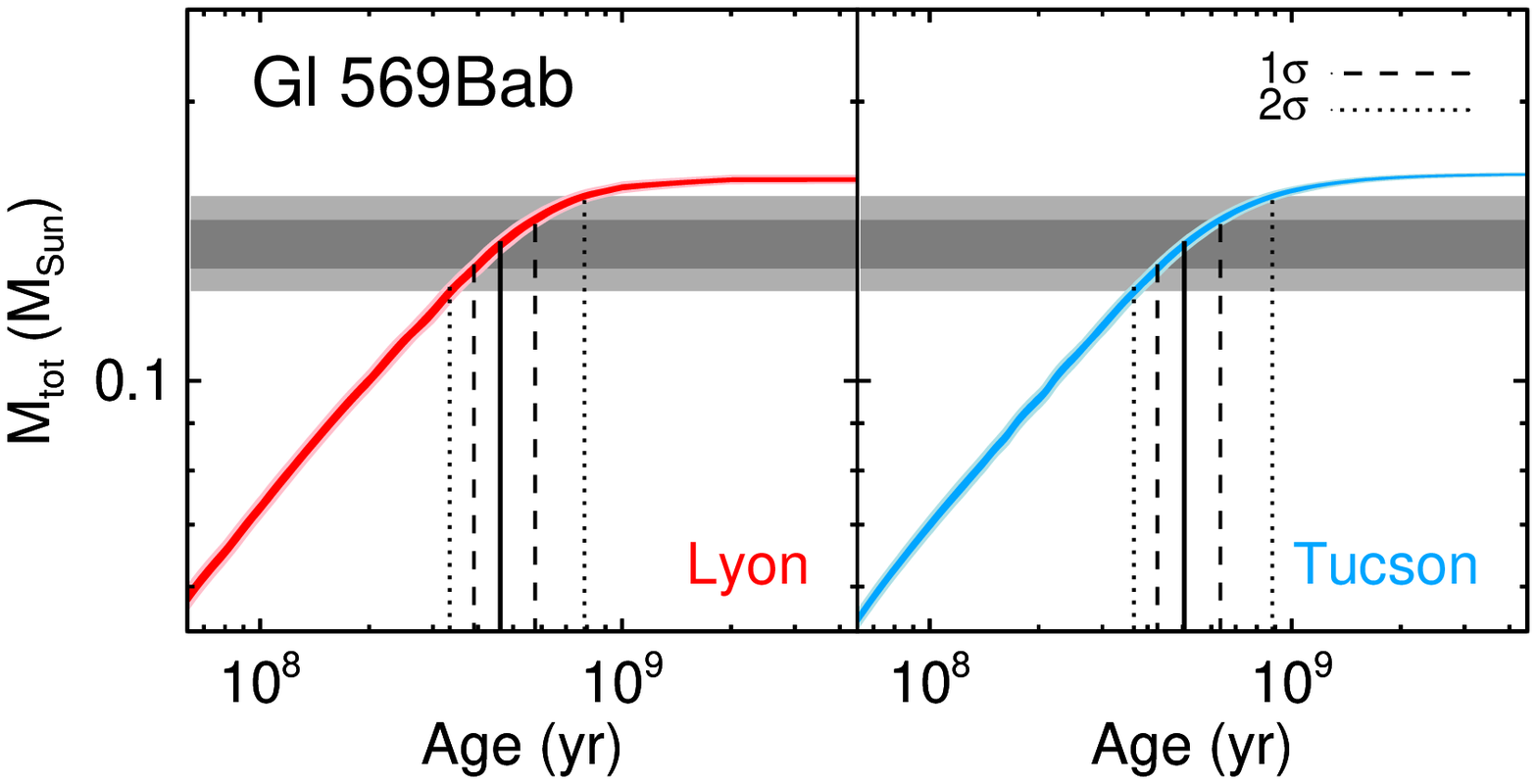}}

\caption{ \normalsize Total mass (\Mtot) of \lpAB\ and \glAB\
  predicted by evolutionary models as a function of age, given the
  observational constraint of the luminosities of the individual
  components. The curved shaded regions show the 1$\sigma$ and
  2$\sigma$ ranges in the model-derived masses, which increase as a
  function of age.  By applying the independent constraint of the
  measured total masses, we used Lyon models (left panel) and Tucson
  models (right panel) to derive ages for these binaries (see Tables
  \ref{tbl:lpmodel} and \ref{tbl:glmodel}).  The horizontal gray bars
  show our 1$\sigma$ and 2$\sigma$ constraints on the total mass, and
  the resulting median, 1$\sigma$, and 2$\sigma$ model-inferred ages
  are shown by solid, dashed, and dotted lines, respectively.  Because
  \lpAB\ has a lower mass than \glAB\ (and higher luminosities), its
  derived age is significantly younger than for
  \glAB.  \label{fig:mtotage}}

\end{figure}

As described in detail by \citet{2008ApJ...689..436L} and
\citet{2009ApJ...692..729D}, the total mass of a binary along with its
individual component luminosities can be used to estimate the age of
the system from evolutionary models. This age estimate can be quite
precise ($\approx$10\%) when both components are substellar, since the
luminosities of brown dwarfs depend very sensitively on age. This is
the case for \lpAB: the Lyon and Tucson models give consistent ages of
$0.127^{+0.021}_{-0.017}$~Gyr and $0.141^{+0.023}_{-0.019}$~Gyr,
respectively (Figure~\ref{fig:mtotage}, Table~\ref{tbl:lpmodel}).  The
model-derived age of the system is consistent with its kinematics
(Section~\ref{sec:age}), and the implications of such a young age for
the \lpAB\ system are addressed in detail in Section~\ref{sec:lpAB}.

Because \lhsAB\ is composed of two very low-mass stars whose
luminosities remain essentially constant over their main-sequence
lifetime, the model-derived age for this system is not very precise.
Lyon and Tucson models give consistent ages of
$0.28^{+9.72}_{-0.08}$~Gyr and $0.37^{+9.63}_{-0.15}$~Gyr,
respectively (Table~\ref{tbl:lhsmodel}).  The large upper limits to
the age distributions are set by the oldest age for which the model
grids are computed (10~Gyr).  These broad age distributions are
consistent with the age constraint that comes from the lack of
chromospheric activity in \lhsint\ ($>6$~Gyr; Section~\ref{sec:age}).
Because the median model-derived ages are quite young, they would be
inconsistent with an age as old as 6~Gyr, but a more precise parallax
measurement is needed to test this. For example, if we fix the
parallax at its presently measured value but improve its precision
from 3.0~mas to 1.0~mas, an attainable goal with seeing-limited
astrometry, this age discrepancy would rise to $>2\sigma$ significance
for the Lyon models.

Gl~569Bab has a higher mass than \lpAB\ but lower component
luminosities, and thus the Lyon and Tucson models give older ages of
$0.46^{+0.11}_{-0.07}$~Gyr and $0.51^{+0.13}_{-0.08}$~Gyr,
respectively (Figure~\ref{fig:mtotage}, Table~\ref{tbl:glmodel}).
These are consistent with the age constraints from Gl~569A discussed
in Section~\ref{sec:age}, with the exception of the age of
100--125~Myr proposed by \citet{2006ApJ...644.1183S} based on its
position on the color--magnitude diagram. Given the unavailability of
such data for M~dwarfs at $\sim$500~Myr, we suggest that Gl~569A may
well be consistent with such an older age. Our model-derived age for
the \glAB\ system is also in agreement with the age estimates from
\citet{2001ApJ...560..390L} and \citet{2004astro.ph..7334O}.

\subsection{Individual Masses \label{sec:qratio}}

All of our target binaries have flux ratios near unity, thus enabling
robust individual mass estimates with only a very weak dependence on
model assumptions.  For \lhsAB, the Lyon and Tucson models give mass
ratios ($q$~$\equiv$~$M_{\rm B}/M_{\rm A}$) of
$0.958^{+0.015}_{-0.014}$ and $0.966^{+0.011}_{-0.016}$, respectively.
Thus, the model-derived individual masses (listed in
Table~\ref{tbl:lhsmodel}) are nearly identical, and their precision is
dominated by the $^{+13}_{-11}\%$ uncertainty in \Mtot.

\begin{figure} 

\centerline{\includegraphics[width=3.2in,angle=0]{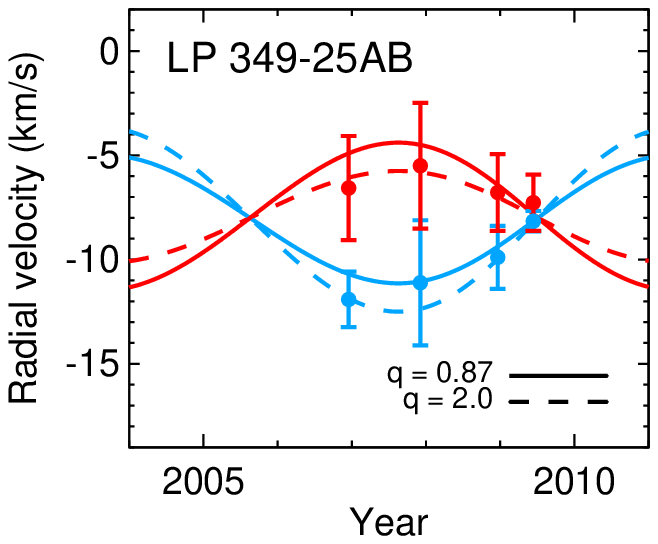}}

\caption{ \normalsize Radial velocities of \lpA\ (blue) and \lpB\
  (red) as measured by \citet{qk10} shown with curves predicted by our
  best-fit orbit, assuming two different mass ratios ($q \equiv
  M_2/M_1$).  Combining our measured luminosity ratio and total mass,
  evolutionary models predict $q = 0.87$, whereas \citet{qk10} found
  $q = 2.0$ from the best fit to their data.  Their 1--3~\kms\ errors
  result in an unrealistically low $\chi^2$ (0.5 for 6 DOF),
  suggesting that the errors are overestimated.  Within these large
  errors, we find that both mass ratios are allowed by their
  data. \label{fig:lp-rv}}

\end{figure}

Because the model-derived ages for \lpAB\ are quite young, the
inferred mass ratios are somewhat further from unity than they would
otherwise be for the measured $0.133\pm0.019$~dex luminosity
ratio. The Lyon and Tucson models give consistent mass ratios of
$0.872^{+0.014}_{-0.018}$ and $0.863^{+0.013}_{-0.019}$,
respectively. This results in individual masses of
$0.064^{+0.005}_{-0.004}$~\Msun\ and $0.056\pm0.004$~\Msun\ for the
components of \lpAB\ (Lyon), placing both below the substellar
boundary \citep[$\sim0.070$~\Msun;][]{2000ApJ...542..464C} and \lpB\
likely below the lithium depletion boundary
\citep[$\approx$0.055--0.065~\Msun;][]{1997A&A...327.1039C}.
\citet{qk10} reported resolved RV measurements for \lpAB\ and,
combined with their astrometric orbit, found $q = 2.0^{+3.0}_{-0.7}$.
This is nominally inconsistent with our model-derived mass ratio at
1.6$\sigma$, with their value corresponding to the secondary being
\emph{more} massive than the primary. In Figure~\ref{fig:lp-rv}, we
show their measurements along with the RV curve predicted for mass
ratios of 0.87 and 2.0, assuming their best-fit center-of-mass
velocity of $-8.0\pm0.5$~\kms. Fixing the orbital elements to the
values from the astrometric orbit, we find $\chi^2$ values for the RV
measurements that are unreasonably small for both values of $q$
($\chi^2 = 0.5$ for $q = 2.0$ and $\chi^2 = 1.7$ for $q = 0.87$; 6
DOF). This implies that the errors on the RV measurements are
overestimated, and within these large errors both values of the mass
ratio (0.87 and 2.0) are acceptable.  We discuss the plausibility of
these different mass ratios in detail in Section~\ref{sec:lpAB}.

For \glAB, the Lyon and Tucson models give consistent mass ratios of
$0.866^{+0.019}_{-0.014}$ and $0.886^{+0.021}_{-0.017}$,
respectively. This agrees with the spectroscopically determined mass
ratio of 0.71$^{+0.19}_{-0.13}$ from \citet{qk10}.
Our slightly higher model-derived mass ratio results in somewhat more
similar masses ($0.075\pm0.004$ and $0.065^{+0.005}_{-0.004}$~\Msun,
Lyon) than implied by the directly measured mass ratio
($0.082^{+0.008}_{-0.009}$ and $0.059^{+0.009}_{-0.007}$~\Msun). In
both cases however, \glB\ is expected to lie near the mass-limit for
lithium burning, which can be tested directly in the future with
resolved optical spectroscopy of the \ion{Li}{1} doublet at 6708~\AA.
In addition, absolute astrometric monitoring of the 2.4-yr orbital
period binary can readily yield a refined measurement of the mass
ratio: 1~mas astrometry would give a 5\% mass ratio error,
$\sim$4$\times$ better than spectroscopy.

\subsection{Temperatures and Surface Gravities \label{sec:teff-logg}}

\begin{figure}
\vskip -0.4in
\centerline{\includegraphics[width=6.2in,angle=0]{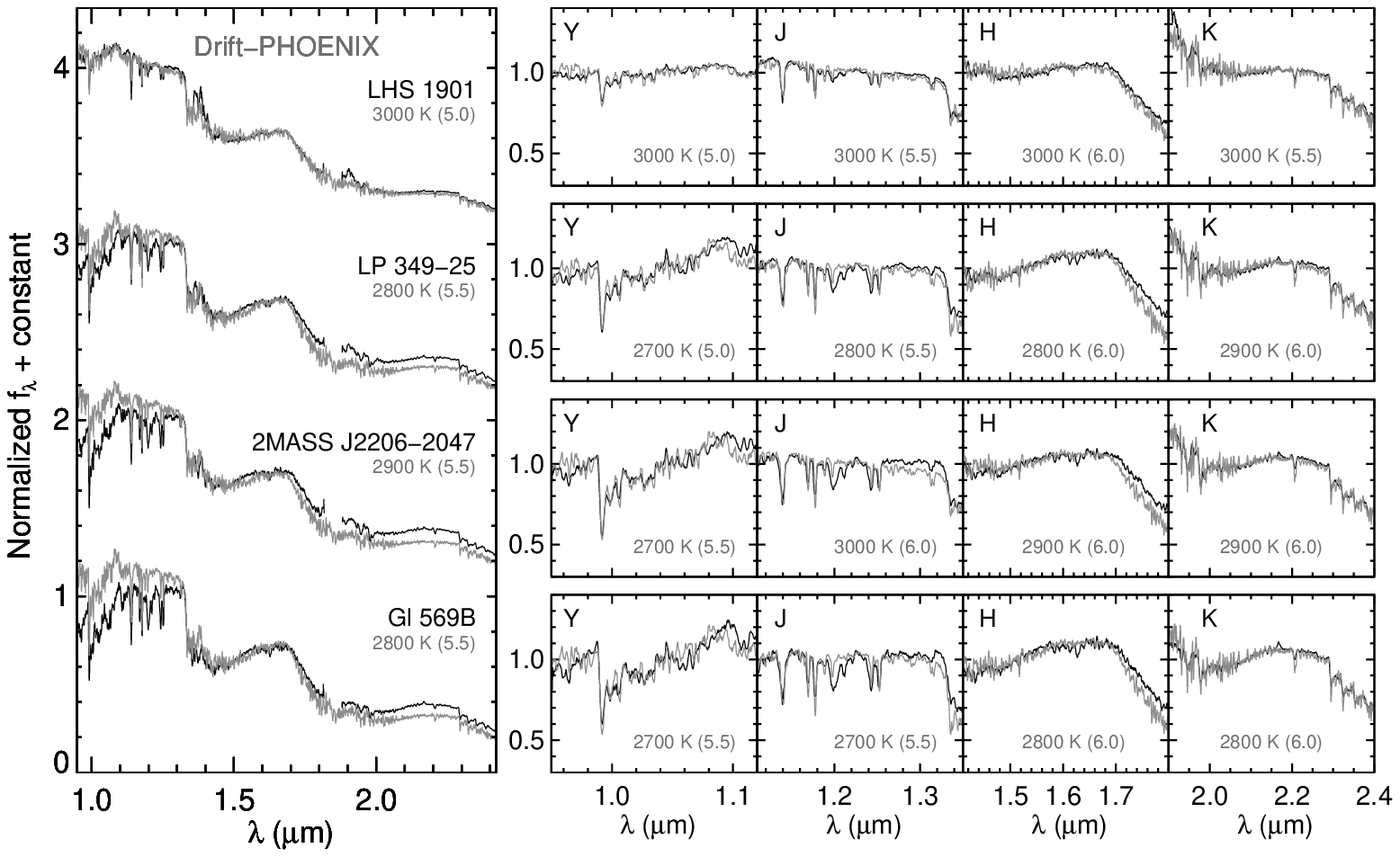}}
\vskip -0.1in
\centerline{\includegraphics[width=6.2in,angle=0]{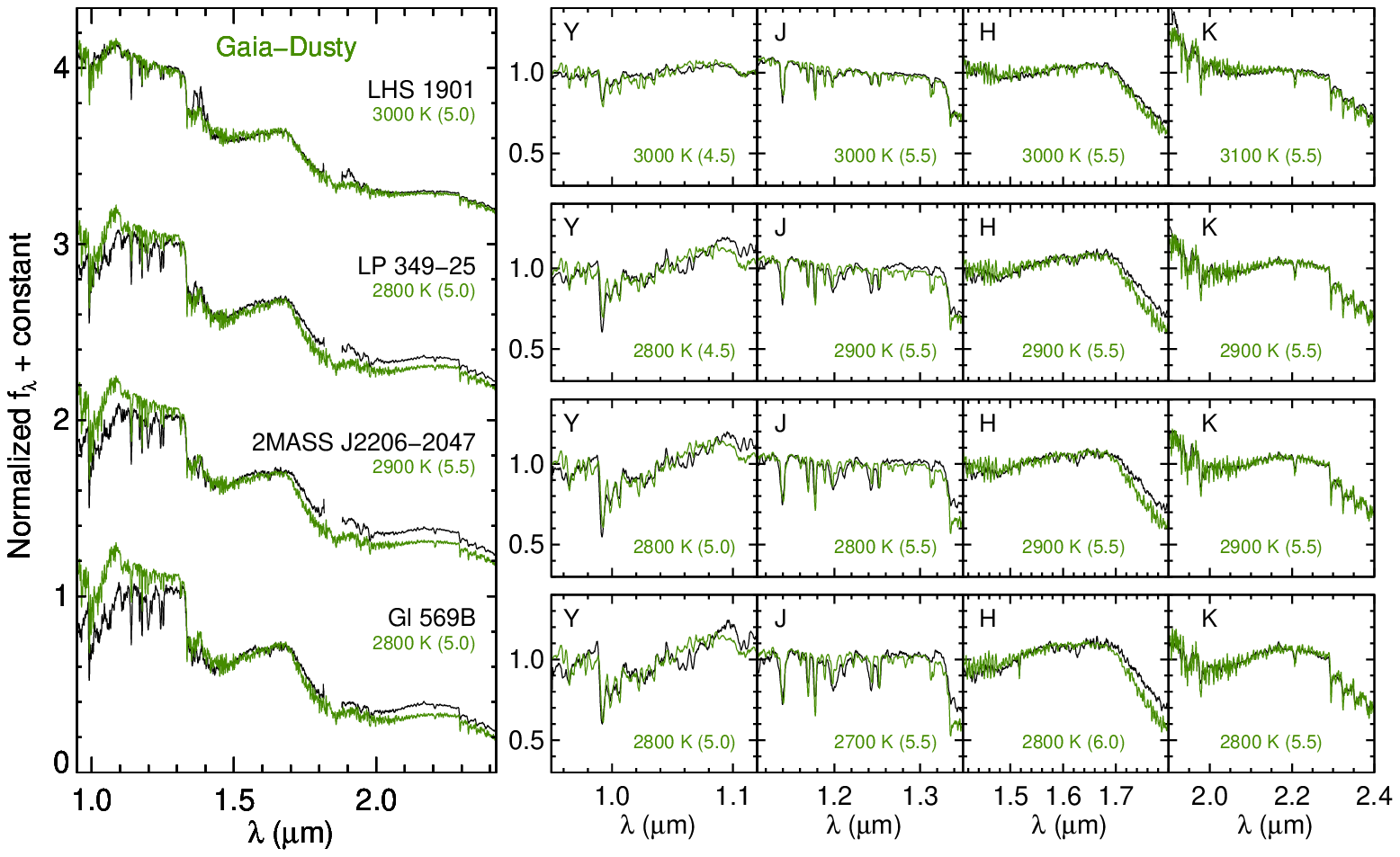}}

\caption{ \normalsize Integrated-light SpeX SXD spectra (black) shown
  with the \emph{best fitting atmospheric model spectra} (top:
  Drift-PHOENIX, gray; bottom: Gaia-Dusty, green).  Each best-fit
  model spectrum is labeled indicating its effective temperature, with
  surface gravity in parentheses.  We performed separate fits for the
  full NIR SED (left) and individual bands (right).  All spectra are
  smoothed by 10~pixels $\approx$ 0.003~\micron\ for display
  only.  \label{fig:atm-fit}}

\end{figure}

Effective temperature (\Teff) is one of the most difficult properties
to measure directly, as it requires a direct measurement of the radius
and luminosity. To date, radius measurements remain elusive for very
low-mass stars and brown dwarfs in the field. Thus, when testing model
predictions we are restricted to consistency checks between different
methods of estimating \Teff. In Section~\ref{sec:atm-fit}, we
estimated the temperatures of the primary components of our sample
binaries by fitting their integrated-light spectra with atmospheric
models. The best-fit Drift-PHOENIX and Gaia-Dusty model spectra are
shown in Figure~\ref{fig:atm-fit} plotted over the observed spectra.
Evolutionary models provide a nearly independent temperature estimate
by providing a prediction of the radius, which yields \Teff\ when
combined with the measured \Lbol.\footnote{Nearly independent because
  evolutionary models use theoretical atmospheres as boundary
  conditions when evaluating hydrostatic equilibrium and the amount of
  energy being released.  This can result in radius deviations of up
  to $\approx$10\% when comparing dustless to extremely dusty model
  atmospheres at low temperatures ($\Teff \lesssim 2000$~K), but the
  effect is $<1\%$ at the warmer temperatures of late-M dwarfs
  \citep[see Figure~2 of][]{2000ApJ...542..464C}.}

For \lhsA\ and \lhsB, Lyon evolutionary models give effective
temperatures of $2860\pm50$~K and $2820^{+50}_{-40}$~K, respectively,
and the Tucson evolutionary models give 100~K warmer temperatures of
$2960\pm30$~K and $2930^{+30}_{-40}$~K. Both sets of models give
consistent surface gravities of \logg = 5.2 (cgs) for both components.
All NIR atmospheric model fits gave systematically higher temperatures
of 3000--3100~K and surface gravities near the predicted value
($5.0\pm0.5$~dex). Given the nominal uncertainty (i.e., the model
atmosphere grid step) of 100~K in the best-fit \Teff\ values, the
largest discrepancy of $\approx$150~K with the Lyon evolutionary
models is marginally significant. To provide an additional point of
comparison for the \Teff\ estimates, we compiled all M6.5 and M7
dwarfs with effective temperatures derived by
\citet{2007ApJ...667..527G} using the nearly model-independent
infrared flux method \citep{1977MNRAS.180..177B}. We found a mean and
standard deviation of $2710\pm30$~K for these objects, which is lower
than \Teff\ estimates from both classes of models by 140--290~K,
agreeing somewhat better with evolutionary models.

For \lpA\ and \lpB, Lyon models give effective temperatures of
$2660\pm30$~K and $2520\pm30$~K, respectively, and the Tucson models
give 120~K warmer temperatures of $2780\pm30$~K and $2640\pm30$~K.
Lyon models predict surface gravities of about 4.9 (cgs) for both
components, and Tucson models predict 5.0 (cgs). In comparison,
atmospheric model fits of \lpA\ gave 2800--3000~K and surface
gravities near the predicted value ($5.0\pm0.5$~dex).\footnote{We note
  that all SED fits including the 0.81 to 0.95~\micron\ optical
  segment gives a surface gravity of 4.0 (cgs) for \lpA, which is
  highly discrepant with the evolutionary model value of 5.0 (cgs).
  This would require an unrealistic radius of $\approx$0.4~\Rsun\
  given the known mass of \lpA. Such an unrealistic surface gravity is
  also found from the Ames-Dusty models for \lhsA.} Again, the Tucson
evolutionary model \Teff\ is consistent with that derived from
atmospheric models, while the Lyon evolutionary models are lower by
$\approx$150~K.

For \glA\ and \glB, Lyon models give effective temperatures of
$2430\pm30$~K and $2210\pm30$~K, respectively, and the Tucson models
give 100~K warmer temperatures of $2530\pm30$~K and $2300\pm30$~K.
Lyon models predict surface gravities of about 5.2 (cgs) for both
components, and Tucson models predict 5.3 (cgs). In comparison,
atmospheric model fits of \glA\ gave 2700--2900~K, with a wide range
of surface gravities ($\logg = 4.5$ to 6.0). These \Teff\ values are
much higher than evolutionary models by 150--250~K, and the broad
range of surface gravities is in reasonable agreement.

\begin{figure}
\vskip -0.4in
\centerline{\includegraphics[width=6.2in,angle=0]{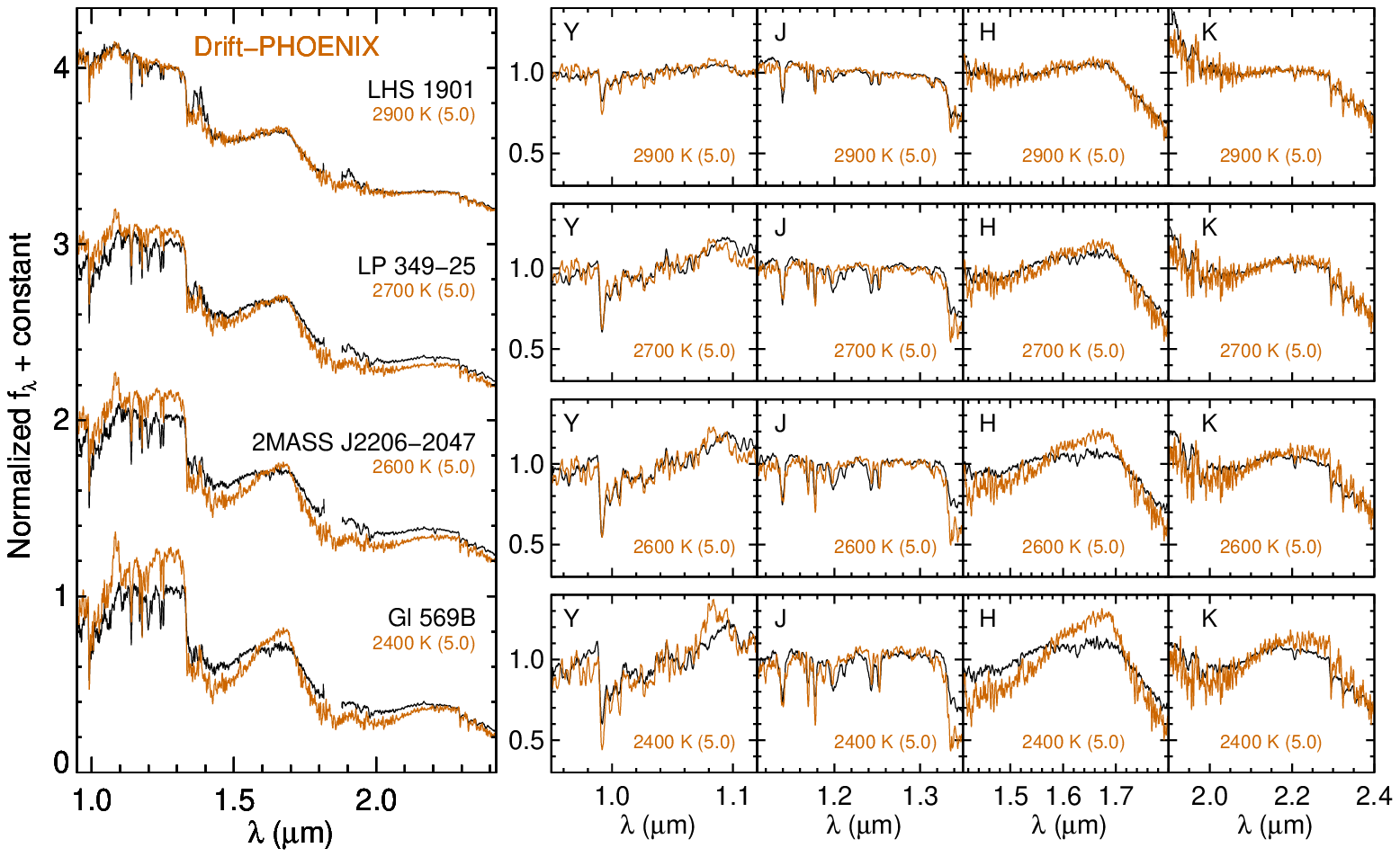}}
\vskip -0.1in
\centerline{\includegraphics[width=6.2in,angle=0]{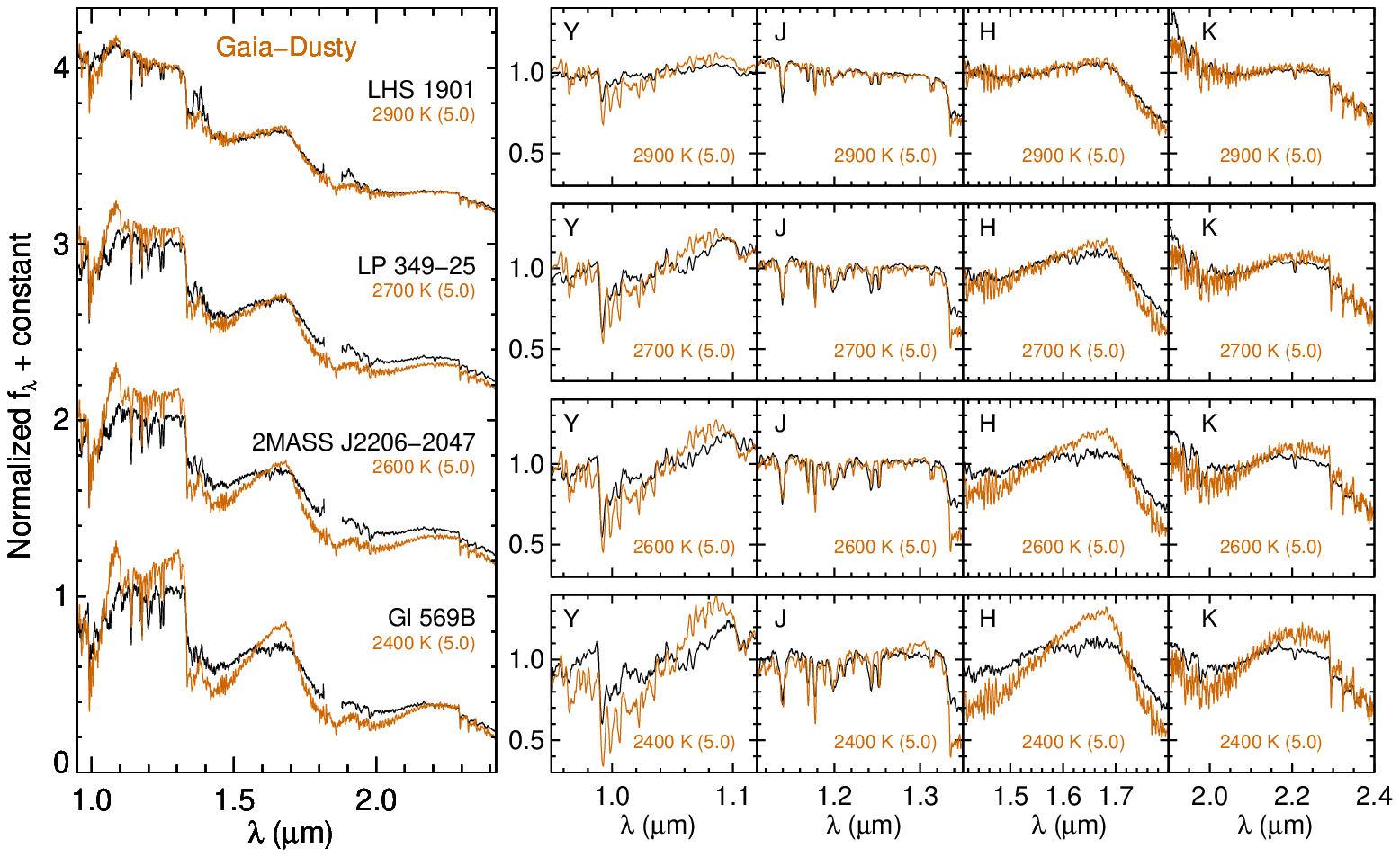}}
\vskip -0.1in

\caption{ \normalsize Integrated-light SpeX SXD spectra (black) shown
  with the \emph{atmospheric model spectra corresponding to the values
    of \Teff\ and \logg\ derived from Lyon Dusty evolutionary models}.
  Each model spectrum (brown) is labeled indicating its effective
  temperature, with surface gravity in parentheses.  All spectra are
  smoothed by 10~pixels $\approx$ 0.003~\micron\ for display only.
  The most prominent deviations in the model spectra are: the
  exaggerated shapes of $H$ and $K$ bands; the overpredicted depth of
  H$_2$O absorption at 1.33~\micron; and the strength of the FeH
  bandhead at 0.99~\micron.  Without directly measured radii we cannot
  determine if the problem lies in the model spectra, the evolutionary
  model radii, or both.  \label{fig:atm-evol}}

\end{figure}

In order to understand why temperature estimates from the two classes
of models typically do not agree, we show our observed spectra in
Figure~\ref{fig:atm-evol} plotted with Drift-PHOENIX and Gaia-Dusty
model spectra that have \Teff\ and \logg\ values closest to the Lyon
Dusty evolutionary model-derived values. This enables us to assess
what spectral features prevented our fitting procedure from selecting
these model spectra, which are very similar to the spectra used for
the atmospheric boundary condition in the Lyon models.\footnote{The
    Ames-Dusty model atmospheres provided the spectra used for the
    boundary condition in the Lyon Dusty evolutionary models.
    However, we choose to show the Gaia-Dusty models for comparison as
    their SEDs are very similar but with updated line lists
    (Table~\ref{tbl:atm-ref}). This enables us to identify remaining
    problems and not dwell on issues that have already been resolved
    in the latest generation of that model grid.}
The NIR SED fits do not seem to be significantly affected by broadband
colors (e.g., the better fitting models do not necessarily have $J-K$
colors closer to what is observed). Rather, the shapes of the spectra,
which are sculpted by broad molecular absorption bands, seem to
determine the best-fit models. 
\begin{itemize}

\item The disagreement in shape is most pronounced at $H$ band, where
  the observed spectrum is much flatter than the low-\Teff,
  low-gravity model spectra that would better correspond to
  evolutionary model-derived properties. This could be due partly to
  missing FeH opacity in the models between $\sim$1.5 and 1.7 \micron\
  \citep{2001AJ....121.1710R, 2001ApJ...559..424W,
    2003ApJ...582.1066C}.

\item The observed $K$-band shape is also flatter (i.e., the band peak
  at 2.2~\micron\ is more suppressed) than in the low-\Teff, low-gravity
  model spectra.

\item In $J$ band, the fit seems to be driven by the deep H$_2$O
  absorption band at 1.33~\micron, which is too deep in the low-\Teff,
  low-gravity model spectra.

\item In $Y$ band, the FeH 0$-$0 bandhead at 0.99~\micron\ is too
  strong in the spectra corresponding to evolutionary model-derived
  properties.  This could alternatively be interpreted as the overall
  continuum at $Y$ band being too high in the atmospheric models
  (e.g., missing opacity from wings of resonant alkali lines such as
  \ion{K}{1}).

\end{itemize}

\begin{figure}

\centerline{\includegraphics[width=6.2in,angle=0]{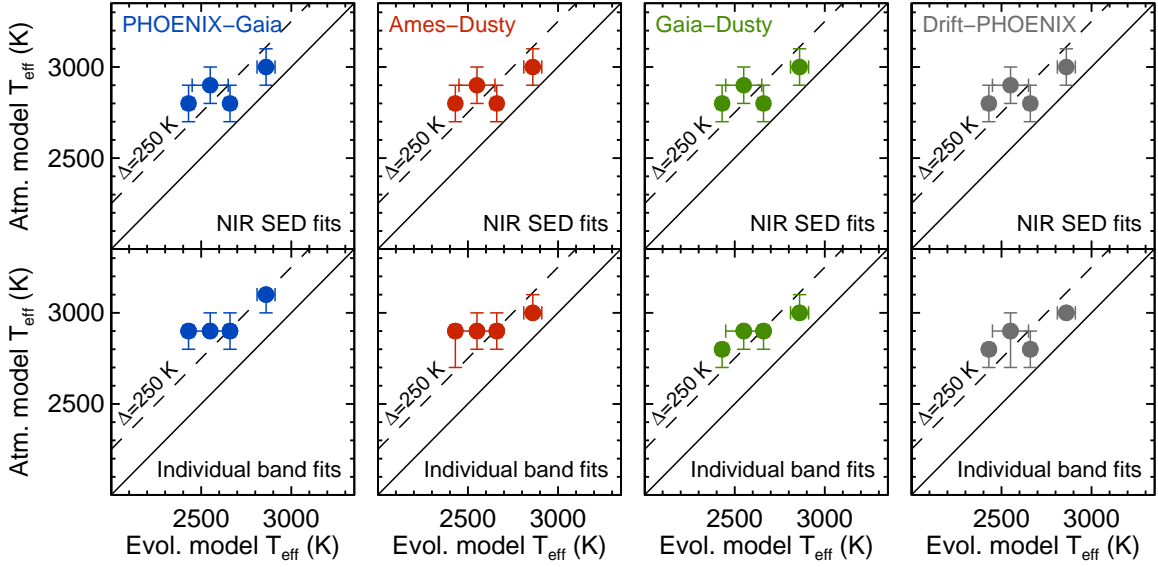}}

\caption{ \normalsize Effective temperatures determined from model
  atmosphere fitting compared to Lyon Dusty evolutionary model-derived
  \Teff\ (i.e., from measured total mass and individual luminosities).
  Results from fitting NIR SEDs (top; 100~K error bars) are shown
  separately from the results of fitting individual bandpass (bottom;
  error bars indicate full range of band fits).  Measurements
  typically lie above (or to the left of) the line of equality.  An
  offset of 250~K can largely account for the observed discrepancies
  (dashed line), implying that either evolutionary model estimates are
  too cool (i.e., radii too large by 15--20\%) or that atmospheric
  model estimates are too warm by 250~K.  \label{fig:dteff}}

\end{figure}

In general, we have found that atmospheric models predict
systematically higher effective temperatures than evolutionary
models.\footnote{\citet{2010ApJS..186...63R} have recently fit model
  atmospheres to high- and medium-resolution $J$-band spectra of mid-
  to late-M dwarfs.  Most of their sample are young M dwarfs
  ($\lesssim10$~Myr), but for their three field M7 to M9 dwarfs they
  find higher \Teff\ values than previously published estimates based
  on measured luminosities and estimated radii for objects of similar
  spectral type.  Thus, their findings are consistent with our
  results.}  This pattern is illustrated in Figure~\ref{fig:dteff},
which shows the \Teff\ estimates from each class of models plotted
against each other. A systematic shift of $\approx$250~K would bring
the models into agreement, with either the atmospheric model estimates
being too warm or the evolutionary model estimates too cool (i.e.,
model radii too large by 15--20\%).\footnote{ Interestingly, the SED
  fits that include the wavelength segment from 0.81 to 0.95~\micron\
  (which are less preferred for reasons described in
  Section~\ref{sec:atm-fit}) give \Teff\ values $\approx$100~K more
  consistent with evolutionary models. This indicates that optical
  diagnostics utilizing TiO bandhead strengths \citep[e.g.,
  see][]{2004ApJ...609..854M} may be needed to provide more accurate
  \Teff\ estimates than can be obtained from near-infrared
  spectroscopy alone.}  We note that the observed discrepancy cannot
simply be due to the fact that our method relies on fitting the
integrated-light spectra of binaries. As discussed in
Section~\ref{sec:atm-fit}, if the secondary component's flux impacted
the best-fit model it would have the effect of lowering the derived
\Teff, whereas we find temperatures that are too warm.  This means
that the discrepancy could only be larger than we observe.  In
addition, even a significant nonsolar metallicity for our binaries
could not explain the observed temperature offset, as
\citet{2009ApJ...706.1114B} showed that models of varying metallicity
gave the same best-fit \Teff\ to within the model grid step of 100~K
for the ultracool subdwarf HD~114762B (d/sdM9).

We have considered what systematic errors could contribute to this
discrepancy between evolutionary and atmospheric models.  Currently
available evolutionary models use much older versions of model
atmospheres as their boundary condition, and this could potentially
result in systematic errors in their output.
\citet{2008ApJ...689.1327S} calculated evolutionary models both with
and without the effects of dust clouds in the atmosphere and found at
worst 6\% differences in the resulting radii, which is much lower than
what is needed to account for our observed discrepancy.  We can also
estimate how other input assumptions impact evolutionary model output
simply by comparing the predictions of the Lyon and Tucson models,
which use different helium abundances, model atmospheres, and dust
treatment.  These two sets of models predict radii different by
7--10\% and temperatures different by $\approx$100~K for the objects
in our sample -- more than a factor of two lower than what is needed
to fully account for the observed discrepancy.  Presently available
model atmospheres are more advanced than evolutionary models, but as
shown in Figure~\ref{fig:atm-fit} they cannot completely match our
observed near-infrared spectra.  Therefore, the resulting temperature
values must harbor systematic errors at some level.  Thus, we conclude
that the model atmospheres must be responsible for at least part of
the observed 250~K discrepancy.  In fact, we suggest that the
atmospheric models are more likely than the evolutionary models to be
the primary source of the discrepancy, since roughly the same \Teff\
offset is observed over a wide range of masses, ages, and activity
levels but the same temperature range.  This proposition will be
readily testable with eclipsing binary measurements given the large
implied radius difference between models (15--20\%).

\subsection{Near-Infrared Colors \label{sec:cmd}}

\begin{figure} 

\centerline{\includegraphics[width=6.5in,angle=0]{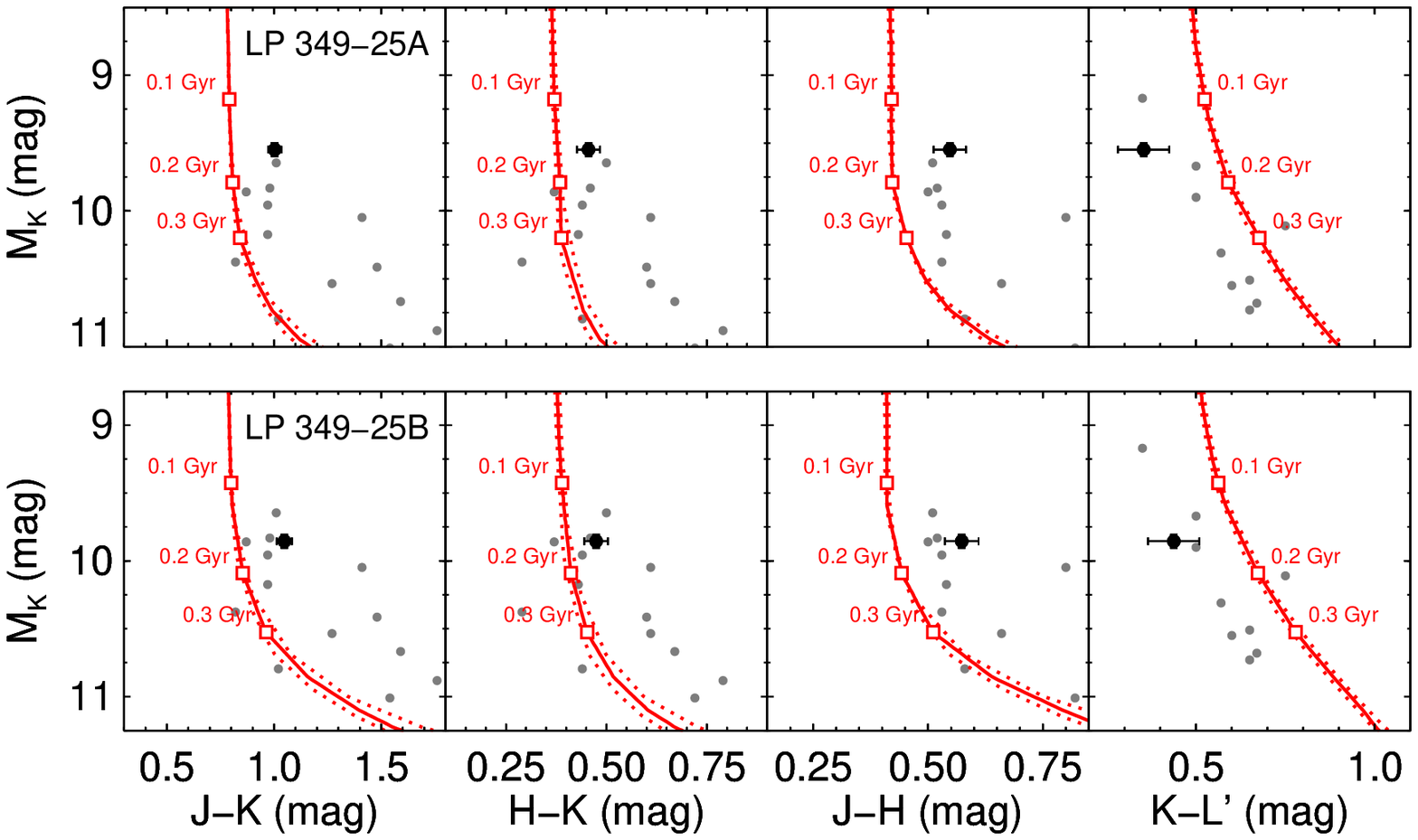}}

\caption{ \normalsize Color-magnitude diagrams showing the measured
  photometry of \lpA\ (top) and \lpB\ (bottom) compared to Lyon
  evolutionary tracks (all photometry on the MKO system).  The solid
  lines are isomass tracks from the Dusty \citep{2000ApJ...542..464C}
  models with dotted lines encompassing the 1$\sigma$ mass
  uncertainties, and open squares demarcate ages of 0.1, 0.2, and
  0.3~Gyr.  Field dwarfs with parallax measurements are shown for
  comparison as filled gray circles.  LP~349-25A and B are somewhat
  redder (0.1--0.2~mag) than evolutionary model tracks for $JHK$
  colors, while their $K-\Lp$ colors are 0.1--0.2~mag bluer than
  models. \label{fig:lp-jhk}}

\end{figure}

\begin{figure} 

\includegraphics[width=6.5in,angle=0]{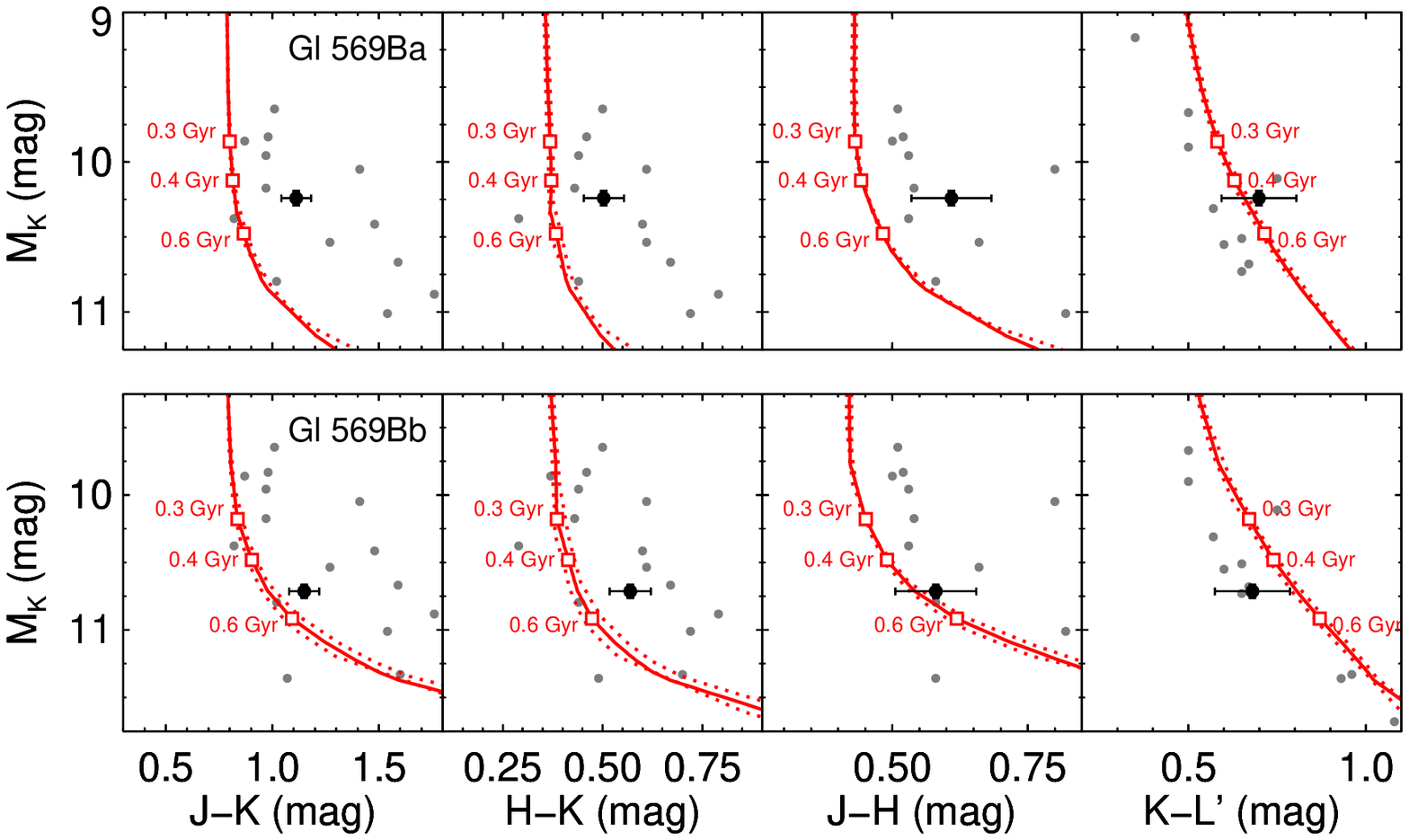}

\caption{ \normalsize Same as Figure~\ref{fig:lp-jhk} but for
  \glAB. \label{fig:gl-jhk}}

\end{figure}

We computed resolved photometry for our sample binaries in the MKO
photometric system to compare to Lyon model predictions.  For \glAB,
we used our integrated-light MKO photometry described in
Section~\ref{sec:keck} along with the best available flux ratios from
Table~\ref{tbl:astrom}.  For \lpint\ and \lhsint, we first converted
their integrated-light 2MASS photometry to the MKO system by deriving
correction terms from our near-infrared spectra.  The resulting
component magnitudes for all three binaries are listed in
Table~\ref{tbl:meas}.  Near-infrared colors on the MKO photometric
system were derived from the Lyon models in the same fashion as other
properties (e.g., \Teff), and the results are listed in
Tables~\ref{tbl:lpmodel}, \ref{tbl:lhsmodel}, and \ref{tbl:glmodel}.
In Figures~\ref{fig:lp-jhk} and \ref{fig:gl-jhk}, we show the
photometry of \lpAB\ and \glAB\ compared to Lyon mass tracks computed
for the individual masses of both components. The observed $JHK\Lp$
photometry is significantly discrepant with model tracks at all ages,
with models typically predicting $\approx$0.1--0.2~mag bluer $JHK$
colors than observed and $\approx$0.1--0.2~mag redder $K-\Lp$
colors. For \lhsAB, the $JHK$ colors predicted by models are in much
better agreement with the observations.  The most discrepant color is
$H-K$, which is only 0.07~mag (2.3$\sigma$) redder than predicted, and
this may be due to missing FeH opacity in the $H$ band.  (We do not
show a corresponding figure for \lhsAB\ as its components are
main-sequence stars that evolve at essentially constant color.)

\subsection{An Alternative Approach: Model-derived Properties Using
  \Lbol\ and \Teff \label{sec:alt}}

\citet{qk10} have recently described an approach to test models of
ultracool dwarfs with dynamical mass measurements that is somewhat
different from ours.  They use the measured individual luminosities
along with \Teff\ estimates for each component to derive any other
property (e.g., \Mtot\ or age) from evolutionary models.  (Their
\Teff\ estimates are derived by fitting model atmospheres to resolved
broadband photometry of the binaries.)  This approach is similar to
ours in that it uses two quantities to determine a third from
evolutionary models, which is necessary because brown dwarfs occupy a
mass--\Lbol--age (or \Teff--\Lbol--age) relation instead of the
simpler mass--\Lbol\ relation for stars on the main sequence. In
\citet{2008ApJ...689..436L}, we considered in detail such model tests
using different combinations of these fundamental parameters, and we
ultimately chose to use mass and \Lbol\ to derive other properties as
these are the most directly measured quantities.  However, any
discrepancies we have found in the models should also manifest
themselves in the \Teff--\Lbol\ approach that has been adopted by
\citet{qk10}.  In this section, we examine the results of analyzing
our late-M dwarf measurements using the \citet{qk10} approach.

\begin{figure} 

\centerline{
\includegraphics[width=3.2in,angle=0]{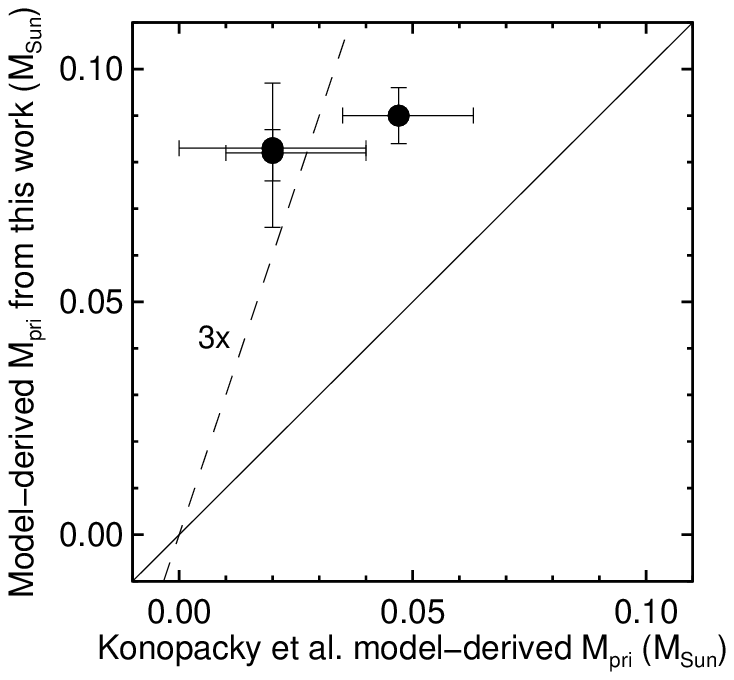}
\includegraphics[width=3.2in,angle=0]{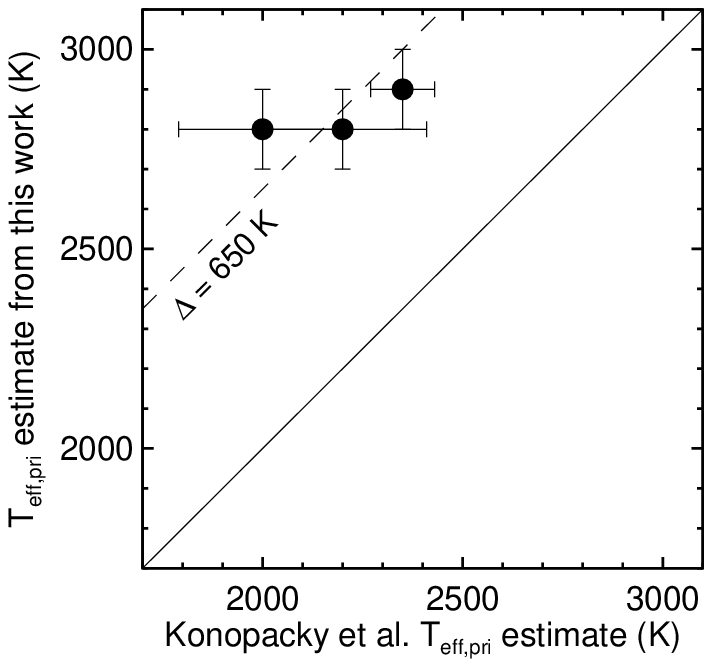}}

\caption{ \normalsize Comparison of the primary component masses
  derived from evolutionary models using the \Lbol--\Teff\ approach by
  \citet{qk10} and by us. The masses we derive using this approach are
  systematically higher by a factor of $\sim$2 to 4 (left) because our
  model atmosphere fitting of integrated-light spectra yielded
  $\approx$650~K higher temperatures than their fitting of resolved
  broadband photometry (right).  We suggest that the sensitivity of
  this model testing approach to the input \Teff\ estimates makes it
  more likely to identify problems with model atmospheres than
  evolutionary models. \label{fig:dmass}}

\end{figure}

We first interpolated the Lyon and Tucson model grids at individual
(\Teff, \Lbol) points in the same fashion as \citet{qk10}, with our
temperatures coming from the Gaia-Dusty atmospheric model fitting of
the NIR SED as described in Section~\ref{sec:atm-fit}.  (This is the
exact same model atmosphere grid as used by \citealp{qk10}.)  In order
to compute uncertainties in the model-derived properties using this
approach, we used $10^3$ luminosity and temperature values randomly
drawn from normal distributions corresponding to the measurement
errors. We assumed 100~K errors in \Teff, equal to the model
atmosphere grid step. The resulting evolutionary model-derived masses
are plotted in Figure~\ref{fig:dmass} against those derived by
\citet{qk10}.

Our masses are significantly higher than theirs by about a factor of
$\sim$2 to 4 because our \Teff\ estimates are systematically higher
while their luminosity measurements are very similar to ours.  In
order for models to match our much higher temperatures at similar
luminosities, the masses must be larger and the ages somewhat
older. The best illustration of this disagreement is the case of
2MASS~J2206$-$2047A, as \citet{qk10} used essentially identical
luminosities and photometry as \citet{me-2206}, and the component
fluxes are nearly identical, removing any ambiguity in fitting its
integrated-light spectrum.  \citet{qk10} found $2350\pm80$~K from
fitting its resolved broadband photometry, while our best-fit spectrum
from the very same grid of model atmospheres has a temperature of
$2900\pm100$~K.  Therefore, despite the other similarities in our
inputs, we found a primary mass of $0.090^{+0.006}_{-0.006}$~\Msun\
using the Lyon Dusty evolutionary models, while they found
$0.047^{+0.016}_{-0.012}$~\Msun.  Neither of these masses agrees with
the measured mass ($0.077^{+0.012}_{-0.017}$~\Msun, Table~7 of
\citealp{me-2206}): our mass is higher, and the \citet{qk10} mass is
much lower.  The differences between our derived masses for \lpA\ and
\glA\ are even larger, as \citet{qk10} found extremely low masses for
these objects ($0.02\pm0.02$~\Msun\ and $0.02^{+0.02}_{-0.01}$~\Msun,
respectively).\footnote{We note that the extremely low model-derived
  masses of $\approx$20~\Mjup\ reported by \citet{qk10} imply very
  young ages that are inconsistent with the lack of any low-gravity
  signatures in these objects' spectra.}

Fundamentally, both the \citet{qk10} approach and our standard method
to test models provide similar information: a consistency check on the
temperatures from atmospheric and evolutionary models. The comparison
presented here highlights how the amplitude, and even the sign, of
discrepancies found between measured and model-predicted masses depend
strongly on the input temperature estimates. Using the exact same
atmospheric model grids, but different methods, our two groups have
arrived at values of \Teff\ different by $\approx$650~K. This is due
to the fact that the fitting method of \citet{qk10} relies solely on
the broadband colors of models, whereas our method is driven by the
shape of specific spectral features. As shown in
Figure~\ref{fig:lp-jhk} and in our previous work
\citep{2008ApJ...689..436L, me-2206, 2009ApJ...692..729D, me-2397a},
the broadband colors of models consistently disagree with observations
of objects of known mass over a broad range of spectral types. We also
note that a significant disadvantage of the \citet{qk10} \Lbol--\Teff\
approach is that evolutionary models occupy a very thin strip in this
parameter space as opposed to \Lbol--$M$. This resulted in many of our
randomly drawn (\Lbol, \Teff) measurements falling outside the range
predicted by models (from 1.2\% to 99.8\% in the worst case), and we
simply excluded these points from the analysis.

\citet{qk10} interpreted their observed discrepancies with theory as
an error in the evolutionary model cooling curves, i.e., a mass
problem. Using their method, we find an opposite mass discrepancy
because our \Teff\ estimates are $\approx$650~K higher than theirs,
resulting in higher masses. We suggest that the strong sensitivity of
this model-testing method to the input \Teff\ estimates means that any
observed discrepancies more readily identify problems in the way
\Teff\ is determined from atmospheric models (i.e., a temperature
problem), rather than a problem with evolutionary models.


\section{The Nature of \lpAB \label{sec:lpAB}}

It is somewhat surprising that \lpAB\ has turned out to be a pair of
young ($\sim$140~Myr) brown dwarfs given the lack of detectable
lithium \citep{2009ApJ...705.1416R} and its M$8.0\pm0.5$ spectral
classification \citep{2000AJ....120.1085G}. With a total mass of
$0.120^{+0.008}_{-0.007}$~\Msun, one or both of the components may
even lie below the theoretically predicted lithium depletion boundary
at $\approx$0.055--0.065~\Msun. In fact, even objects massive enough
to deplete lithium take time to achieve the internal temperature
necessary to do so
\citep[$\approx2\times10^6$~K;][]{1991MmSAI..62..171P}.  For example,
\citet{1996ApJ...459L..91C} show that a 0.070~\Msun\ object takes
220~Myr to destroy 99\% of its initial lithium, while at 145~Myr only
$>0.080$~\Msun\ stars are similarly depleted. Using the Lyon and
Tucson models, we find predicted lithium fractions of
$0.68^{+0.24}_{-0.42}$ and $0.75^{+0.15}_{-0.52}$ for \lpA,
respectively, and $0.95^{+0.04}_{-0.18}$ and $0.96^{+0.03}_{-0.08}$
for \lpB. Thus, both sets of models predict that the primary may be
depleted in lithium, but that the secondary should retain nearly all
of its initial lithium. The presence of lithium in the fainter
secondary may be masked by the primary flux in the integrated-light
spectroscopy of the system, so resolved spectroscopy to detect the
\ion{Li}{1} doublet at 6708~\AA\ is needed to test this prediction of
brown dwarf evolutionary models directly.

Evolutionary models make very precise predictions for the age of
\lpAB\ from its measured total mass and individual luminosities. Taken
together, the Lyon and Tucson models predict an age of $140\pm30$~Myr.
This is consistent with the age of 125~Myr derived for the Pleiades by
\citet{1998ApJ...499L.199S} using the lithium depletion boundary.
Thus, we might expect \lpAB\ to obey roughly the same boundaries in
spectral type and color as found by \citet{1998ApJ...499L.199S}. The
lithium boundary in the Pleiades lies between spectral types of M6.5
and M7, with objects $\ge{\rm M}7$ having detectable lithium. Although
the spectral type determinations for the components of \lpAB\ are
hampered by the lack of resolved spectroscopy, the measured
integrated-light type of M$8.0\pm0.5$ makes it unlikely that \lpA\ is
earlier than M7, yet it is clearly depleted in lithium.
\citet{1998ApJ...499L.199S} used spectroscopically derived $(R-I)$
colors to define the lithium depletion boundary in the Pleiades, with
all objects bluer than 2.20~mag being depleted in lithium. From the
integrated-light optical spectrum of \citet{2003AJ....126.3007R}, we
derived $(R-I) = 2.50$~mag using the same method as
\citet{1998ApJ...499L.199S}.\footnote{The method used by
  \citet{1998ApJ...499L.199S} was calibrated against M dwarfs in the
  Gliese catalog \citep{1991NSC3..C......0G}, and the scatter in (R-I)
  color at a given M subtype is 0.07 to 0.14~mag.  Therefore, we
  estimate the uncertainties in spectroscopically derived values of
  $(R-I)$ to be $\approx$0.10~mag.} At this color index, all brown
dwarfs in the Pleiades sample of \citet{1998ApJ...499L.199S} show
evidence for the presence of lithium.

LP~349-25A thus stands out as potentially anomalous compared to
determination of the lithium boundary in the Pleiades.  (Note that our
comparison is purely empirical in nature and does not depend on any
theoretical assumptions regarding lithium depletion in brown dwarfs.)
This indicates that there may be a problem with the model-derived age
for \lpAB.  At older ages, the lithium depletion boundary moves to
lower masses (i.e., later spectral types), so an older age for \lpAB\
would bring it into better agreement with the Pleiades. The age
derived in our analysis is based solely on the model-predicted \Lbol\
evolution, and for its actual age to be older either: (1) the measured
parallax would have to be smaller (i.e., \lpAB\ more distant); or (2)
the models would need to under-predict the luminosities of \lpAB. In
the former case, a parallax that is 3.5$\sigma$ smaller (placing the
binary at $14.3\pm0.3$~pc) would increase the total mass to
0.152~\Msun, resulting in a model-derived age $190\pm50$~Myr and
individual components predicted to be nearly fully depleted in
lithium.  In the latter case, the luminosities we measure would indeed
be high for their mass, not because the objects are very young but
rather because model \Lbol\ evolution is not correct. Such an error
has previously been suggested by \citet{2009ApJ...692..729D} to
explain the higher-than-predicted luminosities of the benchmark brown
dwarf binary HD~130948BC (L4+L4). An independent measurement of the
parallax for \lpAB\ and resolved optical spectroscopy is needed to
better assess this possible discrepancy with brown dwarfs in the
Pleiades.

The youth of \lpAB\ could potentially be reflected in low-gravity
spectral features such as have been noted for other field ultracool
dwarfs \citep[e.g.,][]{2004ApJ...600.1020M, 2007ApJ...657..511A,
  2008ApJ...689.1295K, 2010ApJS..186...63R}. The integrated-light
optical spectrum from \citet{2003AJ....126.3007R} and our
near-infrared spectrum do not show obvious hallmarks of low surface
gravity. However, near-infrared spectra of M7 and M8 dwarfs in the
Pleiades recently published by \citet{2010arXiv1005.3249B} also show
no distinguishing characteristics relative to field dwarfs of the same
spectral type, at least at low resolution, $R \sim 50$, and modest
$S/N$. We have also checked for possible association of \lpAB\ with
young moving groups using our derived $(U, V, W)$ heliocentric space
velocity (Section \ref{sec:age}). We find no such association with any
of the groups listed by \citet{2004ARA&A..42..685Z} and
\citet{2008hsf2.book..757T}.

LP~349-25AB is one of the relatively few ultracool dwarfs that
displays radio emission \citep[$\sim$10\% occurrence rate for $>$M7
dwarfs;][]{2006ApJ...648..629B}.  \citet{2007ApJ...658..553P}
discovered the radio emission from \lpAB\ at 8.5~GHz but with a beam
size of $9\farcs2 \times 8\farcs0$ were unable to resolve the binary
to determine whether both components are radio-luminous.
\citet{2009ApJ...700.1750O} presented additional unresolved,
multi-frequency, multi-epoch observations to better constrain the
emission mechanism, finding a lack of variability in both the radio
flux and spectral index on both short (hours) and long (months)
timescales. Their observations spanning 10.7~hours would have easily
captured multiple rotations of the emitting component, as the measured
$v\sin(i)$ of $56\pm6$~\kms\ \citep{2010ApJ...710..924R} implies a
maximum rotation period of $2.9/\sin(i)$~hours (assuming an average
model-derived radius of 0.136~\Rsun) or a period of 2.6~hours if the
rotation axis is co-aligned with the binary orbit. \lpAB\ is the only
known radio-luminous ultracool dwarf that does not display some level
of variability. To explain all observed radio properties,
\citet{2009ApJ...700.1750O} favored a long-lived ($\gtrsim0.6$~yr),
high-latitude (i.e., polar) source emitting gyrosynchrotron radiation.
Future observations at higher angular resolution using the Extended
Very Large Array are needed to determine whether only one or both
components are radio-luminous. Our dynamical mass measurement will be
a key input in developing customized models to better understand the
origin of the magnetic structures generating the unique radio emission
of \lpAB.

Finally, we consider the possibility that \lpAB\ is actually a higher
order multiple system. The resolved colors and magnitudes of the two
components are consistent with a simple binary as the secondary is
slightly fainter and redder than the primary. As discussed in
Section~\ref{sec:qratio}, we find that the resolved radial velocity
measurements of \citet{qk10} are consistent with both the
model-derived mass ratio of 0.87 and their best fit value of 2.0, in
which the secondary has twice the mass of the primary. In this
scenario, the B component would have to actually comprise two fainter
brown dwarfs whose combined luminosity is below that of the single A
component.\footnote{Another possibility resulting in $q = 2.0$ that
  can be ruled out is that the secondary component of \lpAB\ is more
  luminous than the primary.  As shown in Figure~1 of \citet{bur01}, a
  secondary with a putative mass of 0.04~\Msun\ would never outshine
  the primary, even during its deuterium-fusing phase at
  $\approx$2.5~Myr.} However, in this case the A component's mass
would only be about 0.04~\Msun, which is well below the lithium
depletion limit and thus \lpint\ should display detectable lithium
absorption. In addition, dividing the binary into a higher order
multiple would necessarily require the age to be younger for such
lower mass component objects to output the same amount of energy. At
younger ages, it would be even less likely that the most massive
component (whatever its mass) would have had time to reach the core
temperature needed to destroy its initial lithium. Thus, we conclude
that given all available constraints \lpAB\ cannot be a higher order
multiple.


\section{Conclusions}

We have determined the orbits of three late-M binaries from Keck NGS
and LGS AO orbital monitoring.  From observations spanning 5.9 years
(75\%) of the 7.8~yr orbit of \lpAB, we have determined a total mass
of $0.120^{+0.008}_{-0.007}$~\Msun. For \lhsAB, we have found a highly
eccentric ($e = 0.830\pm0.005$) 16.1-yr orbit from observations
spanning 5.9~years (36\% of the orbit, including periastron passage),
deriving a dynamical mass of $0.194^{+0.025}_{-0.021}$~\Msun.  We have
also determined a new orbit for \glAB\ based solely on astrometrically
well-calibrated instruments (Keck/NIRC2 and \HST/STIS), finding a
larger semimajor axis and thus larger dynamical mass
($0.140^{+0.009}_{-0.008}$~\Msun) than previous work.  For all
binaries, the most significant contribution to their mass
uncertainties are the errors in their parallaxes.

Despite \lpAB's integrated-light spectral type of M8 \emph{and} lack
of detectable lithium absorption, this binary has turned out to be
pair of brown dwarfs. To match the observed total mass and component
luminosities, evolutionary models predict that the system must be
quite young ($140\pm30$~Myr) and that at least the secondary component
should have retained most of its initial lithium. If the model-derived
age for \lpAB\ is correct, it is nearly the same age as the Pleiades;
however, it appears to be discrepant with the boundary between
lithium-depleted and lithium-bearing objects from
\citet{1998ApJ...499L.199S}. If the parallax measured by
\citet{2009AJ....137..402G} is accurate, this disagreement with the
Pleiades indicates a possible problem with the model-derived age,
which in our analysis is based solely on the predicted \Lbol\
evolution. An older age for \lpAB\ would remedy this problem, as the
lithium depletion boundary moves to later spectral types at older
ages. This solution would require that models under-predict the
luminosities of brown dwarfs, which has previously been suggested by
\citet{2009ApJ...692..729D} to explain the higher-than-predicted
luminosities of the benchmark brown dwarf binary HD~130948BC (L4+L4).
We also find that the lack of lithium depletion in at least the
primary component effectively rules out the possibility that \lpAB\ is
a higher order multiple.

As in our previous work \citep[e.g.,][]{2008ApJ...689..436L,
  2009ApJ...692..729D}, we have used measured dynamical masses to test
the predictions of evolutionary and atmospheric models. Properties
such as \Teff\ and \logg\ are derived from evolutionary models using
only \Mtot\ and the resolved component luminosities. We have also fit
model atmospheres to the integrated-light 0.95--2.42~\micron\ spectra
as an independent method for estimating \Teff. Including our previous
work on 2MASS~J2206$-$2047AB \citep{me-2206}, there is now a sample of
four late-M binaries with precise dynamical masses and near-infrared
spectra enabling a consistency check between evolutionary
model-derived properties and model atmospheres. All four model
atmosphere grids we tested give effective temperatures $\approx$250~K
warmer than predicted for our objects of known mass from Lyon Dusty
evolutionary model radii (including the Ames-Dusty model atmospheres
which were used as the boundary condition for the Dusty evolutionary
models).

Although model atmospheres now incorporate considerably more advanced
theory than the decade-old evolutionary models, we propose that such a
large offset cannot be entirely explained by problems with model
radii. Instead, model atmospheres are likely to be the major source of
this discrepancy given that: (1)~they must harbor systematic errors at
some level, since they do not completely match our observed spectra;
and (2)~roughly the same \Teff\ offset is observed over a narrow range
of \Teff\ but a wide range of masses, ages, and activity levels. Such
a discrepancy would be difficult to explain in the context of
evolutionary models, although they may also harbor systematic errors
that partially contribute to this inconsistency.  Directly measured
radii are needed to determine which set of models correctly predicts
\Teff\ (if either do). This can be readily tested with future
discoveries of late-M eclipsing binaries, as a 250~K offset
corresponds to a substantial radius difference (15--20\%).

In addition to our standard line of analysis, we have also explored
the approach to testing models used by \citet{qk10} for the late-M
binaries common to our two samples. Our model atmosphere fitting of
integrated-light spectra yields higher \Teff\ estimates than their
approach of fitting resolved broadband photometry.  Consequently, we
find evolutionary model-derived masses from their approach that are a
factor of $\sim$2 to 4 higher than theirs, resulting in an opposite
discrepancy with measured dynamical masses.  This illustrates that
their method sensitively relies upon the input \Teff, a property that
cannot be determined directly without radius measurements. Thus, this
approach may have limited utility in testing evolutionary models, as
observed discrepancies are more readily caused by problems with \Teff\
estimates from model atmospheres.  To summarize how our conclusions
contrast with those of \citet{qk10}: (1)~They found estimates for
\Teff\ from model atmospheres that are $\approx$400~K \emph{lower}
than our evolutionary model-derived values, whereas we find \Teff\
estimates that are $\approx$250~K \emph{higher}.  (2)~They interpreted
these results in terms of mass, concluding that systematic errors in
the evolutionary model cooling curves were responsible for observed
discrepancies.  We propose instead that the discrepancy is largely
caused by systematic errors in the model atmospheres.

As the sample of ultracool dwarfs with dynamical mass measurements
grows, we are beginning to probe the diversity of the field
population, which comprises both very low-mass stars and brown dwarfs.
At the cool temperatures of late-M dwarfs, field objects can span
nearly a factor a two in mass depending on their age, which makes mass
measurements crucial for breaking this degeneracy. At cooler
temperatures there are even fewer mass measurements presently
available, and the effects of complex processes such as atmospheric
dust formation become more important.  Therefore, continued orbital
monitoring will be imperative to bring similar tests of models as we
report here into the much cooler regimes of the L and T dwarfs.


\acknowledgments

It is a pleasure to thank Joel Aycock, Randy Campbell, Al Conrad,
Heather Hershley, Jim Lyke, Jason McIlroy, Gary Punawai, Julie
Riviera, Hien Tran, Cynthia Wilburn,
and the Keck Observatory staff for assistance with the Keck
observations.  We are very grateful to Michal Simon and Chad Bender
for providing us with their Keck/NIRC2 images of \glAB.  We also thank
France Allard for providing evolutionary models with near-infrared
photometry on the MKO system, Travis Barman for providing us with the
Gaia-Dusty model spectra, Kelle Cruz for providing the optical
spectrum of \lpint, Adam Burgasser for providing the SpeX prism
spectrum of Gl~569A, Brian Cameron for making available his NIRC2
distortion solution, and C\'{e}line Reyl\'{e} for customized
Besan\c{c}on Galaxy models.
We are indebted to Katelyn Allers for assistance in obtaining
IRTF/SpeX data of \glint.
Our research has employed the 2MASS data products; NASA's
Astrophysical Data System; the SIMBAD database operated at CDS,
Strasbourg, France; and the M, L, and T~dwarf compendium housed at
\texttt{http://www.DwarfArchives.org} and maintained by Chris Gelino,
Davy Kirkpatrick, and Adam Burgasser \citep{2003IAUS..211..189K,
  2004AAS...205.1113G}.
MCL, TJD, and BPB acknowledge support for this work from NSF grants
AST-0507833 and AST-0909222.
Finally, the authors wish to recognize and acknowledge the very
significant cultural role and reverence that the summit of Mauna Kea has
always had within the indigenous Hawaiian community.  We are most
fortunate to have the opportunity to conduct observations from this
mountain.

{\it Facilities:} \facility{Keck II Telescope (LGS AO, NIRC2)},
\facility{CFHT (PUEO, KIR)}, \facility{VLT (NACO)}, \facility{IRTF
  (SpeX)}



\clearpage
\begin{deluxetable}{lccccccc}
\tabletypesize{\scriptsize}
\tablecaption{Keck AO Observations \label{tbl:obs}}
\tablewidth{0pt}
\tablehead{
\colhead{Date} &
\colhead{Time} &
\colhead{Airmass} &
\colhead{Filter} &
\colhead{FWHM\tablenotemark{a}} &
\colhead{Strehl ratio\tablenotemark{a}} \\
\colhead{(UT)} &
\colhead{(UT)} &
\colhead{} &
\colhead{} &
\colhead{(mas)} &
\colhead{}}
\startdata
\multicolumn{6}{c}{\bf \lpAB} \\
\cline{1-6}

 2008 Jan 16\tablenotemark{c} & 05:56 & 1.280 & \Ks\ &   $57\pm3$   &  $0.23\pm0.05$   \\ 

 2008 Jun 30\tablenotemark{c} & 13:20 & 1.314 & $J$  &   $55\pm4$   & $0.026\pm0.009$  \\
                              & 13:29 & 1.274 & $H$  & $48.6\pm0.8$ & $0.123\pm0.012$  \\
                              & 13:12 & 1.350 & \Ks\ &   $58\pm4$   &  $0.19\pm0.04$   \\

 2008 Aug 20\tablenotemark{b} & 14:21 & 1.069 & \Ks\ & $50.1\pm0.5$ & $0.448\pm0.007$  \\ 

 2008 Sep  9\tablenotemark{b} & 13:51 & 1.169 & \Lp\ & $80.2\pm0.9$ &  $0.78\pm0.03$   \\

 2009 Sep 28\tablenotemark{c} & 13:17 & 1.328 & \Ks\ &   $54\pm3$   &  $0.29\pm0.05$   \\

 2009 Dec 15\tablenotemark{c} & 05:22 & 1.479 & $K$  & $55.8\pm0.9$ &  $0.33\pm0.03$   \\

 2010 May 22\tablenotemark{c} & 14:54 & 1.721 & $K$  &   $58\pm3$   &  $0.36\pm0.04$   \\

\cline{1-6}

\multicolumn{6}{c}{} \\
\multicolumn{6}{c}{\bf \lhsAB} \\
\cline{1-6}

 2008 Jan 15\tablenotemark{b} & 11:05 & 1.128 & \Ks\ & $49.0\pm1.9$ & $0.411\pm0.012$  \\

 2008 Sep  9\tablenotemark{b} & 15:13 & 1.411 & $J$  & $39.8\pm1.8$ & $0.087\pm0.015$  \\
                              & 14:59 & 1.475 & $H$  & $42.6\pm1.3$ & $0.217\pm0.014$  \\
                              & 15:17 & 1.393 & \Ks\ & $46.7\pm1.1$ &  $0.47\pm0.03$   \\

 2009 Sep 28\tablenotemark{b} & 15:37 & 1.150 & \Ks\ &   $51\pm3$   &  $0.30\pm0.07$   \\

 2009 Dec 16\tablenotemark{b} & 08:45 & 1.428 & \Ks\ & $53.7\pm1.7$ &  $0.30\pm0.04$   \\ 

 2010 Mar 22\tablenotemark{b} & 05:59 & 1.096 & $J$  &   $41\pm5$   &  $0.05\pm0.02$   \\
                              & 05:55 & 1.094 & $H$  & $40.1\pm0.8$ &  $0.23\pm0.03$   \\
                              & 05:52 & 1.093 & $K$  & $50.3\pm0.2$ &  $0.43\pm0.02$   \\
                              & 06:04 & 1.097 & \Lp\ & $82.7\pm0.4$ &  $0.63\pm0.07$   \\

\cline{1-6}

\multicolumn{6}{c}{} \\
\multicolumn{6}{c}{\bf \glAB} \\
\cline{1-6}

 2004 Dec 24\tablenotemark{b, d} & 16:22 & 1.270 & $H$  & $46.8\pm0.5$ &  $0.17\pm0.04$   \\

 2005 Feb 25\tablenotemark{b, d} & 12:08 & 1.298 & \Hc\ & $39.0\pm1.1$ &  $0.23\pm0.03$   \\

 2008 Jan 16\tablenotemark{b} & 16:33 & 1.036 & \Hc\ & $36.9\pm0.8$ &  $0.31\pm0.05$   \\

 2009 Apr 29\tablenotemark{b} & 12:24 & 1.089 & \Kc\ & $50.0\pm1.3$ &  $0.39\pm0.07$   \\

 2009 May 29\tablenotemark{b} & 10:56 & 1.159 & \Kc\ & $52.0\pm0.7$ &  $0.58\pm0.04$   \\

 2010 Mar 22\tablenotemark{b} & 15:48 & 1.235 & $K$  & $48.6\pm1.2$ &  $0.50\pm0.06$   \\

 2010 May 23\tablenotemark{b} & 11:10 & 1.133 & $J$  &   $37\pm2$   & $0.067\pm0.011$  \\
                              & 11:16 & 1.148 & $H$  & $40.6\pm1.6$ &  $0.19\pm0.06$  \\
                              & 11:23 & 1.166 & $K$  & $49.8\pm0.6$ &  $0.44\pm0.04$  \\
                              & 11:32 & 1.193 & \Lp\ & $82.4\pm0.5$ &  $0.76\pm0.02$  \\

\enddata

\tablenotetext{a}{Strehl ratios and FWHM were computed using the
  publicly available routine \texttt{NIRC2STREHL}.  Errors are the
  rms of individual dithers.}

\tablenotetext{b}{NGS AO observations.}

\tablenotetext{c}{{LGS} AO observations.}

\tablenotetext{d}{Data originally published by
  \citet{2006ApJ...644.1183S} and reanalyzed by us.}

\end{deluxetable}

\clearpage
\begin{deluxetable}{lccccc}
\tabletypesize{\scriptsize}
\tablecaption{Best-Fit Binary Parameters \label{tbl:astrom}}
\tablewidth{0pt}
\tablehead{
\colhead{Epoch (UT)} &
\colhead{Instrument} &
\colhead{Filter} &
\colhead{$\rho$ (mas)} &
\colhead{PA (\degree)} &
\colhead{$\Delta{m}$ (mag)}}
\startdata
\multicolumn{6}{c}{\bf \lpAB} \\
\cline{1-6}

 2004 Jul  3 & CFHT/PUEO\tablenotemark{a,b}& \Kp\ & \phn$125\pm10$   &    \phn$12.7\pm2.0$  &  $0.26\pm0.05$  \\ 

 2004 Sep 26 & VLT/NACO\tablenotemark{a,b} & $H$  & \phn$107\pm10$   & \phn\phn$7.1\pm0.5$  &  $0.38\pm0.05$  \\ 

 2006 Jul 13 & VLT/NACO\tablenotemark{a}   & \Ks\ &   $105.6\pm0.3$  &       $247.3\pm0.2$  &  $0.33\pm0.03$  \\ 
 2006 Sep 21 & VLT/NACO\tablenotemark{a}   & \Ks\ &   $116.6\pm0.6$  &       $240.4\pm0.2$  &  $0.31\pm0.03$  \\ 
 2006 Oct  9 & VLT/NACO\tablenotemark{a}   & \Ks\ &   $118.9\pm0.6$  &       $238.9\pm0.3$  & $0.285\pm0.013$ \\ 
 2006 Nov 11 & VLT/NACO\tablenotemark{a}   & \Ks\ &   $123.3\pm0.4$  &       $236.0\pm0.2$  & $0.315\pm0.019$ \\ 
 2006 Dec 24 & VLT/NACO\tablenotemark{a}   & \Ks\ &   $128.2\pm0.7$  &       $232.7\pm0.3$  & $0.276\pm0.018$ \\ 

 2008 Jan 16 & Keck/NIRC2\tablenotemark{a} & \Ks\ &   $137.3\pm0.4$  &      $207.92\pm0.09$ & $0.315\pm0.011$ \\ 

 2008 Jun 30 & Keck/NIRC2                  & $J$  &   $115.2\pm0.3$  &       $194.4\pm0.4$  &  $0.35\pm0.03$  \\
             & Keck/NIRC2\tablenotemark{a} & $H$  &  $114.87\pm0.12$ &      $194.71\pm0.13$ & $0.326\pm0.011$ \\
             & Keck/NIRC2                  & \Ks\ &   $115.2\pm0.5$  &      $194.64\pm0.13$ & $0.318\pm0.007$ \\

 2008 Aug 20 & Keck/NIRC2\tablenotemark{a} & \Ks\ &   $105.9\pm0.2$  &      $189.71\pm0.11$ & $0.314\pm0.007$ \\ 

 2008 Sep  9 & Keck/NIRC2\tablenotemark{a} & \Lp\ &  $102.45\pm0.19$ &      $187.46\pm0.10$ & $0.222\pm0.005$ \\

 2009 Sep 28 & Keck/NIRC2\tablenotemark{a} & \Ks\ &\phn$71.1\pm0.3$  &    \phn$98.3\pm0.6$  &  $0.24\pm0.03$  \\

 2009 Dec 15 & Keck/NIRC2\tablenotemark{a} & $K$  &\phn$83.4\pm0.3$  &    \phn$81.6\pm0.4$  &  $0.38\pm0.06$  \\

 2010 May 22 & Keck/NIRC2\tablenotemark{a} & $K$  &   $112.6\pm0.4$  &   \phn$59.76\pm0.13$ & $0.307\pm0.008$ \\

\cline{1-6}

\multicolumn{6}{c}{} \\
\multicolumn{6}{c}{\bf \lhsAB} \\
\cline{1-6}

 2004 Jan  8 & CFHT/PUEO\tablenotemark{a,c}& \Kp\ &     $275\pm5$    &       $208.0\pm0.5$  &  $0.13\pm0.03$  \\

 2005 Apr 27 & CFHT/PUEO\tablenotemark{a,c}& \Kp\ &     $204\pm5$    &       $215.0\pm0.5$  &  $0.07\pm0.03$  \\

 2005 Oct 14 & CFHT/PUEO\tablenotemark{a,c}& $H$  &     $174\pm5$    &       $219.6\pm0.5$  &  $0.14\pm0.05$  \\

 2008 Jan 15 & Keck/NIRC2\tablenotemark{a} & \Ks\ &\phn$60.5\pm1.5$  &       $308.6\pm0.8$  &  $0.16\pm0.06$  \\

 2008 Sep  9 & Keck/NIRC2\tablenotemark{a} & $J$  &\phn$57.4\pm0.6$  & \phn\phn$1.0\pm0.8$  &  $0.31\pm0.16$  \\
             & Keck/NIRC2                  & $H$  &\phn$57.2\pm1.3$  & \phn\phn$1.2\pm1.8$  &  $0.28\pm0.13$  \\
             & Keck/NIRC2                  & \Ks\ &\phn$57.8\pm1.0$  & \phn\phn$1.5\pm0.8$  &  $0.15\pm0.04$  \\

 2009 Sep 28 & Keck/NIRC2\tablenotemark{a} & \Ks\ &  $177.90\pm0.14$ &      $179.54\pm0.03$ & $0.098\pm0.007$ \\

 2009 Dec 16 & Keck/NIRC2\tablenotemark{a} & \Ks\ &  $206.78\pm0.13$ &      $181.31\pm0.03$ & $0.094\pm0.003$ \\ 

 2010 Mar 22 & Keck/NIRC2                  & $J$  &   $236.9\pm0.4$  &      $183.12\pm0.03$ & $0.111\pm0.016$ \\
             & Keck/NIRC2                  & $H$  &   $236.9\pm0.4$  &      $183.11\pm0.04$ & $0.115\pm0.009$ \\
             & Keck/NIRC2\tablenotemark{a} & $K$  &   $237.0\pm0.3$  &      $183.10\pm0.04$ & $0.107\pm0.007$ \\
             & Keck/NIRC2                  & \Lp\ &   $237.3\pm0.7$  &      $183.15\pm0.05$ & $0.099\pm0.003$ \\
\cline{1-6}

\multicolumn{6}{c}{} \\
\multicolumn{6}{c}{\bf \glAB} \\
\cline{1-6}

 2002 Jun 26 & \HST/STIS\tablenotemark{a}&F28X50LP&\phn$97.3\pm1.3$  &    \phn$94.0\pm1.3$  &  $0.99\pm0.03$  \\ 

 2004 Dec 24 & Keck/NIRC2\tablenotemark{a} & $H$  &\phn$93.9\pm0.3$  &       $110.8\pm0.4$  &  $0.54\pm0.02$  \\

 2005 Feb 25 & Keck/NIRC2\tablenotemark{a} & \Hc\ &\phn$85.2\pm0.8$  &       $132.9\pm0.7$  &  $0.55\pm0.08$  \\

 2008 Jan 16 & Keck/NIRC2\tablenotemark{a} & \Hc\ &\phn$61.8\pm0.7$  &       $272.5\pm1.4$  &  $0.58\pm0.16$  \\

 2009 Apr 29 & Keck/NIRC2\tablenotemark{a} & \Kc\ &   $100.2\pm0.6$  &    \phn$66.5\pm0.6$  &  $0.55\pm0.07$  \\

 2009 May 29 & Keck/NIRC2\tablenotemark{a} & \Kc\ &   $100.5\pm1.8$  &    \phn$75.0\pm1.7$  &  $0.59\pm0.07$  \\

 2010 Mar 22 & Keck/NIRC2\tablenotemark{a} & $K$  &    $55.8\pm0.3$  &       $206.5\pm1.1$  & $0.473\pm0.010$ \\

 2010 May 23 & Keck/NIRC2                  & $J$  &    $58.5\pm1.1$  &       $265.9\pm1.0$  &  $0.45\pm0.13$  \\
             & Keck/NIRC2                  & $H$  &    $58.8\pm0.5$  &       $267.6\pm0.6$  &  $0.46\pm0.06$  \\
             & Keck/NIRC2\tablenotemark{a} & $K$  &    $59.9\pm0.4$  &       $268.1\pm0.4$  &  $0.49\pm0.04$  \\
             & Keck/NIRC2                  & \Lp\ &    $59.9\pm0.6$  &       $267.6\pm0.5$  &  $0.49\pm0.03$  \\

\enddata

\tablenotetext{a}{Used in the orbit fit.}

\tablenotetext{b}{Measurements from \citet{2005A&A...435L...5F}.}

\tablenotetext{c}{Measurements from \citet{2006A&A...460L..19M}.}

\end{deluxetable}

\clearpage
\begin{deluxetable}{lcccc}
\tabletypesize{\scriptsize}
\tablewidth{0pt}
\tablecaption{Derived Orbital Parameters\tablenotemark{a} \label{tbl:orbit}}

\tablehead{
\colhead{}   &
\multicolumn{3}{c}{MCMC}      &
\colhead{MPFIT} \\
\cline{2-4}
\colhead{Parameter}   &
\colhead{Median}      &
\colhead{68.3\% c.l.} &
\colhead{95.4\% c.l.} &
\colhead{routine} }
\startdata

\multicolumn{5}{c}{\bf \lpAB} \\
\cline{1-5}
Semimajor axis $a$ (mas)                                         &   146.7   &    $-$0.6, 0.6    &    $-$1.1, 1.2    &     $ 146.7\pm0.6   $\phn\phn \\
Orbital period $P$ (yr)                                          &   7.76    &   $-$0.04, 0.04   &   $-$0.07, 0.08   &     $  7.76\pm0.04  $         \\
Eccentricity $e$                                                 &   0.051   &  $-$0.003, 0.003  &  $-$0.005, 0.006  &     $ 0.051\pm0.003 $         \\
Inclination $i$ (\degree)                                        &   117.24  &   $-$0.14, 0.14   &    $-$0.3, 0.3    &     $117.23\pm0.13  $\phn\phn \\
Time of periastron passage $T_0-2454860.5$\tablenotemark{b} (JD) &     0     &     $-$24, 26     &     $-$50, 50     &     $    -2\pm25    $\phs\phn \\ 
PA of the ascending node $\Omega$ (\degree)                      &   35.95   &   $-$0.12, 0.12   &   $-$0.23, 0.23   &     $ 35.95\pm0.11  $\phn     \\
Argument of periastron $\omega$ (\degree)                        &   250     &      $-$3, 4      &      $-$6, 7      &     $   250\pm3     $\phn\phn \\
Total mass (\Msun): fitted\tablenotemark{c}                      &   0.1205  & $-$0.0007, 0.0007 & $-$0.0013, 0.0013 &     $0.1205\pm0.0006$         \\
Total mass (\Msun): final\tablenotemark{d}                       &   0.120   &  $-$0.007, 0.008  &  $-$0.014, 0.017  &     $ 0.120\pm0.008 $         \\
$\chi^2$ (21 degrees of freedom)                                 &  \nodata  &       \nodata     &       \nodata     &           18.1                \\

\cline{1-5}

\multicolumn{5}{c}{} \\
\multicolumn{5}{c}{\bf \lhsAB} \\
\cline{1-5}

Semimajor axis $a$ (mas)                                         &   288    &      $-$5, 5      &      $-$8, 8      &     $   288\pm3     $\phn\phn \\ 
Orbital period $P$ (yr)                                          &   16.1   &    $-$0.5, 0.5    &    $-$0.8, 0.8    &     $  16.1\pm0.3   $\phn     \\ 
Eccentricity $e$                                                 &   0.830  &  $-$0.005, 0.005  &  $-$0.008, 0.008  &     $ 0.830\pm0.003 $         \\ 
Inclination\tablenotemark{a} $i$ (\degree)                       &   72.1   &    $-$0.3, 0.3    &    $-$0.5, 0.5    &     $ 72.12\pm0.18  $\phn     \\ 
Time of periastron passage $T_0-2454765.0$\tablenotemark{e} (JD) &   0.0    &    $-$2.5, 2.5    &      $-$4, 4      &     $   0.0\pm1.6   $\phs     \\ 
PA of the ascending node $\Omega$ (\degree)                      &   182.0  &    $-$0.2, 0.2    &    $-$0.4, 0.4    &     $181.99\pm0.15  $\phn\phn \\ 
Argument of periastron $\omega$ (\degree)                        &   224.8  &    $-$0.6, 0.6    &    $-$1.1, 1.1    &     $ 224.8\pm0.4   $\phn\phn \\ 
Total mass (\Msun): fitted\tablenotemark{c}                      &   0.1944 & $-$0.0028, 0.0028 &  $-$0.005, 0.005  &     $0.1944\pm0.0018$         \\ 
Total mass (\Msun): final\tablenotemark{d}                       &   0.194  &  $-$0.021, 0.025  &  $-$0.039, 0.053  &     $ 0.194\pm0.023 $         \\ 
$\chi^2$ (9 degrees of freedom)                                  &  \nodata &       \nodata     &       \nodata     &            6.8                \\

\cline{1-5}

\multicolumn{5}{c}{} \\
\multicolumn{5}{c}{\bf \glAB} \\
\cline{1-5}

Semimajor axis $a$ (mas)                                         &   95.6   &    $-$1.0, 1.1    &    $-$1.8, 2.0    &     $  95.7\pm0.7   $\phn     \\ 
Orbital period $P$ (yr)                                          &  2.367   &  $-$0.003, 0.003  &  $-$0.006, 0.006  &     $ 2.367\pm0.002 $\phn     \\ 
Eccentricity $e$                                                 &   0.316  &  $-$0.005, 0.005  &  $-$0.009, 0.009  &     $ 0.316\pm0.003 $         \\ 
Inclination\tablenotemark{a} $i$ (\degree)                       &   35.0   &    $-$1.1, 1.1    &    $-$2.2, 2.0    &     $  35.1\pm0.7   $\phn     \\ 
Time of periastron passage $T_0-2455290.5$\tablenotemark{f} (JD) &   0.0    &    $-$1.5, 1.5    &    $-$2.5, 2.5    &     $   0.0\pm0.9   $         \\ 
PA of the ascending node $\Omega$ (\degree)                      &   324.8  &    $-$2.0, 2.0    &    $-$3.3, 3.3    &     $ 324.9\pm1.2   $\phn\phn \\ 
Argument of periastron $\omega$ (\degree)                        &   257.9  &    $-$2.0, 2.0    &    $-$3.0, 3.3    &     $ 257.8\pm1.2   $\phn\phn \\ 
Total mass (\Msun): fitted\tablenotemark{c}                      &   0.140  &  $-$0.004, 0.005  &  $-$0.007, 0.009  &     $0.1408\pm0.0028$         \\ 
Total mass (\Msun): final\tablenotemark{d}                       &   0.140  &  $-$0.008, 0.009  &  $-$0.015, 0.018  &     $ 0.141\pm0.008 $         \\ 
$\chi^2$ (9 degrees of freedom)                                  &  \nodata &       \nodata     &       \nodata     &            6.3                \\

\enddata

\tablenotetext{a}{Note that the uncertainties quoted here are ``single
  parameter'' errors (i.e., $\Delta\chi^2 = 1$) and thus are only
  valid when a single parameter is of interest.}

\tablenotetext{b}{2009 January 27 00:00:00.0 UT} 

\tablenotetext{c}{The ``fitted'' total mass represents the direct
  results from fitting the observed orbital motion without accounting
  for the parallax error.}

\tablenotetext{d}{The ``final'' total mass includes the additional
  error in the mass due to the error in the parallax.}

\tablenotetext{e}{2008 October 25 12:00:00.0 UT} 

\tablenotetext{e}{2010 April 4 00:00:00.0 UT} 

\end{deluxetable}

\clearpage
\begin{deluxetable}{lccccccccccc}
\tabletypesize{\scriptsize}
\rotate
\tablewidth{0pt}
\tablecaption{Measured Properties of Target Binaries\tablenotemark{a} \label{tbl:meas}}
\tablehead{
\colhead{}                    &
\multicolumn{2}{c}{LP~349-25} &
\colhead{}                    &
\colhead{}                    &
\multicolumn{2}{c}{LHS~1901}  &
\colhead{}                    &
\colhead{}                    &
\multicolumn{2}{c}{Gl~569B}   &
\colhead{}                    \\
\cline{2-4}
\cline{6-8}
\cline{10-12}
\colhead{Property} &
\colhead{A}        &
\colhead{B}        &
\colhead{Ref.}     &
\colhead{}         &
\colhead{A}        &
\colhead{B}        &
\colhead{Ref.}     &
\colhead{}         &
\colhead{Ba}       &
\colhead{Bb}       &
\colhead{Ref.}     }

\startdata
\Mtot\ (\Msun)             &             \multicolumn{2}{c}{ $0.120^{+0.008}_{-0.007}$ }     &  1,2  & &          \multicolumn{2}{c}{$0.194^{+0.025}_{-0.021}$}\phn  &  1,3  & &       \multicolumn{2}{c}{$0.140^{+0.009}_{-0.008}$}\phn     &  1,4  \\
Semimajor axis (AU)        &             \multicolumn{2}{c}{    $ 1.94\pm0.04 $    }         &  1,2  & &            \multicolumn{2}{c}{    $3.70\pm0.16$    }        &  1,3  & &        \multicolumn{2}{c}{    $0.923\pm0.018$      }        &  1,4  \\
$d$ (pc)                   &             \multicolumn{2}{c}{    $ 13.2\pm0.3  $\phn}         &   2   & &            \multicolumn{2}{c}{    $12.9\pm0.5 $\phn}        &   3   & &        \multicolumn{2}{c}{     $9.65\pm0.16$       }        &   4   \\
Spectral type              & \phs    ${\rm M7.5}\pm1.0$ & \phs    ${\rm M8.0}^{+1.5}_{-1.0}$ &   1   & &\phs   ${\rm M6.5}^{+1.0}_{-0.5}$ & ${\rm M6.5}^{+1.0}_{-0.5}$ &  1  & &\phs    ${\rm M8.5}\pm0.5$  &\phs    ${\rm M9.0}\pm0.5$        &   5,6    \\
$J$ (mag)                  & \phs    $   11.154\pm0.025 $ & \phs    $   11.506\pm0.028$      &  1,7  & &\phs     $  10.631\pm0.020 $&\phs     $  10.743\pm0.020     $&  1,7  & &\phs     $   11.27\pm0.06  $&\phs     $   11.78\pm0.06      $&   1   \\
$H$ (mag)                  & \phs    $   10.606\pm0.023 $ & \phs    $   10.932\pm0.023$      &  1,7  & &\phs     $  10.197\pm0.016 $&\phs     $  10.312\pm0.016     $&  1,7  & &\phs     $   10.67\pm0.04  $&\phs     $   11.21\pm0.04      $&   1   \\
$K$ (mag)                  & \phs    $   10.151\pm0.018 $ & \phs    $   10.458\pm0.018$      &  1,7  & &\phs     $   9.796\pm0.019 $&\phs     $   9.904\pm0.019     $&  1,7  & &\phs     $   10.16\pm0.03  $&\phs     $   10.64\pm0.03      $&   1   \\
\Lp\ (mag)                 & \phs\phn$     9.80\pm0.07  $ & \phs    $    10.02\pm0.07 $      &  1,8  & &\phs       \nodata          &\phs        \nodata             &       & &\phs\phn $    9.46\pm0.10  $&\phs\phn $    9.95\pm0.10      $&  1,9  \\
$J-K$ (mag)                & \phs\phn$     1.00\pm0.03  $ & \phs\phn$     1.05\pm0.03 $      &  1,7  & &\phs\phn $    0.83\pm0.03  $&\phs\phn $    0.84\pm0.03      $&  1,7  & &\phs\phn $    1.11\pm0.07  $&\phs\phn $    1.15\pm0.07      $&   1   \\
$H-K$ (mag)                & \phs\phn$     0.46\pm0.03  $ & \phs\phn$     0.47\pm0.03 $      &  1,7  & &\phs\phn $    0.40\pm0.02  $&\phs\phn $    0.41\pm0.03      $&  1,7  & &\phs\phn $    0.50\pm0.05  $&\phs\phn $    0.57\pm0.05      $&   1   \\
$J-H$ (mag)                & \phs\phn$     0.55\pm0.03  $ & \phs\phn$     0.57\pm0.04 $      &  1,7  & &\phs\phn $    0.43\pm0.03  $&\phs\phn $    0.43\pm0.03      $&  1,7  & &\phs\phn $    0.61\pm0.07  $&\phs\phn $    0.58\pm0.07      $&   1   \\
$K-\Lp$ (mag)              & \phs\phn$     0.35\pm0.07  $ & \phs\phn$     0.44\pm0.07 $      & 1,7,8 & &\phs       \nodata          &\phs        \nodata             &       & &\phs\phn $    0.70\pm0.10  $&\phs\phn $    0.68\pm0.11      $&  1,9  \\
$M_J$ (mag)                & \phs    $    10.55\pm0.05  $ & \phs    $    10.90\pm0.05 $      & 1,2,7 & &\phs     $   10.09\pm0.09  $&\phs     $   10.20\pm0.09      $& 1,3,7 & &\phs     $   11.35\pm0.07  $&\phs     $   11.86\pm0.07      $&  1,4  \\
$M_H$ (mag)                & \phs    $    10.01\pm0.05  $ & \phs    $    10.33\pm0.05 $      & 1,2,7 & &\phs     $    9.65\pm0.09  $&\phs     $    9.77\pm0.09      $& 1,3,7 & &\phs     $   10.74\pm0.06  $&\phs     $   11.28\pm0.05      $&  1,4  \\
$M_K$ (mag)                & \phs\phn$     9.55\pm0.05  $ & \phs\phn$     9.86\pm0.05 $      & 1,2,7 & &\phs\phn $    9.25\pm0.09  $&\phs\phn $    9.36\pm0.09      $& 1,3,7 & &\phs\phn $   10.24\pm0.05  $&\phs\phn $   10.71\pm0.05      $&  1,4  \\
$M_{L^{\prime}}$ (mag)     & \phs\phn$     9.20\pm0.08  $ & \phs\phn$     9.42\pm0.08 $      & 1,2,8 & &\phs       \nodata          &\phs        \nodata             &       & &\phs\phn $    9.54\pm0.10  $&\phs     $   10.03\pm0.11      $& 1,4,9 \\
$\log$(\Lbol/\Lsun)        & \phn    $   -3.041\pm0.024 $ & \phn    $   -3.175\pm0.026$      &   1   & &\phn     $   -2.95\pm0.04  $&\phn     $   -2.99\pm0.04      $&   1   & &\phn     $  -3.424\pm0.019 $&\phn     $  -3.623\pm0.020     $&   1   \\
$\Delta\log$(\Lbol)        &               \multicolumn{2}{c}{$0.133\pm0.019$}               &   1   & &              \multicolumn{2}{c}{$ 0.044\pm0.015 $}          &   1   & &              \multicolumn{2}{c}{$ 0.199\pm0.016 $}          &   1   \\
\enddata                                                   

\tablenotetext{a}{All near-infrared photometry on the MKO system.}

\tablerefs{(1)~This work; (2)~\citet{2009AJ....137..402G};
  (3)~\citet{2009AJ....137.4109L}; (4)~\citet{2007hnrr.book.....V};
  (5)~\citet{1990ApJ...354L..29H}; (6)~\citet{2004astro.ph..7334O};
  (7)~\citet{2mass}; (8)~\citet{gol04};
  (9)~\citet{1988ApJ...330L.119F}.}

\end{deluxetable}

\clearpage
\begin{deluxetable}{lcccccccl}
\tabletypesize{\scriptsize}
\tablewidth{0pt}
\tablecaption{ Description of Model Atmosphere Grids \label{tbl:atm-ref}}
\tablehead{
\colhead{Grid name} &
\colhead{Grid} &
\colhead{}   &
\multicolumn{4}{c}{Molecular line lists} &
\colhead{Element} &
\colhead{Dust treatment} \\
\cline{3-6}
\colhead{}   &
\colhead{Ref.}   &
\colhead{}   &
\colhead{H$_2$O}   &
\colhead{TiO}   &
\colhead{FeH}   &
\colhead{CrH}   &
\colhead{abundance}   &
\colhead{} }
\startdata

PHOENIX-Gaia        & BH05 &  &  Ames  &  Ames  & DA99 &  F99  &  GN93  &  dust disappears after forming  (``Cond'') \\
                                                            
Ames-Dusty          & A01  &  &  Ames  &  Ames  & PD93 &  F99  &  GN93  &  dust stays where it forms (``Dusty'') \\
                                                            
Gaia-Dusty          & B10  &  &  BT06  &  Ames  & D03  &  B02  &  A05  &  dust stays where it forms (``Dusty'') \\
                                                            
Drift-PHOENIX       & W09  &  &  Ames  &  Ames  & B06  &  B06  &  GN93  &  nonequilibrium cloud model (``Drift'') \\

\enddata

\tablerefs{Ames: \citet{1997JChPh.106.4618P},
  \citet{1998FaDi..109..321S}; A01: \citet{2001ApJ...556..357A}; A05:
  \citet{2005ASPC..336...25A}; B02: \citet{2002ApJ...577..986B}; B06:
  \citet{2006AIPC..855..143B}; B10: T.\ Barman (2010, private
  communication); BH05: \citet{2005ESASP.576..565B}; BT06:
  \citet{2006MNRAS.368.1087B}; D03: \citet{2003ApJ...594..651D}; DA99:
  Davis \& Allard (1999, private communication); F99: R.\ D.\ Freedman
  (1999, private communication); GN93: \citet{1993oee..conf...15G};
  PD93: \citet{1993ApJ...409..860P}; W09:
  \citet{2009A&A...506.1367W}.}

\end{deluxetable}

\clearpage
\begin{deluxetable}{rccccccccccc}
\tabletypesize{\footnotesize}
\tablewidth{0pt}
\tablecaption{ Best-fit Atmospheric Models \label{tbl:atm-fit}}
\tablehead{
\colhead{}   &
\multicolumn{2}{c}{PHOENIX-Gaia}    &
\colhead{}   &
\multicolumn{2}{c}{Ames-Dusty}      &
\colhead{}   &
\multicolumn{2}{c}{Gaia-Dusty}      &
\colhead{}   &
\multicolumn{2}{c}{Drift-PHOENIX}   \\
\cline{2-3}
\cline{5-6}
\cline{8-9}
\cline{11-12}
\colhead{Spectral range}    &
\colhead{\Teff\ (K)}      &
\colhead{\logg\ } &
\colhead{}   &
\colhead{\Teff\ (K)}      &
\colhead{\logg\ } &
\colhead{}   &
\colhead{\Teff\ (K)}      &
\colhead{\logg\ } &
\colhead{}   &
\colhead{\Teff\ (K)}      &
\colhead{\logg\ } }
\startdata

\multicolumn{12}{c}{\bf \lpint} \\
\cline{1-12}
{\bf NIR (0.95--2.42~\micron)\tablenotemark{a}}   & {\bf 2800} & {\bf 4.5}                  &   & {\bf 2800} & {\bf 5.5}                  &   & {\bf 2800} & {\bf 5.0}                  &   & {\bf 2800} & {\bf 5.5}                   \\
All (0.81--2.42~\micron)   & 2600 & 4.0                  &   & 2600 & 4.0\tablenotemark{b} &   & 2600 & 4.0                  &   & 2600 & 4.0                   \\
$Y$ (0.95--1.12~\micron)   & 2800 & 4.5                  &   & 2800 & 5.5                  &   & 2800 & 4.5                  &   & 2700 & 5.0                   \\
$J$ (1.10--1.34~\micron)   & 2900 & 5.5\tablenotemark{b} &   & 2900 & 5.5                  &   & 2900 & 5.5                  &   & 2800 & 5.5                   \\
$H$ (1.40--1.80~\micron)   & 2900 & 5.5\tablenotemark{b} &   & 2900 & 5.5                  &   & 2900 & 5.5                  &   & 2800 & 6.0\tablenotemark{b}  \\
$K$ (1.90--2.40~\micron)   & 3000 & 5.5\tablenotemark{b} &   & 3000 & 5.5                  &   & 2900 & 5.5                  &   & 2900 & 6.0\tablenotemark{b}  \\
\cline{1-12}                                                                                                                                                    
                                                                                                                                                               
\multicolumn{12}{c}{} \\                                                                                                                                        
\multicolumn{12}{c}{\bf \lhsint} \\                                                                                                                             
\cline{1-12}                                                                                                                                                    
{\bf NIR (0.95--2.42~\micron)\tablenotemark{a}}   & {\bf 3000} & {\bf 5.0}                  &   & {\bf 3000} & {\bf 5.0}                  &   & {\bf 3000} & {\bf 5.0}                  &   & {\bf 3000} & {\bf 5.0}                   \\
All (0.81--2.42~\micron)   & 2900 & 4.5                  &   & 2900 & 4.0\tablenotemark{b} &   & 2900 & 4.5                  &   & 2900 & 5.0                   \\
$Y$ (0.95--1.12~\micron)   & 3100 & 4.5                  &   & 3000 & 5.0                  &   & 3000 & 4.5                  &   & 3000 & 5.0                   \\
$J$ (1.10--1.34~\micron)   & 3000 & 5.5\tablenotemark{b} &   & 3000 & 5.5                  &   & 3000 & 5.5                  &   & 3000 & 5.5                   \\
$H$ (1.40--1.80~\micron)   & 3000 & 5.5\tablenotemark{b} &   & 3000 & 5.5                  &   & 3000 & 5.5                  &   & 3000 & 6.0\tablenotemark{b}  \\
$K$ (1.90--2.40~\micron)   & 3100 & 5.5\tablenotemark{b} &   & 3100 & 5.5                  &   & 3100 & 5.5                  &   & 3000 & 5.5                   \\
\cline{1-12}                                                                                                                                                    
                                                                                                                                                               
\multicolumn{12}{c}{} \\                                                                                                                                        
\multicolumn{12}{c}{\bf \glint} \\                                                                                                                              
\cline{1-12}                                                                                                                                                    
{\bf NIR (0.95--2.42~\micron)\tablenotemark{a}}   & {\bf 2800} & {\bf 4.5}                  &   & {\bf 2800} & {\bf 5.5}                  &   & {\bf 2800} & {\bf 5.0}                  &   & {\bf 2800} & {\bf 5.5\tablenotemark{c}}  \\
All (0.81--2.42~\micron)   & 2700 & 4.5                  &   & 2800 & 5.5                  &   & 2700 & 5.0                  &   & 2700 & 5.0\tablenotemark{c}  \\
$Y$ (0.95--1.12~\micron)   & 2900 & 5.0                  &   & 2900 & 6.0\tablenotemark{b} &   & 2800 & 5.0                  &   & 2700 & 5.5                   \\
$J$ (1.10--1.34~\micron)   & 2800 & 5.5\tablenotemark{b} &   & 2700 & 5.5                  &   & 2700 & 5.5                  &   & 2700 & 5.5                   \\
$H$ (1.40--1.80~\micron)   & 2900 & 5.5\tablenotemark{b} &   & 2800 & 6.0\tablenotemark{b} &   & 2800 & 6.0\tablenotemark{b} &   & 2800 & 6.0\tablenotemark{b}  \\
$K$ (1.90--2.40~\micron)   & 2900 & 5.5\tablenotemark{b} &   & 2900 & 5.5                  &   & 2800 & 5.5                  &   & 2800 & 6.0\tablenotemark{b}  \\
\cline{1-12}                                                                                                                                                    
                                                                                                                                                               
\multicolumn{12}{c}{} \\                                                                                                                                        
\multicolumn{12}{c}{\bf 2MASS~J2206$-$2047} \\                                                                               
\cline{1-12}                                                                                                                                                    
{\bf NIR (0.95--2.42~\micron)\tablenotemark{a}}   & {\bf 2900} & {\bf 5.0\tablenotemark{c}} &   & {\bf 2900} & {\bf 6.0\tablenotemark{b}} &   & {\bf 2900} & {\bf 5.5}                  &   & {\bf 2900} & {\bf 5.5}                   \\
All (0.81--2.42~\micron)   & 2800 & 4.5                  &   & 2800 & 6.0\tablenotemark{b} &   & 2700 & 5.0                  &   & 2800 & 5.5                   \\
$Y$ (0.95--1.12~\micron)   & 2900 & 5.0                  &   & 2900 & 6.0\tablenotemark{b} &   & 2800 & 5.0                  &   & 2700 & 5.5                   \\
$J$ (1.10--1.34~\micron)   & 2900 & 5.5\tablenotemark{b} &   & 2800 & 5.5                  &   & 2800 & 5.5                  &   & 3000 & 6.0\tablenotemark{b}  \\
$H$ (1.40--1.80~\micron)   & 2900 & 5.5\tablenotemark{b} &   & 2900 & 6.0\tablenotemark{b} &   & 2900 & 5.5                  &   & 2900 & 6.0\tablenotemark{b}  \\
$K$ (1.90--2.40~\micron)   & 3000 & 5.5\tablenotemark{b} &   & 3000 & 5.5                  &   & 2900 & 5.5                  &   & 2900 & 6.0\tablenotemark{b}  \\

\enddata

\tablenotetext{a}{The NIR results are our preferred choice for model
  atmosphere fitting, as discussed in Section~\ref{sec:atm-fit}.}

\tablenotetext{b}{The best-fit value is at the edge of the model
  grid.}

\tablenotetext{c}{In these rare cases ($\approx$3\% of all the
  tabulated results), a significant $\chi^2$ minimum developed that
  was very discrepant from the prevailing best-fit models at
  $\approx$2700--2900~K.  Thus, we selected the second best fitting
  model rather than: 2000~K/6.0 for \glint\ Drift-PHOENIX; 3300~K/3.5
  for 2MASS~J2206$-$2047 PHOENIX-Gaia.}

\end{deluxetable}

\clearpage
\begin{deluxetable}{lccc}
\tabletypesize{\scriptsize}
\tablewidth{0pt}
\tablecaption{Evolutionary Model-derived Properties of \lpAB \label{tbl:lpmodel}}
\tablehead{
\colhead{Property}    &
\colhead{Median}      &
\colhead{68.3\% c.l.} &
\colhead{95.4\% c.l.} }
\startdata
\multicolumn{4}{c}{\bf Tucson models \citep{1997ApJ...491..856B}} \\
\cline{1-4}
\multicolumn{4}{c}{\lpAB\ system} \\
\cline{1-4}
Age (Gyr)                     & 0.141   & $-  0.019, 0.023  $ & $-   0.04, 0.05   $ \\
$q$ ($M_{\rm B}/M_{\rm A}$)   & 0.863   & $-  0.019, 0.013  $ & $-   0.04, 0.03   $ \\
$\Delta$\Teff\ (K)            & 137     & $-     15, 21     $ & $-     40, 50     $ \\
\cline{1-4}

\multicolumn{4}{c}{\lpA} \\
\cline{1-4}
$M_{\rm A}$ (\Msun)           & 0.065   & $-  0.004, 0.004  $ & $-  0.008, 0.008  $ \\
$T_{\rm eff,A}$ (K)           & 2780    & $-     30, 30     $ & $-     50, 50     $ \\
$\log(g_{\rm A})$ (cgs)       & 5.02    & $-   0.04, 0.04   $ & $-   0.08, 0.08   $ \\
$R_{\rm A}$ (\Rsun)           & 0.130   & $-  0.003, 0.003  $ & $-  0.005, 0.006  $ \\
Li$_{\rm A}$/Li$_0$\tablenotemark{a}           & 0.75    & $-   0.52, 0.15   $ & $-   0.74, 0.22   $ \\
\cline{1-4}

\multicolumn{4}{c}{\lpB} \\
\cline{1-4}
$M_{\rm B}$ (\Msun)           & 0.056   & $-  0.003, 0.004  $ & $-  0.006, 0.007  $ \\
$T_{\rm eff,B}$ (K)           & 2640    & $-     30, 30     $ & $-     60, 60     $ \\
$\log(g_{\rm B})$ (cgs)       & 5.00    & $-   0.04, 0.04   $ & $-   0.08, 0.08   $ \\
$R_{\rm B}$ (\Rsun)           & 0.124   & $-  0.003, 0.003  $ & $-  0.005, 0.005  $ \\
Li$_{\rm B}$/Li$_0$\tablenotemark{a}           & 0.96    & $-   0.08, 0.03   $ & $-   0.28, 0.04   $ \\
\cline{1-4}

\multicolumn{4}{c}{} \\
\multicolumn{4}{c}{\bf Lyon models \citep[Dusty;][]{2000ApJ...542..464C}} \\
\cline{1-4}
\multicolumn{4}{c}{\lpAB\ system} \\
\cline{1-4}
Age (Gyr)\tablenotemark{a}    & 0.127   & $-  0.017, 0.021  $ & $-   0.03, 0.04   $ \\
$q$ ($M_{\rm B}/M_{\rm A}$)   & 0.872   & $-  0.018, 0.014  $ & $-   0.04, 0.03   $ \\
$\Delta$\Teff\ (K)            & 134     & $-     15, 20     $ & $-     40, 50     $ \\

\cline{1-4}
\multicolumn{4}{c}{\lpA} \\
\cline{1-4}
$M_{\rm A}$ (\Msun)           & 0.064   & $-  0.004, 0.005  $ & $-  0.007, 0.009  $ \\
$T_{\rm eff,A}$ (K)           & 2660    & $-     30, 30     $ & $-     50, 50     $ \\
$\log(g_{\rm A})$ (cgs)       & 4.93    & $-   0.04, 0.04   $ & $-   0.08, 0.09   $ \\
$R_{\rm A}$ (\Rsun)           & 0.144   & $-  0.003, 0.003  $ & $-  0.006, 0.006  $ \\
Li$_{\rm A}$/Li$_0$\tablenotemark{a}           & 0.68    & $-   0.42, 0.24   $ & $-   0.64, 0.30   $ \\
$(J-K)_{\rm A}$ (mag)         & 0.795   & $-  0.002, 0.002  $ & $-  0.003, 0.004  $ \\
$(H-K)_{\rm A}$ (mag)         & 0.375   & $-  0.006, 0.005  $ & $-  0.011, 0.011  $ \\
$(J-H)_{\rm A}$ (mag)         & 0.420   & $-  0.005, 0.004  $ & $-  0.008, 0.008  $ \\
$(K-\Lp)_{\rm A}$ (mag)       & 0.544   & $-  0.009, 0.008  $ & $-  0.017, 0.016  $ \\
\cline{1-4}

\multicolumn{4}{c}{\lpB} \\
\cline{1-4}
$M_{\rm B}$ (\Msun)           & 0.056   & $-  0.004, 0.004  $ & $-  0.007, 0.007  $ \\
$T_{\rm eff,B}$ (K)           & 2520    & $-     30, 30     $ & $-     60, 60     $ \\
$\log(g_{\rm B})$ (cgs)       & 4.91    & $-   0.04, 0.04   $ & $-   0.08, 0.08   $ \\
$R_{\rm B}$ (\Rsun)           & 0.137   & $-  0.003, 0.003  $ & $-  0.005, 0.006  $ \\
Li$_{\rm B}$/Li$_0$\tablenotemark{a}           & 0.95    & $-   0.18, 0.04   $ & $-   0.54, 0.05   $ \\
$(J-K)_{\rm B}$ (mag)         & 0.810   & $-  0.005, 0.006  $ & $-  0.008, 0.013  $ \\
$(H-K)_{\rm B}$ (mag)         & 0.396   & $-  0.009, 0.006  $ & $-  0.013, 0.013  $ \\
$(J-H)_{\rm B}$ (mag)         & 0.415   & $-  0.004, 0.005  $ & $-  0.007, 0.007  $ \\
$(K-\Lp)_{\rm B}$ (mag)       & 0.594   & $-  0.011, 0.012  $ & $-  0.021, 0.025  $ \\
\enddata

\tablenotetext{a}{Abundance of lithium relative to the initial amount
    (Li$_0$).}

\end{deluxetable}

\clearpage
\begin{deluxetable}{lccc}
\tabletypesize{\scriptsize}
\tablewidth{0pt}
\tablecaption{Evolutionary Model-derived Properties of \lhsAB \label{tbl:lhsmodel}}
\tablehead{
\colhead{Property}    &
\colhead{Median}      &
\colhead{68.3\% c.l.} &
\colhead{95.4\% c.l.} }
\startdata
\multicolumn{4}{c}{\bf Tucson models \citep{1997ApJ...491..856B}} \\
\cline{1-4}
\multicolumn{4}{c}{\lhsAB\ system} \\
\cline{1-4}
Age (Gyr)\tablenotemark{a}    & 0.37    & $-  0.15, 9.63  $ & $-  0.21, 9.63  $ \\
$q$ ($M_{\rm B}/M_{\rm A}$)   & 0.966   & $- 0.016, 0.011 $ & $-  0.04, 0.02  $ \\
$\Delta$\Teff\ (K)            & 35      & $-    8, 10     $ & $-     24, 32     $ \\

\cline{1-4}
\multicolumn{4}{c}{\lhsA} \\
\cline{1-4}
$M_{\rm A}$ (\Msun)           & 0.098   & $-  0.010, 0.004  $ & $-  0.018, 0.008  $ \\
$T_{\rm eff,A}$ (K)           & 2960    & $-     30, 30     $ & $-     70, 60     $ \\
$\log(g_{\rm A})$ (cgs)       & 5.22    & $-   0.07, 0.03   $ & $-   0.14, 0.04   $ \\
$R_{\rm A}$ (\Rsun)           & 0.128   & $-  0.004, 0.004  $ & $-  0.007, 0.009  $ \\

\cline{1-4}
\multicolumn{4}{c}{\lhsB} \\
\cline{1-4}
$M_{\rm B}$ (\Msun)           & 0.095   & $-  0.011, 0.005  $ & $-  0.019, 0.007  $ \\
$T_{\rm eff,B}$ (K)           & 2930    & $-     40, 30     $ & $-     80, 60     $ \\
$\log(g_{\rm B})$ (cgs)       & 5.23    & $-   0.08, 0.03   $ & $-   0.15, 0.04   $ \\
$R_{\rm B}$ (\Rsun)           & 0.124   & $-  0.003, 0.004  $ & $-  0.006, 0.010  $ \\

\cline{1-4}
\multicolumn{4}{c}{} \\
\multicolumn{4}{c}{\bf Lyon models \citep[Dusty;][]{2000ApJ...542..464C}} \\
\cline{1-4}
\multicolumn{4}{c}{\lhsAB\ system} \\
\cline{1-4}
Age (Gyr)\tablenotemark{a}    & 0.28    & $-  0.08, 9.72  $ & $-  0.13, 9.72  $ \\
$q$ ($M_{\rm B}/M_{\rm A}$)   & 0.958   & $-  0.014, 0.015  $ & $-  0.04, 0.03  $ \\
$\Delta$\Teff\ (K)            & 39      & $-      9, 10     $ & $-     27, 35     $ \\

\cline{1-4}
\multicolumn{4}{c}{\lhsA} \\
\cline{1-4}
$M_{\rm A}$ (\Msun)           & 0.099   & $-  0.011, 0.008  $ & $-  0.020, 0.013  $ \\
$T_{\rm eff,A}$ (K)           & 2860    & $-     50, 50     $ & $-     90, 80     $ \\
$\log(g_{\rm A})$ (cgs)       & 5.15    & $-   0.07, 0.07   $ & $-   0.15, 0.08   $ \\
$R_{\rm A}$ (\Rsun)           & 0.138   & $-  0.005, 0.006  $ & $-  0.008, 0.012  $ \\
$(J-K)_{\rm A}$ (mag)         & 0.787   & $-  0.001, 0.001  $ & $-  0.002, 0.002  $ \\
$(H-K)_{\rm A}$ (mag)         & 0.336   & $-  0.009, 0.009  $ & $-  0.013, 0.018  $ \\
$(J-H)_{\rm A}$ (mag)         & 0.451   & $-  0.009, 0.008  $ & $-  0.017, 0.010  $ \\

\cline{1-4}
\multicolumn{4}{c}{\lhsB} \\
\cline{1-4}
$M_{\rm B}$ (\Msun)           & 0.094   & $-  0.010, 0.010  $ & $-  0.019, 0.014  $ \\
$T_{\rm eff,B}$ (K)           & 2820    & $-     40, 50     $ & $-     90, 90     $ \\
$\log(g_{\rm B})$ (cgs)       & 5.15    & $-   0.07, 0.08   $ & $-   0.15, 0.09   $ \\
$R_{\rm B}$ (\Rsun)           & 0.135   & $-  0.005, 0.006  $ & $-  0.008, 0.012  $ \\
$(J-K)_{\rm B}$ (mag)         & 0.788   & $-  0.001, 0.001  $ & $-  0.002, 0.002  $ \\
$(H-K)_{\rm B}$ (mag)         & 0.341   & $-  0.010, 0.009  $ & $-  0.014, 0.019  $ \\
$(J-H)_{\rm B}$ (mag)         & 0.447   & $-  0.008, 0.009  $ & $-  0.016, 0.012  $ \\

\enddata

\tablenotetext{a}{Both sets of evolutionary models are only computed
  up to an age of 10~Gyr; therefore, this defines the upper limit on
  the model-derived ages.}

\end{deluxetable}

\clearpage
\begin{deluxetable}{lccc}
\tabletypesize{\scriptsize}
\tablewidth{0pt}
\tablecaption{Evolutionary Model-derived Properties of Gl~569Bab \label{tbl:glmodel}}
\tablehead{
\colhead{Property}    &
\colhead{Median}      &
\colhead{68.3\% c.l.} &
\colhead{95.4\% c.l.} }
\startdata
\multicolumn{4}{c}{\bf Tucson models \citep{1997ApJ...491..856B}} \\
\cline{1-4}
\multicolumn{4}{c}{Gl~569Bab system} \\
\cline{1-4}
Age (Gyr)                     & 0.51    & $-   0.08, 0.13   $ & $-   0.14, 0.38   $ \\
$q$ ($M_{\rm Bb}/M_{\rm Ba}$)   & 0.886   & $-  0.017, 0.021  $ & $-   0.03, 0.05   $ \\
$\Delta$\Teff\ (K)            & 229     & $-     19, 19     $ & $-     37, 39     $ \\

\cline{1-4}
\multicolumn{4}{c}{Gl~569Ba} \\
\cline{1-4}
$M_{\rm Ba}$ (\Msun)           & 0.074   & $-  0.004, 0.004  $ & $-  0.007, 0.007  $ \\
$T_{\rm eff,Ba}$ (K)           & 2530    & $-     30, 30     $ & $-     50, 50     $ \\
$\log(g_{\rm Ba})$ (cgs)       & 5.30    & $-   0.04, 0.03   $ & $-   0.07, 0.06   $ \\
$R_{\rm Ba}$ (\Rsun)           & 0.101   & $-  0.002, 0.002  $ & $-  0.003, 0.003  $ \\
Li$_{\rm Ba}$/Li$_0$\tablenotemark{a}           & 0.00    & $-   0.00, 0.00   $ & $-   0.00, 0.00   $ \\

\cline{1-4}
\multicolumn{4}{c}{Gl~569Bb} \\
\cline{1-4}
$M_{\rm Bb}$ (\Msun)           & 0.066   & $-  0.004, 0.005  $ & $-  0.008, 0.010  $ \\
$T_{\rm eff,Bb}$ (K)           & 2300    & $-     30, 30     $ & $-     60, 60     $ \\
$\log(g_{\rm Bb})$ (cgs)       & 5.28    & $-   0.05, 0.05   $ & $-   0.09, 0.10   $ \\
$R_{\rm Bb}$ (\Rsun)           & 0.097   & $-  0.002, 0.002  $ & $-  0.004, 0.004  $ \\
Li$_{\rm Bb}$/Li$_0$\tablenotemark{a}           & 0.02    & $-   0.02, 0.49   $ & $-   0.02, 0.82   $ \\

\cline{1-4}
\multicolumn{4}{c}{} \\
\multicolumn{4}{c}{\bf Lyon models \citep[Dusty;][]{2000ApJ...542..464C}} \\
\cline{1-4}
\multicolumn{4}{c}{Gl~569Bab system} \\
\cline{1-4}
Age (Gyr)                     & 0.46    & $-   0.07, 0.11   $ & $-   0.13, 0.33   $ \\
$q$ ($M_{\rm Bb}/M_{\rm Ba}$)   & 0.866   & $-  0.014, 0.019  $ & $-   0.03, 0.05   $ \\
$\Delta$\Teff\ (K)            & 221     & $-     18, 19     $ & $-     36, 38     $ \\

\cline{1-4}
\multicolumn{4}{c}{Gl~569Ba} \\
\cline{1-4}
$M_{\rm Ba}$ (\Msun)           & 0.075   & $-  0.004, 0.004  $ & $-  0.008, 0.008  $ \\
$T_{\rm eff,Ba}$ (K)           & 2430    & $-     30, 30     $ & $-     50, 50     $ \\
$\log(g_{\rm Ba})$ (cgs)       & 5.23    & $-   0.04, 0.04   $ & $-   0.08, 0.06   $ \\
$R_{\rm Ba}$ (\Rsun)           & 0.110   & $-  0.002, 0.002  $ & $-  0.003, 0.005  $ \\
Li$_{\rm Ba}$/Li$_0$\tablenotemark{a}           & 0.00    & $-   0.00, 0.00   $ & $-   0.00, 0.00   $ \\
$(J-K)_{\rm Ba}$ (mag)         & 0.827   & $-  0.007, 0.010  $ & $-  0.013, 0.021  $ \\
$(H-K)_{\rm Ba}$ (mag)         & 0.371   & $-  0.005, 0.007  $ & $-  0.009, 0.018  $ \\
$(J-H)_{\rm Ba}$ (mag)         & 0.456   & $-  0.004, 0.004  $ & $-  0.008, 0.009  $ \\
$(K-\Lp)_{\rm Ba}$ (mag)       & 0.660   & $-  0.014, 0.014  $ & $-   0.03, 0.03   $ \\

\cline{1-4}
\multicolumn{4}{c}{Gl~569Bb} \\
\cline{1-4}
$M_{\rm Bb}$ (\Msun)           & 0.065   & $-  0.004, 0.005  $ & $-  0.008, 0.010  $ \\
$T_{\rm eff,Bb}$ (K)           & 2210    & $-     30, 30     $ & $-     60, 60     $ \\
$\log(g_{\rm Bb})$ (cgs)       & 5.20    & $-   0.05, 0.05   $ & $-   0.09, 0.11   $ \\
$R_{\rm Bb}$ (\Rsun)           & 0.106   & $-  0.003, 0.002  $ & $-  0.005, 0.004  $ \\
Li$_{\rm Bb}$/Li$_0$\tablenotemark{a}           & 0.00    & $-   0.00, 0.02   $ & $-   0.00, 0.43   $ \\
$(J-K)_{\rm Bb}$ (mag)         & 0.96    & $-   0.03, 0.03   $ & $-   0.05, 0.06   $ \\
$(H-K)_{\rm Bb}$ (mag)         & 0.432   & $-  0.017, 0.014  $ & $-   0.03, 0.04   $ \\
$(J-H)_{\rm Bb}$ (mag)         & 0.525   & $-  0.011, 0.013  $ & $-   0.02, 0.03   $ \\
$(K-\Lp)_{\rm Bb}$ (mag)       & 0.786   & $-  0.016, 0.015  $ & $-   0.03, 0.03   $ \\

\enddata

\tablenotetext{a}{Abundance of lithium relative to the initial amount
  (Li$_0$).}

\end{deluxetable}

\end{document}